%% file: locuss_WL-20160610.tex
\documentclass[useAMS,usenatbib,usegraphicx]{mn2e}

\usepackage{color,ulem,natbib,aas_macros,amsmath}
\title[Weak-lensing mass calibration of galaxy clusters]
      {LoCuSS: Weak-lensing mass calibration of galaxy clusters}
\author[Okabe and Smith]
       {Nobuhiro Okabe,$^{1,2,3}$\thanks{E-mail:
      okabe@hiroshima-u.ac.jp} and
        Graham P.\ Smith$^4$\thanks{E-mail: gps@star.sr.bham.ac.uk}\\\\
        $^1$ Department of Physical Science, Hiroshima University, 1-3-1
        Kagamiyama, Higashi-Hiroshima, Hiroshima 739-8526, Japan\\
        $^2$ Hiroshima Astrophysical Science Center, Hiroshima University, Higashi-Hiroshima, Kagamiyama 1-3-1, 739-8526, Japan\\
        $^3$ Kavli Institute for the Physics and Mathematics of the Universe (WPI), Todai Institutes for Advanced Study, University of Tokyo, 5-1-5 Kashiwanoha, Kashiwa, Chiba 277-8583, Japan \\
        $^4$ School of Physics and Astronomy, University of Birmingham,
      Birmingham, B15 2TT, England.
}

\begin{document}

\date{\today}

\pagerange{\pageref{firstpage}--\pageref{lastpage}} \pubyear{2015}

\maketitle

\label{firstpage}

\newcommand{\simgt}{\lower.5ex\hbox{$\; \buildrel > \over \sim \;$}}
\newcommand{\simlt}{\lower.5ex\hbox{$\; \buildrel < \over \sim \;$}}
\newcommand{\tp}{\hspace{-1mm}+\hspace{-1mm}}
\newcommand{\tm}{\hspace{-1mm}-\hspace{-1mm}}
\newcommand{\gIII}{I\hspace{-.3mm}I\hspace{-.3mm}I}
\newcommand{\bmf}[1]{\mbox{\boldmath$#1$}}
\def\bbeta{\mbox{\boldmath $\beta$}}
\def\btheta{\mbox{\boldmath $\theta$}}
\def\bnabla{\mbox{\boldmath $\nabla$}}
\def\bk{\mbox{\boldmath $k$}}
\newcommand{\cA}{{\cal A}}
\newcommand{\cD}{{\cal D}}
\newcommand{\cF}{{\cal F}}
\newcommand{\cG}{{\cal G}}
\newcommand{\trQ}{{\rm tr}Q}
\newcommand{\Real}[1]{{\rm Re}\left[ #1 \right]}
\newcommand{\paren}[1]{\left( #1 \right)}
\newcommand{\red}{\textcolor{red}}
\newcommand{\blue}{\textcolor{blue}}
\newcommand{\norv}[1]{{\textcolor{red}{#1}}}
\newcommand{\norvnew}[1]{{\textcolor{blue}{#1}}}

\def\hkpc{\mathrel{h^{-1}{\rm kpc}}}
\def\hMpc{\mathrel{h^{-1}{\rm Mpc}}}
\def\Mpc{\mathrel{\rm Mpc}}
\def\Mvir{\mathrel{M_{\rm vir}}}
\def\cvir{\mathrel{c_{\rm vir}}}
\def\rvir{\mathrel{r_{\rm vir}}}
\def\Dvir{\mathrel{\Delta_{\rm vir}}}
\def\rsc{\mathrel{r_{\rm sc}}}
\def\rhoc{\mathrel{\rho_{\rm crit}}}
\def\Msol{\mathrel{M_\odot}}
\def\hMsol{\mathrel{h^{-1}M_\odot}}
\def\h70Msol{\mathrel{h_{70}^{-1}M_\odot}}
\def\ergs{\mathrel{\rm erg\,s^{-1}}}
\def\Om{\mathrel{\Omega_{\rm M}}}
\def\Ol{\mathrel{\Omega_\Lambda}}
\def\keV{\mathrel{\rm keV}}
\def\kpc{\mathrel{\rm kpc}}
\def\ls{\mathrel{\hbox{\rlap{\hbox{\lower4pt\hbox{$\sim$}}}\hbox{$<$}}}}
\def\gs{\mathrel{\hbox{\rlap{\hbox{\lower4pt\hbox{$\sim$}}}\hbox{$>$}}}}
\def\ds{\mathrel{D_{\rm s}}}
\def\dls{\mathrel{D_{\rm ls}}}
\def\dsi{\mathrel{D_{{\rm s},i}}}
\def\dlsi{\mathrel{D_{{\rm ls},i}}}
\def\dsj{\mathrel{D_{{\rm s},j}}}
\def\dlsj{\mathrel{D_{{\rm ls},j}}}
\def\dsi{\mathrel{D_{{\rm s},i}}}
\def\zs{\mathrel{z_{s}}}

\addtolength\topmargin{-12mm}

\begin{abstract}
We present weak-lensing mass measurements of 50 X-ray luminous galaxy
clusters at $0.15\le z\le0.3$, based on uniform high quality
observations with Suprime-Cam mounted on the 8.2-m Subaru telescope.
We pay close attention to possible systematic biases, aiming to
control them at the $\ls4$ per cent level.  The dominant source of
systematic bias in weak-lensing measurements of the mass of individual
galaxy clusters is contamination of background galaxy catalogues by
faint cluster and foreground galaxies.  We extend our conservative
method for selecting background galaxies with $(V-i')$ colours redder
than the red sequence of cluster members to use a colour-cut that
depends on cluster-centric radius.  This allows us to define
background galaxy samples that suffer $\le1$ per cent contamination,
and comprise $13$ galaxies per square arcminute.  Thanks to the purity
of our background galaxy catalogue, the largest systematic that
  we identify in our analysis is a shape measurement bias of $3$ per
cent, that we measure using simulations that probe weak shears
upto $g=0.3$.  Our individual cluster mass and concentration
measurements are in excellent agreement with predictions of the
mass-concentration relation.  Equally, our stacked shear profile is in
excellent agreement with the Navarro Frenk and White profile.  Our new
LoCuSS mass measurements are consistent with the CCCP and CLASH
surveys, and in tension with the Weighing the Giants at
$\sim1-2\sigma$ significance.  Overall, the consensus at
  $z\le0.3$ that is emerging from these complementary surveys
  represents important progress for cluster mass calibration, and
  augurs well for cluster cosmology.
\end{abstract}


\makeatletter

\def\doi{\begingroup
  % The following isn't just \dospecials, because that includes \ , \{, and \}
  \let\do\@makeother \do\\\do\$\do\&\do\#\do\^\do\_\do\%\do\~
  \@ifnextchar[%]
    {\@doi}
    {\@doi[]}}
\def\@doi[#1]#2{%
  \def\@tempa{#1}%
  \ifx\@tempa\@empty
    \href{http://dx.doi.org/#2}{doi:#2}%
  \else
    \href{http://dx.doi.org/#2}{#1}%
  \fi
  \endgroup
}

%
\def\eprint#1#2{%
  \@eprint#1:#2::\@nil}
\def\@eprint@arXiv#1{\href{http://arxiv.org/abs/#1}{{\tt arXiv:#1}}}
\def\@eprint@dblp#1{\href{http://dblp.uni-trier.de/rec/bibtex/#1.xml}{dblp:#1}}
\def\@eprint#1:#2:#3:#4\@nil{%
  \def\@tempa{#1}%
  \def\@tempb{#2}%
  \def\@tempc{#3}%
  \ifx\@tempc\@empty
    \let\@tempc\@tempb
    \let\@tempb\@tempa
  \fi
  \ifx\@tempb\@empty
    \def\@tempb{arXiv}%
  \fi
  \@ifundefined{@eprint@\@tempb}
    {\@tempb:\@tempc}
    {
      \expandafter\expandafter\csname @eprint@\@tempb\endcsname\expandafter{\@tempc}}%
}

%
\def\mniiiauthor#1#2#3{%
  \@ifundefined{mniiiauth@#1}
    {\global\expandafter\let\csname mniiiauth@#1\endcsname\null #2}
    {#3}}

\makeatother

\section{Introduction}

Accurate measurements of the mass and internal structure of dark
matter halos that host galaxy clusters and groups are central to a
broad range of fundamental research spanning cosmological parameters,
the nature of dark matter, the spectrum of primordial density
fluctuations, testing gravity theory, and the formation/evolution of
galaxies and the intergalactic medium. The requirement for accuracy is
most stringent for studies that aim to probe dark energy, e.g. via
evolution of the cluster mass function \citep[e.g.][]{Vikhlinin09b,Allen11}.
Upcoming surveys will discover $\sim10^5$ clusters and intend to infer
the mass of the majority of these systems from scaling relations
between mass and the observable properties of clusters
\citep[e.g.][]{Pillepich12,Sartoris15}.  Notwithstanding the forecast
accuracy and precision of other cosmological probes, the sheer number
of clusters upon which future cosmological results will rely implies
that per cent level control of systematic biases in the ensemble mass
calibration of clusters will ultimately be required.

The challenge of calibrating systematic biases in the ensemble
  galaxy cluster mass calibration at this level of accuracy is
amplified by the fact that the normalization of the calibration is
necessary but not sufficient for accurate cluster cosmology.  Poorly
constrained knowledge of the intrinsic scatter between observable mass
proxies (including all ``masses'' measured from data) and the
underlying mass of dark matter halos that host galaxy clusters is a
source of bias in cluster-based cosmological constraints
\citep[e.g.][]{Smith03}.  Intrinsic scatter between the relevant
observable properties of clusters and between mass measurements and
underlying halo mass is therefore a key parameter that many studies
attempt to constrain \citep[e.g.][]{Okabe10c,Okabe14b, Becker11,
  Bahe12, Marrone12, Mahdavi13, Sifon13, Mulroy14, Rozo15,
  Saliwanchik15}.  Cluster mass measurement methods used for
calibration studies must therefore permit measurements of
individual cluster masses in order to characterise the full
distribution of cluster mass around the mean relation between mass and
observable mass proxy.  Moreover, whilst stacked mass measurements are
powerful probes of the population mean, they offer no useful
constraints on the scatter around the mean.

An increasing number of galaxy cluster mass calibration studies use
weak gravitational lensing measurements to constrain cluster masses
\citep[e.g.][]{Smith05, Bardeau07, Okabe10b,
Okabe10a,Okabe11,Okabe13,Okabe14a,Okabe15b,Okabe15a,Hoekstra12,
    Applegate14, Umetsu14, Hoekstra15}.  This is because
interpretation of the gravitational lensing signal does not require
assumptions about the physical nature or state of the gravitating mass
of the cluster.  Therefore, despite the fact that individual cluster
mass measurements can suffer appreciable biases that correlate with
the observer's viewing angle through asymmetric cluster mass
distributions \citep[e.g.][]{Corless07,Meneghetti10}, gravitational
lensing can yield an accurate mean mass calibration of galaxy
clusters, supported by knowledge of the scatter between true halo
  mass and weak-lensing mass measurements \citep{Becker11,Bahe12}.

The largest samples of clusters for which weak-lensing observations
are available are currently drawn from large-scale X-ray surveys and
number of order 50 clusters.  These surveys are the Local Cluster
Substructure Survey (LoCuSS; \citealt{Okabe13, Martino14}; Smith et
al. in prep.), the Canadian Cluster Cosmology Project (CCCP;
\citealt{Mahdavi13, Hoekstra15}), and the Weighing the Giants
programme (WtG; \citealt{vonderLinden14, Applegate14}).  In the
parlance of the Dark Energy Task Force, these are Stage II studies
that examine the systematic uncertainties inherent in using galaxy
clusters as probes of Dark Energy \citep{Albrecht06}.  The LoCuSS
sample is an $L_X$-limited sub-set of clusters from {\it ROSAT}
All-sky Survey (RASS) at $0.15<z<0.3$; the CCCP sample is a mixture of
X-ray luminous clusters for which optical data are available from the
CFHT archive and a temperature-selected sub-set of clusters from the
{\it ASCA} survey spanning $0.15<z<0.55$; the WtG sample is a
representative flux-limited sub-set of the RASS clusters at
$0.15<z<0.7$.  Smaller, generally heterogeneous, samples of X-ray
clusters are also studied, for example, by the the Cluster Lensing And
Supernova Survey with Hubble (CLASH; \citealt{Postman12,Umetsu14}) and
the 400SD surveys \citep{Israel12}.  Whilst samples of Sunyaev
Zeldovich (SZ) Effect detected clusters are growing rapidly, the
weak-lensing studies of SZ samples currently number handfuls of
clusters \citep[e.g.][]{High12,Gruen14}.  

 Currently, the target accuracy on controlling systematic biases
  in the ensemble cluster mass calibration is therefore set by the
  size of the LoCuSS, CCCP and WtG samples, the typical statistical
  measurement error of a weak-lensing mass measurement of an
  individual cluster, and the intrinsic scatter of weak-lensing masses
  around the true underlying halo masses.  Given that our sample is
  not mass-selected, the intrinsic scatter on $M_{\rm WL}-M_{\rm
    true}$ for our sample is not known a priori.  In setting a nominal
  goal for control of systematic biases, we therefore ignore the
  intrinsic scatter and simply adopt a typical statistical measurement
  error of $30$ per cent as the uncertainty on the ensemble cluster
  mass calibration that would be achieved from studying one cluster.
  This motivates a goal of $\sim30/\sqrt{50}=4$ per cent for control
  of systematic biases.  Our goal in this article is to achieve this
  level of accuracy for the LoCuSS galaxy cluster mass calibration.
  Note that, by ignoring the intrinsic scatter in $M_{\rm WL}-M_{\rm
    true}$, this goal is more challenging than is justified by the
  statistics of cluster mass measurement discussed above.

The principal systematic biases that can affect an weak-lensing
cluster mass measurement relate to (1) the measurement of faint galaxy
shapes, (2) accurate placement of faint galaxies along the line of
sight such that the sample of background galaxies suffers negligible
contamination by faint cluster members and that the inferred redshift
distribution of the background galaxies is accurate, and (3) modelling
of the shear signal in order to infer the cluster mass.  In the brief
review of these biases that follows, a key theme is that the approach
taken to addressing one source of bias can have consequences for how
well other biases are controlled.  We also briefly outline our
approach to these biases -- a unifying theme of which is to
minimize the number of strong assumptions in our analysis.  The
summary that follows intends to help non-experts to understand some of
the more technical aspects of this article.

It is common to calibrate faint galaxy shape measurement codes on the
STEP and STEP2 simulations \citep{Heymans06,Massey07}, however the
gravitational shear signal of clusters typically exceeds the shear
signals injected into these simulations, and therefore these tests are
only relevant to the cluster outskirts.  For example, WtG calibrate
their shape measurement code on the STEP2 simulations, which in part
motivates them to restrict the range of the WtG shear profiles to
projected clustercentric radii of $0.75-1.5h_{70}^{-1}\Mpc$ --
i.e.\ attempting to avoid regions of the clusters at which the
measured shear exceeds that injected into the STEP simulations
\citep{Applegate14}.  \citet{Hoekstra15} recently emphasised the
importance of carefully matching the properties of the simulated data
used for such tests to the observational data.  In this article we
further develop the shape measurement methods that we developed in
\citet{Okabe13} and extend our tests of these methods to include
realistic galaxies.  As in \citet{Okabe13}, we test our code on shear
values upto $g=0.3$, i.e.\ appropriate to the full range of cluster
centric radii relevant to weak-lensing -- clustercentric radii as
small as $\sim200\hkpc$.

Contamination of background galaxy samples by unlensed faint cluster
galaxies dilutes the measured lensing signal and causes a systematic
underestimate of the shear \citep[e.g.][]{Broadhurst05,
    Limousin07a1689}.  It is therefore of prime importance to make a
secure selection of background galaxies.  The number density of
cluster members is a declining function of clustercentric radius, and
thus a radial trend in the number density of galaxies selected as
being in the background is interpreted as evidence for contamination.
Whilst this is qualitatively true, the quantitative details depend on
how the gravitational magnification of the cluster modifies the
observed distribution of background galaxies.  After falling into
disuse for a decade since \citet{Kneib03} first proposed the method,
boosting the measured shear signal to correct statistically for
contamination has enjoyed a renaissance of late
\citep[e.g.][]{Applegate14,Hoekstra15}.  This method is applied to
both red and blue galaxies, either by excluding the red sequence
galaxies or simply selecting faint galaxies, as per \citet{Kneib03}.
Due to imperfect background selection, the number density profile of
these colour-selected galaxies is found to increase at small
cluster-centric radii.  Assuming that the number density profile of a
pure background galaxy sample is independent of radius, that is
ignoring gravitational magnification, the lensing signal is corrected
as a radial function of the galaxy-count excess.  This correction
method is referred to as ``boost correction''.

We also note that photometric redshifts based on upto five photometric
bands are becoming more common as a method for selecting background
galaxies \citep{Limousin07a1689, Gavazzi09, Gruen13, Applegate14,
  McCleary15, Melchior15}.  However photometric redshifts based on a
small number of filters are problematic for galaxies with blue
observed colours because their spectral energy distribution is
relatively featureless.  This leads to the well known degeneracy
between photometric redshifts of $z\ls0.5$ and $z\gs1.5$ for blue
galaxies \citep[e.g.][]{Bolzonella00}.  This is a critical issue for
cluster weak-lensing studies that use blue galaxies \citep{Ziparo15}. 
Furthermore, the requirement for photometric
  redshift accuracy is more stringent for cluster lensing studies than
  for most other fields, because the number density of cluster
  galaxies -- that contaminate background galaxy samples -- is a
  function of clustercentric radius.

We have previously developed a method to select red background that
does not assume the radial distribution of background galaxies and
thus does not require a boost correction to the measured shear
signal \citep{Okabe13}.  Our method also yields a direct measurement
of the fraction of galaxies in the background galaxy sample that are
contaminants.  In \citet{Ziparo15} we considered how to extend
this method to include blue galaxies and concluded that additional
uncertainties of including blue galaxies do not justify the small
number of additional galaxies that we would gain.  We therefore extend
our red galaxy selection methods in this article, and achieve a
$2.6$-fold increase in number density of background galaxies over
\citeauthor{Okabe13} -- i.e.\ sufficient to measure individual cluster
masses, whilst retaining our conservative requirement that
contamination is not greater than 1 per cent.

Despite the intrinsic asphericity of galaxy clusters, it has been
shown that modeling cluster mass distributions as spherical and
following a \citet{Navarro97} profile yields mass measurements that
are accurate in the mean across a sample \citep{Becker11,Bahe12}.
These results are based on numerical dark matter only simulations,
make (well motivated) assumptions about the observational data
available to an individual study, and stress the importance of fitting
the model to the data across a well-defined radial range.  Some
observational studies implement directly the method described by
\citet{Becker11} in their analysis \citep[e.g.][]{Applegate14}.  We
prefer to test our mass modelling scheme directly on
  simulations.  Moreover, parameters that describe the shape of the
density profile (generally, a ``halo concentration parameter'') are at
the same time a nuisance parameter for the mass measurement, and a
physically interesting parameter to extract from the data.  We
  therefore let concentration be a free parameter with a flat prior,
and marginalise over concentration when measuring cluster mass.
We also use the constraints that we derive on concentration to examine
the mass-concentration relation.  Other studies adopt more restrictive
assumptions about halo concentration, in part as a consequence of
seeking to minimise contamination and shear calibration issues (see
preceding discussion) by excluding the central cluster region from
their analysis and modeling.

We describe the observations and data analysis in
Section~\ref{sec:data}, including photometry, shape measurements, and
the selection of background galaxies.  The mass measurements for
individual clusters, the mass concentration relation and stacked
lensing analysis are presented in Section~\ref{sec:results}.  We
discuss several systematics and compare with previous weak lensing
studies in Section~\ref{sec:discuss}, and summarize our conclusions in
Section~\ref{sec:summary}.  We assume $H_0=100h~{\rm
  kms^{-1}Mpc^{-1}}$, $\Omega_{m0}=0.3$ and $\Omega_{\Lambda}=0.7$
through the paper.  We occasionally use the alternative definition of
the Hubble parameter $h_{70}=H_0/70$.

\section{Data and analysis}\label{sec:data}

\subsection{Sample}\label{sec:sample}

\input{tab-sample.tex}

The sample comprises 50 clusters (Table~\ref{tab:data}) drawn from the
{\it ROSAT} All Sky Survey cluster catalogues
\citep{Ebeling98,Ebeling00,Boehringer04} that satisfy the criteria:
$-25^\circ<\delta<+65^\circ$, $n_H\le7\times10^{20}{\rm cm^{-2}}$,
$0.15\le z \le0.3$, $L_X/E(z)>4.1\times10^{44}\ergs$ where $L_X$ is in
the $0.1-2.4\keV$ band and $E(z)=\sqrt{\Om(1+z)^3+\Ol}$.  The sample
is {X-ray luminosity-limited, and therefore approximately
  mass-limited}.  Full details of the selection function are available
in Smith et al.\ (2016, in prep.).

\subsection{Observations}\label{sec:obs}

The clusters were observed with Suprime-Cam \citep{Miyazaki02} on the
8.2-m Subaru Telescope\footnote{Based in part on observations obtained
  at the Subaru Observatory under the Time Exchange program operated
  between the Gemini Observatory and the Subaru
  Observatory.}\footnote{Based in part on data collected at Subaru
  Telescope and obtained from the SMOKA, which is operated by the
  Astronomy Data Center, National Astronomical Observatory of Japan.}
on Mauna Kea.  We observed in both $V$- and $i'$-bands for 28 and 36
minutes respectively, splitting the integration times up into
individual four minute exposures.  The best overhead conditions were
reserved for the $i'$-band observations because we use these data to
measure the shapes of faint galaxies.  The full width half maximum
(FWHM) of point sources was routinely sub-arcsecond, with individual
exposures often enjoying ${\rm FWHM}\ls0.6''$.  The 50 final stacked
and reduced $i'$-band frames have median seeing of ${\rm
  FWHM}=0.71''$, with 38 of the 50 frames having ${\rm FWHM}<0.8''$
(Table~\ref{tab:data}).  Note that we use archival $g$- and $B$-band
data instead of $V$-band data for two clusters in common with
\cite{Okabe08}.  Hereafter we refer to the redder filter in which we
measure faint galaxy shapes as the $i'$-band, and the bluer filter
used for colour measurements as the $V$-band.

\subsection{Data Reduction}\label{sec:reduction}

We reduced all data using a processing pipeline based on the the
standard reduction tasks for Suprime-Cam, {\sc SDFRED}
\citep{Yagi02,Ouchi04}, and described by \cite{Okabe10b}.  The
pipeline includes bias and dark frame subtraction, flat-fielding,
instrumental distortion correction, differential refraction, point
spread function (PSF) matching, sky subtraction and stacking.  The
astrometric solution for the final stacked frames was calibrated
relative to 2MASS \citep{2MASS06} to sub-pixel root mean square (rms)
precision.  Photometric zero-points were calibrated to stellar
photometry from the Sloan Digital Sky Survey \citep[SDSS]{SDSSDR8},
taking into account foreground galactic extinction \citep{Schlafly11},
to a rms precision of $\simlt0.1{\rm mag}$.  To cross-check the
validity of the photometric calibration, we measured the redshift
dependence of the colour of early-type member galaxies, that lie on the
so-called cluster red-sequence, within $10\,{\rm arcmin}$ of each
brightest cluster galaxy (BCG).  The colour of the red sequence
increases from $(V-i')\simeq0.8$ to $(V-i')\simeq1.2$ as cluster
redshift increases from $z=0.15$ to $z=0.3$, in agreement with
\cite{SDSSDR8}.

\subsection{Shape measurement pipeline}\label{sec:shape}

We analyse the $i'$-band frames with methods introduced by
\citet[][the ``KSB$+$'' method]{KSB}, using the {\sc imcat}
  package with our modifications \citep{Okabe13, Okabe14a}.  We first
measure the image ellipticity, $e_\alpha$, from the weighted
quadrupole moments of the surface brightness of objects, and then
correct the PSF anisotropy by solving
\begin{equation} 
e'_{\alpha}(\btheta) = e_{\alpha}(\btheta) - P_{\rm sm}^{\alpha \beta}(\btheta) q^*_{\beta}(\btheta), 
\label{eq:qstar}
\end{equation}
where $P_{\alpha\beta}$ is the smear polarizablity tensor and
$q^*_{\alpha}(\btheta) = (P^*_{{\rm sm}})^{-1}_{\alpha
  \beta}e_*^{\beta}$; quantities with an asterisk denote those for
stellar objects.  The details of anisotropic PSF correction is
described in \citet[][see the Appendix]{Okabe14a}.  In brief, we
  selected bright, unsaturated stars in the half-light radius, $r_h$,
  and magnitude plane to estimate the stellar anisotropy kernel,
  $q^*_{\alpha}$.  Note that the stars and galaxies can be clearly
  discriminated using the half-light radius.  We modeled the variation
  of this kernel across sub-regions of the field of view by fitting
  second-order bi-polynomial functions to the vector $\btheta$ with
  iterative $\sigma$-clipping
  \citep[e.g.][]{Okabe08,Okabe10b,Okabe14a}. Although
  distortions at the corners of the field-of-view are larger than
  those at the centers, modelling across sub-regions is sufficiently
  flexible to correct the anisotropic PSF pattern in our data.

\begin{figure}
\includegraphics[width=\hsize]{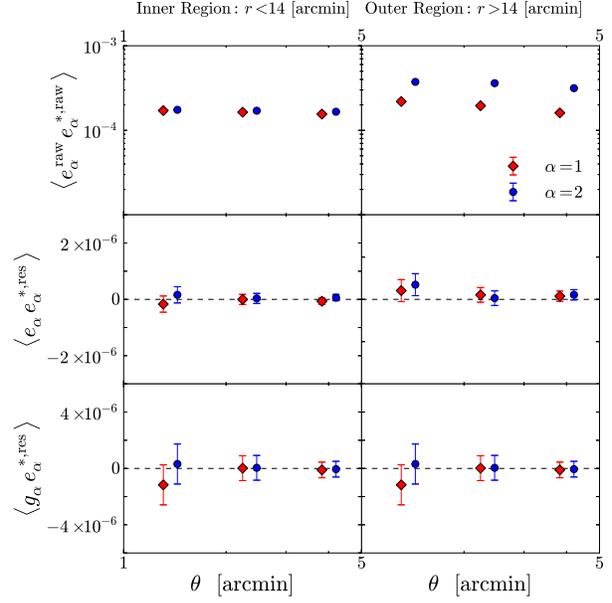}
\caption{{\sc Top} -- The cross-correlation between raw ellipticities
  for stars an galaxies before the correction.  The correlation at the
  outer region (right) is larger than that at the inner region (left).
  {\sc Middle} -- The cross-correlation between residual stellar
  ellipticities and corrected galaxy ellipticities.  {\sc Bottom} --
  The cross-correlation between residual stellar ellipticities and
  reduced ellipticities.  }
\label{fig:xi}
\end{figure}

We tested the validity of our anisotropic PSF correction by measuring
the auto-correlation function between stellar ellipticities and the
cross-correlation function between stellar and galaxy ellipticities,
before and after the correction.  We found that $\langle
e_\alpha^{*,\rm raw}e_\alpha^{*,\rm raw}\rangle$ and $\langle e_\alpha
e_\alpha^{*,\rm raw}\rangle$ before the correction show large positive
correlations ($\mathcal{O}(10^{-4})$), and that $\langle
e_\alpha^{*,\rm res}e_\alpha^{*,\rm res}\rangle$ and $\langle
e_\alpha^{\rm cor} e_\alpha^{*,\rm res}\rangle$ after the correction
are consistent with null correlation in individual cluster
  fields.  Note that $e_\alpha^{\rm cor}$ is the l.h.s. of Equation
  \ref{eq:qstar}.  In order to confirm how well the anisotropic
  PSF correction works at both the corners and centers, we divide the
  regions into the inner ($r<14'$) and outer regions ($r>14'$) of the
  fields of view with respect to the respective BCGs
  (Figure~\ref{fig:xi}).  Note that the BCGs are located close to the
  center of the field of view in all cases. The top-left and
  top-right panels show the cross-correlation function between stellar
  and galaxy ellipticities before the correction at the inner and
  outer regions, respectively.  The raw distortions at the outer
  region is indeed larger than those at the inner region.  The middle
  panel shows the resulting $\langle e_\alpha^{\rm cor}
  e_\alpha^{*,\rm res}\rangle$ after the correction, which are
  consistent with null correlation  both at $r<14'$ and $r>14'$.
 The cross correlation between the residual stellar ellipticities
  and the reduced ellipticities for galaxies which are described in
  next paragraph is shown in the bottom panel.  We again found null
  correlation at $r<14'$ and $r>14'$.  

Next, we correct the isotropic smearing effect of galaxy shapes due to
seeing and the Gaussian window function used for the shape
measurements.  The reduced ellipticity for each galaxy, $g_\alpha$, is
defined by
\begin{eqnarray}
g_\alpha&=& (P_g^{-1})_{\alpha\beta} e'_{\beta}, \label{eq:raw_g}
\end{eqnarray}
where $(P_g)_{\alpha\beta}$ is the pre-seeing shear polarizability
tensor.  The measurement of $(P_g)_{\alpha\beta}$ is very noisy for
individual faint galaxies because of its non-linearity
\citep{Bartelmann01}.  The relationship between noise and biases
  in the measurement of faint galaxy shapes, using a variety of shape
  measurement algorithms, has been considered for both cosmic shear
  and cluster lensing studies \citep[e.g.][]{Hirata04, Kacprzak12,
    Melchior12, Refregier12, Applegate14, Hoekstra15}.  The general
  feature of this relationship is that the bias in faint galaxy shape
  measurements typically increases as the size of galaxies decreases,
  i.e. as signal-to-noise ratio of galaxies decreases.  In common with
  several authors \citep[e.g.][]{Umetsu10, Oguri12, Okura12, Umetsu15}
  we have found that this dependence can be reduced significantly for
  KSB shape measurement methods if the galaxies upon which the
  isotropic PSF correction is based are limited to those detected at
  high signal-to-noise ratio \citep{Okabe13}.  We therefore calibrate
the isotropic PSF correction, using galaxies detected at very high
significance, i.e.\ a signal-to-noise ratio of $\nu>30$.  This
selection acts to suppress the measurement uncertainty of
$(P_g)_{\alpha\beta}$ that is caused by low signal-to-noise ratio in
the objects used for the isotropic PSF correction in other studies.
The polarizability tensor is first estimated by the scalar correction
approximation $(P_{g})_{\alpha\beta}=\frac{1}{2}{\rm
  tr}[P_g]\delta_{\alpha\beta}$.  We then compute the median of
$(P_{g})_{\alpha\beta}$ in $r_g$, with an adaptive grid to assemble as
uniformly as possible, where $r_g$ is the Gaussian smoothing radius
used in the KSB method.  We employ a size condition of
$r_h>\bar{r}_h^*+\sigma(r_h^*)$ and $r_g>\bar{r}_g^*+\sigma(r_g^*)$
and a positive raw $P_g$.  Here, $\bar{r}_h^*$ ($\sigma(r_h^*)$) and
$\bar{r}_g^*$ ($\sigma(r_g^*)$) are the median (rms dispersion) of
half-light radii and Gaussian smoothing radii for the stars used for
the anisotropic PSF correction described above.  Although
  galaxies and stars are well separated in $r_h$, we applied the cut
  in $r_g$ so as to exclude negative values of $P_g$ that are obtained
  for very small galaxies.  We also checked the level of stellar
  contamination that galaxy catalogues, selected based on the size
  cuts described here, might suffer.  We found that the level of
  contamination is below 1 per cent, mainly due to the fact that the
  number density of faint galaxies is an increasing function of
  apparent magnitude.  The faint galaxies therefore out far out-number
  possible stellar contaminants. 

We interpolate the polarizability tensor for individual galaxies with
$\nu>10$ as a function of $r_g$ and the absolute value of the
ellipticity, $|e|$.  Here, we used $|e|$ instead of each
  component $e_\alpha$ because the isotropic PSF correction is
  performed by the half of the trace, $\frac{1}{2}{\rm
    tr}[P_g]\delta_{\alpha\beta}$ Then, in a major departure from
\citet{Okabe13}, we applied a similar interpolation for the
signal-to-noise ratio, $\nu$.  An rms error of the ellipticity
estimates, $\sigma_g$, is estimated from 50 neighbours in the
magnitude-$r_g$ plane.  We also experimented with smaller and
  larger numbers of neighbours and found that our results are
  unchanged.

\subsection{Shape measurement tests}\label{sec:shapetest}

\begin{figure*}
\hspace{-8mm}\includegraphics[width=170mm]{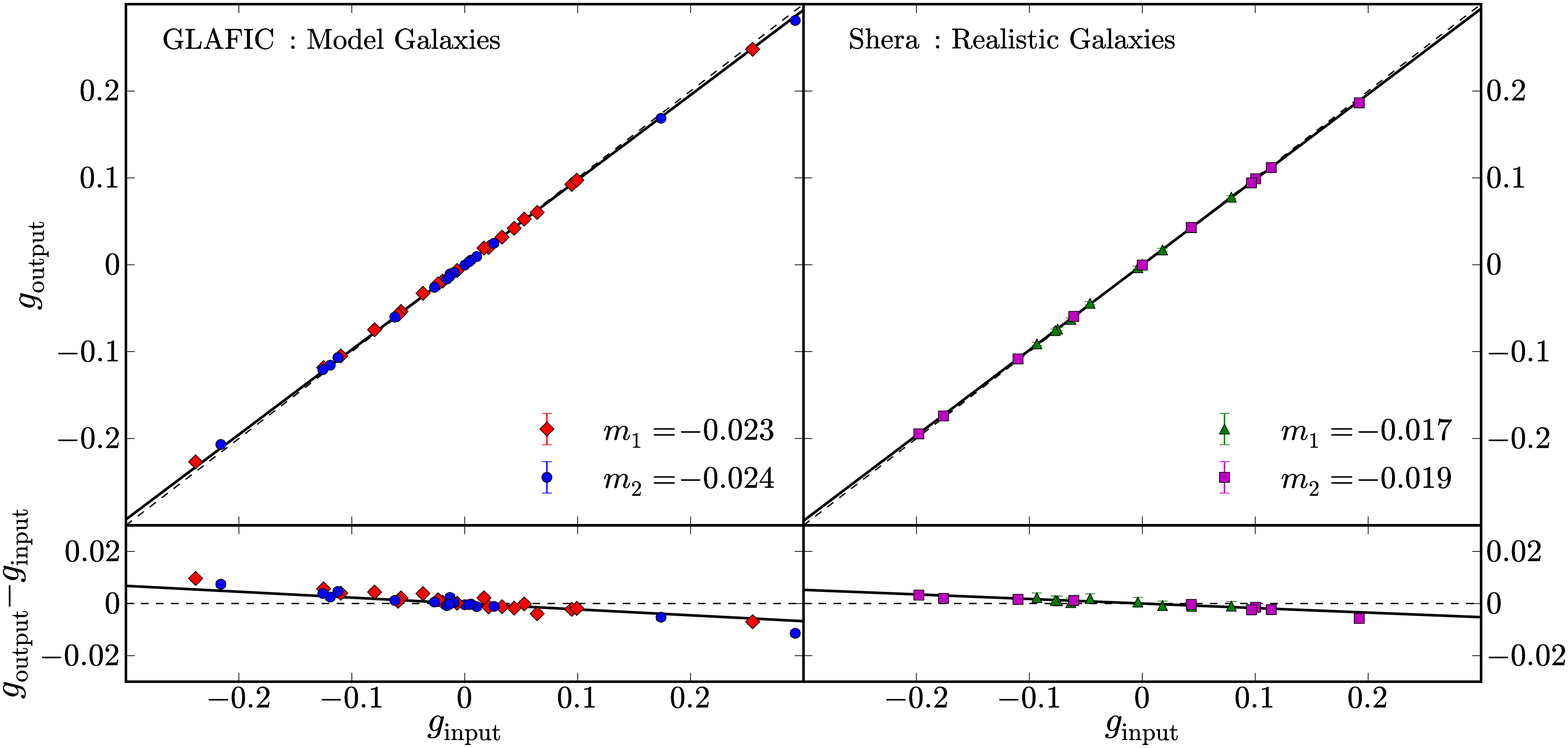}
\vspace{-7mm}
\hspace{-30mm}\includegraphics[width=160mm]{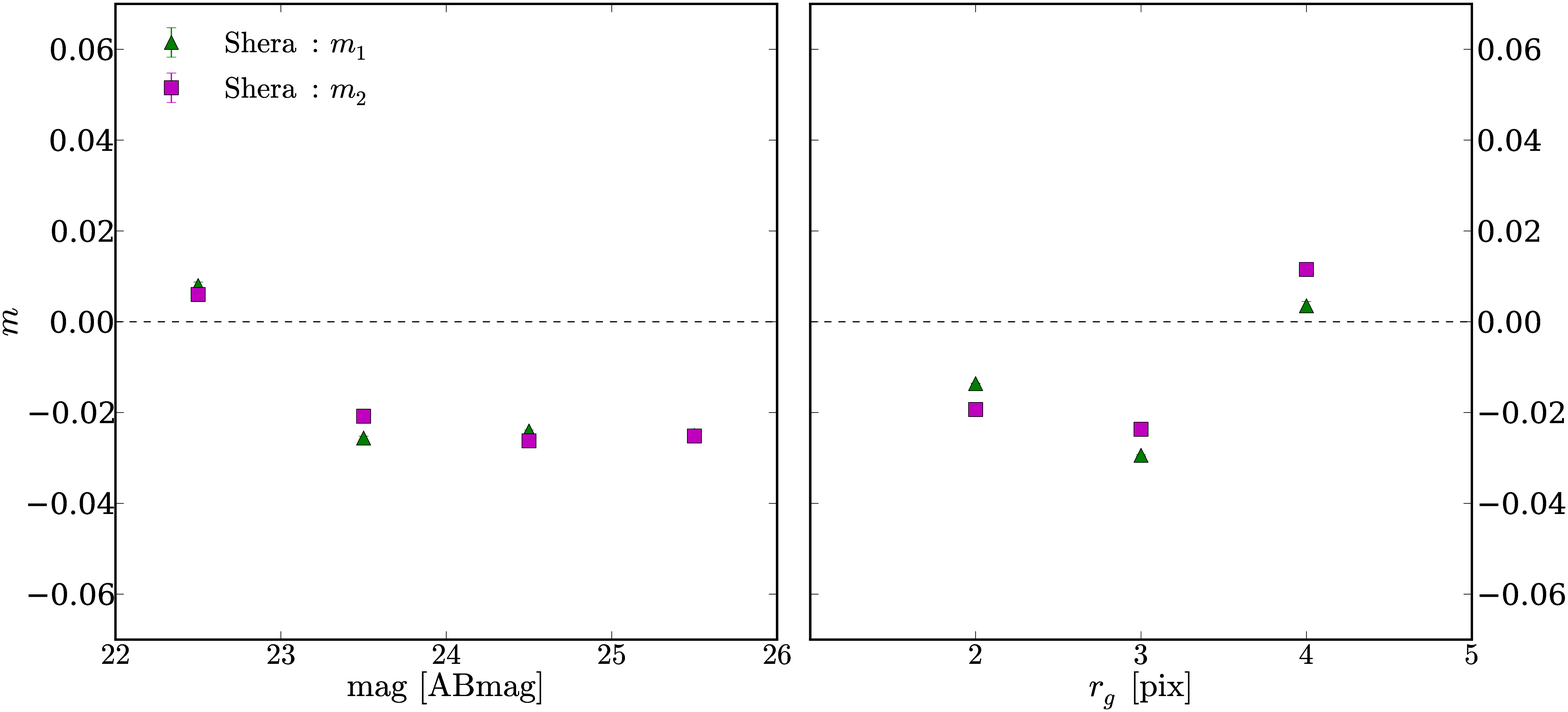}
\caption{{\sc Top} -- Results of the shear measurement tests,
  showing the multiplicative bias, $m_\alpha$, for simulated
   observations ($0.7''$ seeing) of model galaxies ({\sc GLAFIC;
    Left}) and realistic galaxies ({\sc SHERA; Right}). The
  signal-to-noise ratio for the detection are $\nu>10$, matching the
  sample of galaxies that we select for our analysis.  Red diamonds,
  and blue circles, green triangles and magenta squares denote $m_1$
  and $m_2$ for GLAFIC and $m_1$ and $m_2$ for {\sc SHERA},
  respectively.  The lower subpanels show the deviations from
    the input values. The dashed line shows $g_{\rm
      output}=g_{\rm input}$, and the solid line shows the result of
    the test described in the Section 2.5, in both upper and lower
    subpanels.  {\sc Bottom} -- The shear measurement bias,
  $m_\alpha$, as a function of the magnitude and the Gaussian size
  using the imaging simulations of realistic galaxies ({\sc SHERA}).
  Green triangles and magenta squares denote $m_1$ and $m_2$,
  respectively.  The bias is not a strong function of galaxy size
    or magnitude.}
\label{fig:step}
\end{figure*}

We use two simulated datasets to test the reliability of our faint
galaxy shape measurements, broadly following the approach introduced
by the STEP programme \citep{Heymans06,Massey07} with important
modifications compared to STEP and the recent cluster weak-lensing
literature.  These modifications are designed to match our data,
science goals, and our intention to use the weak-shear signal on
scales of a few hundred kpc to constrain the shape of the matter
density profile.  The first modification is to test our ability to
measure reduced shears upto $g\simeq0.3$, as seen in the inner regions
of clusters.  The second modification is to produce simulated fits
frames that match the angular size of the Suprime-Cam data; this
allows us to include sufficient galaxies with $\nu>30$ that we can
test our approach to the isotropic PSF correction.  This is critical
to validating that our shape measurement pipeline delivers accurate
shapes at faint flux levels.  We express the results of the tests
outlined below following the STEP convention of:
\begin{equation}
g_\alpha-g_\alpha^{\rm input}=m_\alpha g_\alpha^{\rm input}+c_\alpha
\end{equation}
where $g_\alpha$ and $g_\alpha^{\rm input}$ are the measured and input
ellipticities, respectively; $m_\alpha$ is the multiplicative bias and
$c_\alpha$ is a residual additive term. 

  Note that in both of the tests described below, we apply a
  constant shear to all of the simulated galaxies, and thus ignore
  higher order lensing effects that are present close to the Einstein
  radius \citep[e.g.][]{Okura07}.  Higher order effects are negligible
  on the scale that we measure and fit the shear profile of clusters.
  The distribution of Einstein radii for a background redshift of
  $z=2$ for clusters from our sample that are known strong lenses is
  lognormal, peaking at $\theta_{\rm E}(z_{\rm S}=2)=14.5\,{\rm
    arcsec}$ \citep{Richard10}.  Rescaling this to the typical
  redshift of $z\simeq0.8$ for the red background galaxies that we use
  in this analysis, we estimate $\theta_{\rm E}(z_{\rm
    S}=0.8)\simeq12\,{\rm arcsec}$.  Converting to physical projected
  distances, we therefore estimate an upper limit of $30h^{-1}{\rm
    kpc}$ on the typical Einstein radius at the median redshift of our
  cluster sample.  Note that we regard this as an upper limit because
  half of our cluster sample have not been identified as strong
  lenses, and thus likely have smaller Einstein radii than the
  clusters discussed by \citet{Richard10}.  For comparison, the
  innermost radius to which we typically fit the shear profile in
  Section~\ref{sec:mass} is $150h^{-1}{\rm kpc}$.  Therefore, our
  shear analysis begins at clustercentric radii of $\gs5\,\theta_{\rm
    E}$, i.e.\ on scales where higher order lensing effects are
  negligible.

The first simulated dataset follows \citet[][note that these authors
  also tested their code upto $g\simeq0.3$]{Okabe13}, and is based on
simulated images, kindly provided by M. Oguri, that are generated with
toy models using the software {\sc stuff} \citep{Bertin09}.  Each
galaxy is characterized by bulge and disc components, with Sersic
profiles indices of $n=4$ and $1$, respectively.  Galaxy images are
convolved with a PSF model based on the Moffat profile
$\Sigma(R)\propto\left(1+(R/a)^2\right)^{-\beta}$, with seeing in the
range $0.5\le{\rm FWHM}\le1$arcsec and the Moffat profile with power
slopes $3<\beta<12$, as described in \cite{Oguri12}.  A number of fits
frames matching the Suprime-Cam field of view were produced and
analysed using the pipeline described in \S\ref{sec:shape}.  We obtain
a shear calibration bias of $m_\alpha\simeq-0.02$ and
$c_\alpha\simeq10^{-4}$ (Fig.~\ref{fig:step}).

We extend \citeauthor{Okabe13}'s tests with a second simulated
dataset, using the {\sc SHERA} software \citep{Mandelbaum12} to
generate simulated ground-based observations that match the properties
of our observational data.  The galaxies images included in these
simulations are from the COSMOS {\it Hubble Space Telescope}
observations, as described by \citeauthor{Mandelbaum12} in detail.
We convolve the simulated data with a Moffat profile that matches our
observational data: $\beta=4.6$ and ${\rm FWHM}=0.7\arcsec$.  We
generated both non-rotated and 90-degree rotated images to extract a
shear estimate from a galaxy pair because the intrinsic ellipticity
cancels out.  The magnitude, size, and signal-to-noise ratio ($\nu$)
distributions of the simulated galaxies match those of our Subaru
data.  We analyse these data using the same pipeline as above,
obtaining again $m_\alpha\simeq-0.02$ and $c_\alpha\simeq10^{-4}$ --
i.e.\ consistent results from two methods of simulating the Subaru
data (Fig.~\ref{fig:step}).

We also checked the magnitude and size dependence of the shear
calibration using the {\sc SHERA}-based simulations.  The low level of shear
bias detected above does not show a strong trend with size and
magnitude, with $m_\alpha\simeq-0.02$ and $c_\alpha\simeq10^{-4}$
(Fig.~\ref{fig:step}) down to apparent magnitudes of $i'=26$.  This
result is achieved because of the high signal-to-noise threshold that
we apply to galaxies used for the isotropic PSF correction described
in \S\ref{sec:shape}.  

  In summary, our shape measurement bias is below our 4 per cent
  goal and does not depend on the size of galaxies.  However we note
  that due to the finite number of galaxies used in our tests, in
  particular in the {\sc SHERA}-based test (due to reliance on the
  COSMOS dataset), we cannot rule out the possibility that our shape
  measurement biases are different from those obtained here.

\subsection{Photometry and redshift estimates}\label{sec:photometry}

We will select faint background galaxies based on their location in
the $(V-i')/i'$ colour magnitude plane in \S\ref{sec:bkg}.  We
therefore analyse the data using SExtractor \citep{SExtractor},
adopting {\sc mag\_auto} as the total $i'$-band magnitude of each
object.  $(V-i')$ colours are measured in seeing matched frames,
within an aperture of $1.5\times$ the FWHM of point sources in the
poorer resolution of the two reduced frames for each cluster.
Hereafter for convenience we often denote colour as $C=(V-i')$.

Early-type cluster galaxies occupy a narrow well-defined relation in
the colour-magnitude diagram -- the so-called red-sequence.  We fit a
linear model of the form 
\begin{equation}
(V-i')_{\rm E/S0}=a\,i' + b. 
\label{eq:redsq}
\end{equation}
to galaxies at $i'\le22$ in order to define the colour of the red
sequence as a function of $i'$-band magnitude for the purpose of
selecting galaxies relative to the red sequence in \S\ref{sec:bkg}.

Our overall strategy is to combine location in the colour-magnitude
plane with redshift estimates and reduced shear measurements to
identify a low contamination sample of background galaxies.  The next
step is therefore to estimate the redshift of each galaxy in the
photometric catalogues.  We base these estimates on the COSMOS
UltraVISTA photometric redshift catalogue (McCracken et al. 2012;
Ilbert et al. 2013) that is limited at $i'< 27.5$, and benefits from
four deep near-infrared filters $Y$, $J$, $H$ and $K_S$. This filter
coverage enables more robust photometric redshifts for galaxies at
$z>1.3$ than earlier versions of the COSMOS catalogue, since the
Balmer break is redshifted to the near-infrared for these
galaxies. Furthermore, the COSMOS UltraVISTA photometric redshifts are
tested against almost 35,000 new spectra with galaxies at $z>1.5$ (for
more details see Ilbert et al. 2013).  This catalogue provides
currently the most reliable redshift distribution for a
magnitude-limited galaxy sample that reaches $i'\simeq26$ with
Suprime-Cam on Subaru.  We emphasise that the COSMOS dataset includes
observations through the same filters with the same camera mounted on
the same telescope that we use in this study, thus matching LoCuSS and
COSMOS photometry is straightforward.

The lensing kernel for the $i$-th galaxy in our photometric catalogues,
$\beta_i\equiv{D_{ls,i}}/{D_{s,i}}$, is estimated by an ensemble average of
the $N$ nearest neighbours in colour-magnitude space of the $i$-th
galaxy in the COSMOS catalogue:
\begin{eqnarray}
  \beta_i=\langle{D_{ls}}/{D_s}\rangle_{\rm COSMOS}=\frac{1}{N}\sum_j^N{D_{ls,j}}({z_s})/{D_{s,j}}({z_s}). \label{eq:beta}
\end{eqnarray}
Here, $D_s$ and $D_{ls}$ are the angular diameter distances from the
observer to the sources and from the lens to the sources,
respectively.  The source redshift, $z_s$, is the median of the
likelihood distribution for the photometric redshift of each COSMOS
galaxy.  We estimate the uncertainty on $\beta_i$ as
the sum of the individual photometric errors and standard errors of
the sample:
\begin{eqnarray}
 \sigma_\beta^2=\frac{1}{N}\sum_j^N\left[
  \sigma_{\beta^{\rm COSMOS},j}^{2}+\frac{1}{N-1}(\beta_j^{\rm COSMOS}-\beta)^2\right]
\end{eqnarray}
The typical uncertainties are $\sigma_\beta\simeq13-28\%$.  We adopt
$N=100$, and check that our results are insensitive to whether we
adopt $N=50$ or $N=200$ finding that the redshift estimates for
individual galaxies are randomly changed by a few per cent.  We
  include this small uncertainty in the redshift uncertainties that
  are incorporated in to the error bars on the cluster shear profiles
  in Section~\ref{sec:mass}.

\subsection{Selection of background galaxies}\label{sec:bkg}

Contamination of background galaxy catalogues by unlensed
member/foreground galaxies leads to a systematic underestimation of
reduced shear signal.  This is often referred to as a dilution effect,
because the contaminants dilute the signal \citep{Broadhurst05,
  Umetsu10, Okabe10b, Okabe13}. 
The dominant source of contaminant
galaxies is faint cluster members, the number density of which
increases towards the cluster core.

In \cite{Okabe13} we quantified the contamination level as a function
of the colour offset of faint galaxies from the red-sequence, defined
by $\Delta C\equiv(V-i')-(V-i')_{\rm ES0}$, and concentrated on
$\Delta C>0$, i.e.\ galaxies redder than the red sequence.  We here
briefly summarize the method.  The mean tangential distortion strength
is averaged over all galaxies satisfying a given colour cut,
across all 50 clusters and all cluster-centric radii.  The mean
lensing signal increases strongly as a function of $\Delta C$ close to
the red sequence, flattening to a shallower trend at larger colour
offsets (see Fig.~1 of Okabe et al. 2013).  We interpreted this
behaviour with a two component model comprising contamination by
cluster members and the redshift dependence of shear signal --
i.e.\ redder galaxies are on average more distant and thus present a
stronger lensing signal.  The model includes a parameter that
describes the fraction of the total population of galaxies that are
contaminants from the cluster.  Thus by fitting this model we were
able to measure the contamination level and adopt a colour cut that
gives a desired level of contamination without relying on any
assumptions about the distribution of mass in the clusters, and about
the run of number density of background galaxies with cluster-centric
radius.  We adopted $1\%$ contamination limit which is less than the
statistical error of the average mass measurement for the sample of 50
clusters.  However, this conservative approach yields a mean number
density of background galaxies of just $\langle n_{\rm
  bkg}\rangle\sim5.3\pm1.9{\rm arcmin^{-2}}$, which makes it difficult
to measure individual cluster masses without invoking assumptions
about the shape of the cluster mass density profile.

\begin{figure*}
  \centerline{
    \includegraphics[width=190mm]{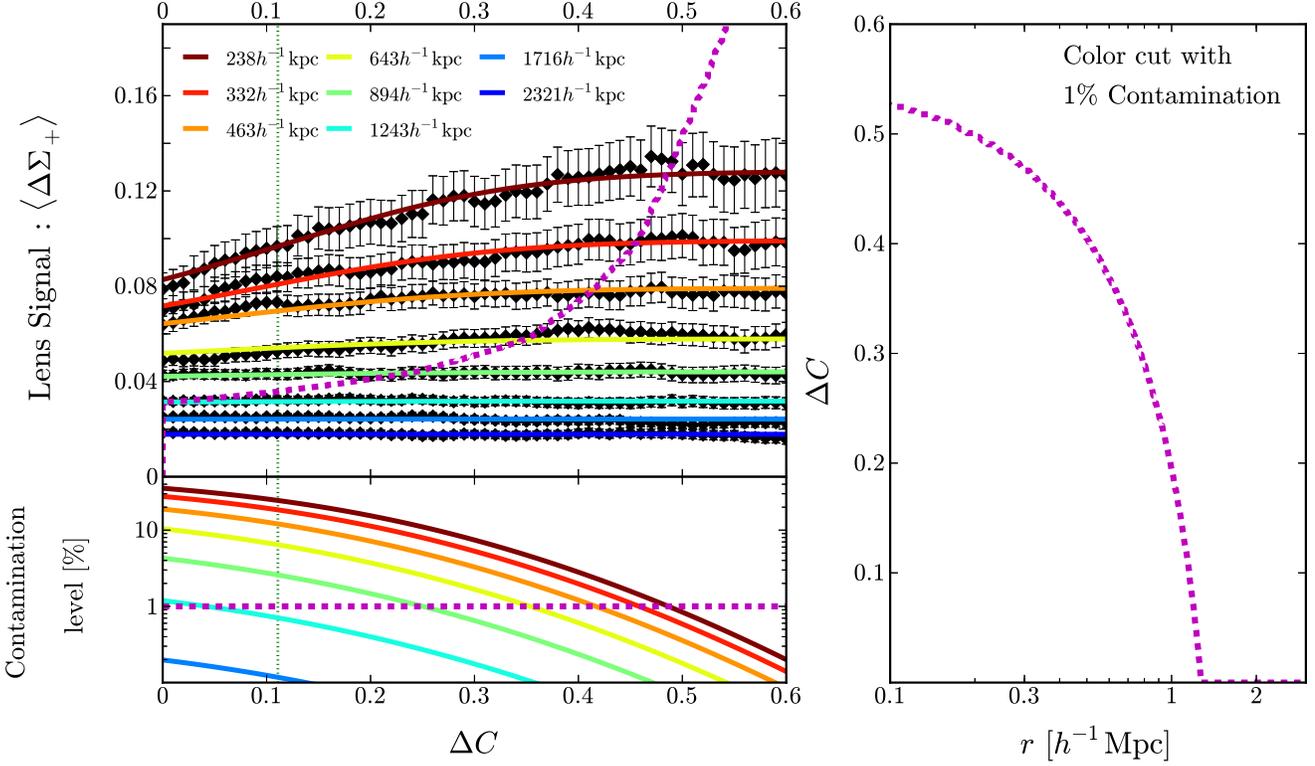}
  }
  \caption{{\sc Upper left}: The mean lensing signal for all 50
    clusters as a function of the colour offset $\Delta C$ in
    cluster-centric radial bins.  The innermost bin has the largest
    shear signal (upper most set of points overlaid with brown curve),
    with progressively smaller shear signals seen at larger
    cluster-centric radii (lower sets of points overlaid with red
    through yellow, green and blue curves).  All of the curves are
    flat at $\Delta C\gs0.5$ -- i.e.\ at galaxy colours used for our
    analysis in Okabe et al. (2013).  At smaller values of $\Delta C$
    the curves relating to smaller cluster-centric radii steepen
    whilst those relating to larger cluster-centric radii are much
    flatter.  The best-fit models are represented by the solid
    coloured curves.  The legend states the typical
    weighted harmonic mean clustercentric radius of each radial
    bin.  The dashed magenta curve denotes the location of the $1\%$
    dilution level; the intersection between this curve and the solid
    curves indicates the colour cut required in each radial bin to
    achieve $1\%$ dilution.  The vertical dotted green line is the
    $1\sigma$ width of the red sequence determined by the bright
    member galaxies.  {\sc Lower left:} The best-fit models of the
    dilution level of member galaxies as a function of the colour
    offset. The colours are the same as those in the upper left
    panel. The dilution level at innermost bin (brown) shows
    $\sim40\%$ at $\Delta C=0$.  The horizontal
      dashed line (magenta) is our requirement for the dilution level.
      {\sc Right}: The colour-cut for background selection with $1\%$
      dilution as a function of clustercentric radius.  The colour cut
      is $\Delta C\sim0.5$ at $r\sim100~h^{-1}{\rm kpc}$ and $\Delta
      C=0$ beyond $\simgt1.3~h^{-1}{\rm Mpc}$, respectively.  }
  \label{fig:dilution}
\end{figure*}

We develop a new method to select securely a larger sample of
background galaxies than achieved by \citet{Okabe13}.  In the new
method, we average the mean tangential distortion strength over all
galaxies satisfying each colour cut ($>\Delta C$) and all 50 clusters,
in several radial bins.  The stacked lensing strength in the $i$-th
radial bin is estimated as:
\begin{eqnarray}
\langle \Delta \Sigma_{+}\rangle(r_i)= \frac{\sum_n g_{+,n} \Sigma_{{\rm
 cr},n} w_{n}}{\sum_n w_{n}}, \label{eq:g+}
\end{eqnarray}
where 
and $\Delta \Sigma_{+}$ is the dimensional reduced shear and
the subscript $n$ denotes the $n$-th galaxy located in the
annulus spanning $r_1<r_i<r_2$.  Note that the reduced
  shear $\Delta \Sigma_+$ is different from the definition of,
for example, \cite{Mandelbaum06}, because we define $\Delta
  \Sigma_+$ in terms of the reduced shear, $g$, and not in terms of
  shear, $\gamma$.  The reduced tangential distortion component,
$g_+$, with respect to the cluster center is defined by
\begin{eqnarray}
g_{+}&=&-(g_{1}\cos2\varphi+g_{2}\sin2\varphi),
\label{eq:gt}
\end{eqnarray}
where $\varphi$ is the position angle between the first coordinate
axis on the sky and the vector connecting the cluster center and the
galaxy position.  The reduced tangential shear is expressed by
\begin{eqnarray}
 g_+=\frac{\gamma_+}{1-\kappa},
\end{eqnarray}
where $\gamma_+$ is the tangential shear and $\kappa$ is the
dimensionless surface mass density.  The weighting functions are taken
into account for both the statistical weights, $w_n$, and the critical
projected mass density, $\Sigma_{{\rm cr},n}$, describing the lensing
efficiency.  The weighting $w_{n}$
\citep[e.g. ][]{Hoekstra00,Hamana03,Okabe10b,Umetsu10,Oguri12} is used
to down-weight galaxies whose shapes are less reliably measured, based
on the uncertainty in the shape measurement, $\sigma_{g,n}$, given by
\begin{eqnarray}
w_n=\frac{1}{\alpha^2+\sigma_{g,n}^2}\frac{1}{\Sigma_{{\rm cr},n}^2}\label{eq:weight}.
\end{eqnarray}
We choose $\alpha=0.4$ throughout this paper.
The critical projected mass density for individual clusters is defined by
\begin{eqnarray}
 \Sigma_{{\rm cr}}= \frac{c^2}{4\pi GD_l}\beta^{-1}, 
\end{eqnarray}
where $D_l$ is the angular diameter distance to a cluster and $\beta$
is the lensing depth (equation \ref{eq:beta}).

 To clarify the relationship between $\Delta\Sigma_+$ and $g_+$ in
  the presence of contamination of background galaxy catalogues, we
  write an equation to describe the contamination effect, as
  follows:
\begin{eqnarray}
 \langle\Delta\Sigma_+\rangle &=& \frac{\sum_n (g_{+,n} \Sigma_{{\rm cr},n} w_{{\rm
  bkg},n} + 0 \times w_{{\rm non-bkg},n} )}{\sum (w_{{\rm bkg},n} + w_{{\rm non-bkg},n})}
  \nonumber \\
&=& \frac{\sum \Delta \Sigma_{+,n} w_{{\rm bkg},n}}{\sum (w_{{\rm bkg},n} + w_{{\rm
 non-bkg},n})}, \label{eq:dilution_basic}
\end{eqnarray}
where $\Delta\Sigma_+=g_+ \Sigma_{cr}$ is the dimensional shear for
each background galaxy and the subscripts ``bkg''
  and``non-bkg'' denote background galaxies and non-background
  galaxies, respectively. Since non-background galaxies are not
  lensed by the cluster, the second term in the numerator is zero.
  Therefore, contamination by non-background galaxies causes the
  observed lensing signal to be underestimated. The
  underestimation of the ensemble average is simply determined by the
  fraction of non-background galaxies (mainly member galaxies) to
  background galaxies. This is even clearer if, for the sake of
  illustration, one assumes that the galaxies have uniform weights,
  $w=1$, in which case Equation~\ref{eq:dilution_basic} can be
  rewritten as:
\begin{eqnarray}
\langle\Delta\Sigma_+\rangle =\langle\Delta\Sigma_+\rangle_{\rm bkg} \frac{1}{1+n_{\rm non-bkg}/n_{\rm bkg}}, 
\end{eqnarray}
where $\langle\Delta\Sigma_+\rangle_{\rm bkg}$ is the average
distortion strength using a pure sample of background
  galaxies. The correction factor, $(1+n_{\rm non-bkg}/n_{\rm bkg})$
  is equivalent to the {\it boost factor}, under the assumption
  of a radially uniform distribution of background galaxies
  \citep[e.g.][]{Applegate14,Hoekstra15}, that we discuss in Section
  \ref{sec:boost}.  Equation \ref{eq:dilution_basic} shows that the
  average lensing signal obtained from our formalism is simply
  underestimated by the fraction of non-background to background
  galaxies.  Therefore also note that the non-linear term in the
  reduced shear $g_+=\gamma_+/(1-\kappa) \propto \Delta \Sigma_+$ does
  not impact on our estimated levels of contamination.

The stacked lensing signal is a decreasing function of clustercentric
radius and an increasing function of $\Delta C$ for small
clustercentric radii, reminiscent of \citeauthor{Okabe13}'s (2013)
analysis of all clustercentric radii as a single bin (upper left panel
of Fig.~\ref{fig:dilution}).  However the increase in lensing signal
at moderate values of $\Delta C$ becomes progressively less pronounced
as one considers radial bins at larger clustercentric radii.  This is
qualitatively consistent with faint cluster galaxies being the
dominant source of contamination, given that the number density of
galaxies in clusters is a declining function of clustercentric radius.

To interpret quantitatively the mean lensing strength as a function of
cluster-centric radius, we parameterise the galaxy distribution in
terms of the projected clustercentric radius $r$ and the colour offset
$\Delta C$, as follows:
\begin{eqnarray}
  n(r,\Delta C)=n_b(r)\left[1+B f(\Delta C) n_{\rm m}(r)\right],\label{eq:n_dil}
\end{eqnarray}
where $n_b(r)$ is the radial distribution of background galaxies.  We
do not assume any specific functions of $n_b$ because the background
distribution is not constant, and may be depleted or boosted by a
magnification bias
\citep[e.g.][]{Broadhurst95,Umetsu11,Umetsu14,Coupon13}.  The second
term in the bracket denotes the member galaxy distribution; $B$ is the
fraction, $f(\Delta C)$ is the colour distribution and $n_{\rm m}(r)$
is the radial distribution.  The effective lensing strengths for red
galaxies are obtained by integrating over the projected radius and the
colour offset:
\begin{eqnarray}
\langle \Delta\Sigma_{+,i} \rangle (r_1<r_i<r_2,\Delta C<) &=&
 \frac{\int^\infty_{\Delta C}d(\Delta C)\int^{r_2}_{r_1} dr\Delta \Sigma_+
 n_b(r) r }{\int^\infty_{\Delta C}d(\Delta C) \int^{r_2}_{r_1}dr n(r)r} \nonumber \\
&=& \frac{A_i}{1+BF(\Delta C)N_i}. \label{eq:dil}
\end{eqnarray}
Here, $A_i\equiv\Delta\Sigma_{+0,i}$ is the lensing signal estimated
from the pure background galaxies and is thereby determined by the
cluster mass distribution.  The contamination levels in the colour and
radial distributions are described by $F(\Delta C)=\int^\infty_{\Delta
  C}d(\Delta C) f(\Delta C)$ and $N_i=\int^{r_2}_{r_1}dr n_m(r)r$,
respectively. As in \cite{Okabe13}, we employ a Gaussian distribution
centering at $\Delta C=0$ as the colour distribution of member
galaxies,
\begin{eqnarray}
 F(\Delta C)=[1-{\rm erf}(\Delta C/\sqrt{2}\sigma)]/2,
\end{eqnarray}
where $\sigma$ is the width of colour distribution composed of the
intrinsic scatter in the colour distribution and the photometric error.
We assume that $\sigma$ is radius-independent.  When we average
equation (\ref{eq:dil}) over all radial bins, the formulation in
\cite{Okabe13} is recovered:
\begin{eqnarray}
\langle\langle\Delta \Sigma_{+} \rangle\rangle&=&\sum_i \langle \Delta\Sigma_{+,i}
 \rangle \simeq \tilde{A}\left[1-\tilde{B}F(\Delta C)\right].
\end{eqnarray}
In this paper, we simultaneously take into account the colour
distribution and the radial distribution for member galaxies.  We
employ $n_m(r)=\exp(-r/r_0)$ as the radial distribution of member
galaxies \citep{Applegate14}.  Subsequently, fitting is performed to
obtain the colour and radial distribution of member galaxies.  

In summary, the fitting parameters are $A_i$, $B$, $\sigma$ and $r_0$.
We stress that this method does not assume any specific mass models,
which is important to interpret the results after defining the
background sample.  As shown in the upper left panel of Figure
\ref{fig:dilution}, the best-fit model (solid lines with different
colours) well describes the data.  The best-fit colour width,
$\sigma=0.21$, is higher than the mean width expected from the
intrinsic scatter determined by the cluster bright galaxies, which is
consistent with \cite{Okabe13}.  This large value of $\sigma$ is
driven by the statistical scatter of the faint galaxies included in
the calculation -- i.e.\ the photometric uncertainties at
$i'\simeq25$.  We note that our method assumes that the colour
distribution of galaxies redward of the red sequence is Gaussian; the
large value of $\sigma$ therefore helps to ammeliorate any concerns
that the wings of the actual distribution contain an excess of
galaxies over the assumed Gaussian form.  The characteristic radius of
the member galaxy distribution is $r_0=258~\hkpc$.  As expected based
on the previous qualitative discussion, the highest level of
contamination occurs at $\Delta C=0$ at the smallest clustercentric
radii, with $F\simeq0.4$ (Fig.~\ref{fig:dilution}), and the level of
contamination declines significantly with increasing clustercentric
radius.  Contamination is negligible in the cluster outskirts.  We
conservatively adopt a limit of $1\%$ on contaminating fraction, and
use this to define a radially dependent colour cut (right panel of
Fig.~\ref{fig:dilution}).  Note that at $r>1.3~\hMpc$ the
contamination level is so low that we adopt $\Delta C>0$ in this
region.  We achieve a number density of background galaxies of $n_{\rm
  bkg}\simeq5-20~{\rm arcmin}^{-2}$ (Table~\ref{tab:data}), with a
mean of $\langle n_{\rm bkg}\rangle\simeq12.8~{\rm arcmin^{-2}}$ that
is more than double that of \cite{Okabe13}.

  We also use our mid- and far-infrared observations with {\it
    Spitzer} and {\it Herschel} as a sanity check on the possible
  impact of heavily dust-obscured galaxies on our red galaxy selection
  and on our COSMOS-based estimates of $\beta$ in the previous
  Section.  Specifically, we consider whether dusty cluster members
  might leak into the red background galaxy samples and whether
  $\beta$ might be biased due to the presence of optically faint and
  heavily dust-obscured galaxies -- i.e.\ Luminous Infrared Galaxies
  (LIRGs) and Ultra-Luminous Infrared Galaxies (ULIRGs).  On the
  latter point, the COSMOS photometry extends to $8\mu{\rm m}$,
  whereas we have observed half of the cluster sample discussed in
  this article with {\it Spitzer}/MIPS at $24\mu{\rm m}$
  \citep{Haines15} and with {\it Herschel}/PACS and SPIRE at
  $100-500\mu{\rm m}$ \citep{Smith10}, albeit only to a depth
  corresponding to a bolometric infrared luminosity of $L_{\rm
    IR}\simeq5\times10^{10}L_\odot$ at $z\simeq0.2$.  The
  spectroscopic completeness of follow-up observations with Hectospec
  \citep{Fabricant05} is $96$ per cent for objects detected with {\it
    Spitzer} down to $0.5{\rm mJy}$ and and $\sim80$ per cent down to
  $i=20$ \citep{Mulroy14,Haines15}.  Clearly these data are not
  sensitive enough to obtain definitive estimates of the number and
  redshifts of dusty galaxies that satisfy our faint red optical
  background galaxy selection.  However they provide a useful sanity
  check based on bright galaxies.  We select galaxies from the
  catalogues discussed by \citet{Haines15} at $i<20$ that have $(V-i)$
  colours that would place them in our red background galaxy
  catalogues if they were faint enough.  We find that $\sim1$ per cent
  of these bright optically red galaxies are LIRGs, and $\sim1$ per
  cent of the same bright optically red galaxies are cluster members,
  and there is no overlap between these two populations.  We therefore
  conclude that the bright optically red galaxy population seen along
  lines of sight through our cluster sample appear to be consistent
  with (1) LIRGs and ULIRGs not being a significant population in our
  red background galaxy samples, and thus not being a concern in terms
  of the accuracy of $\beta$, and (2) our red background galaxy
  catalogues suffering just $1$ per cent contamination by cluster
  members.

\section{Modelling and results}\label{sec:results}

\subsection{Model Fitting Methods}\label{sec:fitting}

We describe how we compute the reduced tangential shear profile of
each cluster and the model fitting procedure.  We apply the methods
described in this section to measure cluster masses in the next section.  

 We centre each cluster shear profile on the centroid of the
  optical emission from the cluster's BCG, following numerous previous
  studies that have shown BCGs to be a reliable cluster centre for
  weak-lensing studies of massive galaxy clusters, including
  \citet{Okabe10b,Okabe13}, \citet{vonderLinden14}, and
  \citet{Hoekstra15}.  For example, we derived an upper limit of
  $32\,h^{-1}{\rm kpc}$ on the mean offset of BCGs from the underlying
  centre of the cluster mass distribution for the sample studied here,
  in \citet{Okabe13}.  This upper limit is a factor of 5 smaller than
  the typical innermost radius of the shear profiles upon which our
  mass measurements are ultimately based in Section~\ref{sec:mass}.
  Any bias caused by centring our shear profiles on the BCGs is
  therefore negligible.  

The reduced shear in a given annulus centered on a given cluster is
computed by azimuthally averaging the measured galaxy ellipticities,
as defined by Equation~\ref{eq:g+}.  The mean redshift of the
background galaxies is a function of cluster centric radius, due to
our radially-dependent colour cut (Section~\ref{sec:bkg}).  The
formulation in equation~\ref{eq:g+} takes account of these differences
by expressing the reduced shear in physics units.

We employ a maximum-likelihood method to model the shear profiles, and
write the log-likelihood as follows:
\begin{eqnarray}
-2\ln {\mathcal L}&=&\ln(\det(C_{ij})) +  \label{eq:likelihood} \\
 &&\sum_{i,j}(\Delta \Sigma_{+,i} - f_{{\rm model}}(r_i))C_{ij}^{-1} (\Delta
 \Sigma_{+,j} - f_{{\rm model}}(r_j)), \nonumber
\end{eqnarray}
where the subscripts $i$ and $j$ are the $i-$ and $j-$th radial bins.
Here, $f_{\rm model}$ is the reduced shear prediction for a specific
mass model,
\begin{eqnarray}
 f_{{\rm model}}(r_i)=\frac{\Delta \tilde{\Sigma}_{\rm model}(r_i)}{1 - K_i
  \Sigma_{\rm model}(r_i)) }, \label{eq:g+model}
\end{eqnarray}
where $\Sigma=\Sigma_{\rm cr}\kappa$ and $\Delta
\tilde{\Sigma}=\Sigma_{\rm cr}\gamma$ are the convergence and the
shear in physical units, respectively, and $\kappa$ and $\gamma$ are
the dimensionless convergence and shear, respectively.  The factor,
$K_i$ for the $i$-th bin is given by:
\begin{eqnarray}
 K_i = \frac{\sum_n \Sigma_{\rm cr,n}^{-1} w_n }{\sum_n w_n}.
\end{eqnarray}
Note that $K_i$ is computed separately for each bin due to the radial
dependence of the redshift of the background galaxies.

The covariance matrix, $C$, in equation~\ref{eq:likelihood} is given by:
\begin{eqnarray} C&=&C_g+C_s+C_{{\rm LSS}}.  \end{eqnarray}
Where the shape noise, $C_g$, in each radial bin is estimated as
\begin{eqnarray}
 C_{g,ij}=\frac{1}{2}\frac{\sum_n \sigma_{g,n}^2 \Sigma_{\rm cr,n}^2
   w_n^2}{(\sum_n w_n)^2}\delta_{ij},
\end{eqnarray}
where $\delta_{ij}$ is the Kronecker delta and the factor of $1/2$
accounts for the rms noise, $\sigma_{g,n}$, of two distortion
components.  The photometric redshift error matrix, $C_s$, is computed
from:
\begin{eqnarray}
C_{s,ij}=\tilde{C}_{s,ij}+\frac{\Delta \tilde{\Sigma}_{\rm
    model}^2\Sigma_{\rm model}^2 \sigma_{K,i}^2}{(1-K_i \Sigma_{\rm
    model}(r_i))^4}\delta_{ij}, \label{eq:Cs}
\end{eqnarray}
where the first term is given by
\begin{eqnarray}
\tilde{C}_{s,ij}=\left[\frac{\sum_n (\Delta \Sigma_{+,n}-2\Delta \Sigma_{+})^2 w_n^2 (\sigma_{\Sigma,n}/\Sigma_{{\rm cr},n})^{2}}{(\sum_n w_n)^{2}}\right] \delta_{ij}.\nonumber
\end{eqnarray}
The second term of the equation (\ref{eq:Cs}) is the photometric
redshift errors through an error of the conversion factor, $\sigma_K$,
in the mass model (\ref{eq:g+model}).  The covariance matrix of
uncorrelated large-scale structure (LSS), $C_{\rm LSS}$, along the
line-of-sight \citep{Schneider98} at an angular separation between
$\theta_i=r_i/D_l$ and $\theta_j$ is given by
\begin{eqnarray}
 C_{{\rm LSS},ij}=\int \frac{ldl}{2\pi}P_{\kappa}(l) J_2(l\theta_i) J_2(l\theta_j),
\end{eqnarray}
where $P_\kappa(l)$ is the weak-lensing power spectrum
\citep[e.g.][]{Schneider98,Hoekstra03}, calculated by multipole $l$,
the source redshift, and a given cosmology. We employ the redshift,
${\rm min}(z_{s,i},z_{s,j})$, and WMAP9 cosmology \citep{WMAP09}.
And, $J_2(l\theta_i)$ is the Bessel function of the first kind and
second order at the $i$-th annulus \citep{Hoekstra03}.

It is also important to compute the radius of each radial bin
correctly, because systematic errors in the placement of the binned
shear measurements on the radial axis can cause systematic errors in
the mass measurement when a model is fitted.  This is
  particularly important in practice, because the number density of
  background galaxies is neither uniform nor infinite.  As described
  in detail in Appendix \ref{sec:app0}, we found that the best radius
  at which to place the measurement of mean tangential shear in a
  radial bin is the weighted harmonic mean:
\begin{eqnarray}
  \label{eq:meanr}
  \langle r \rangle_i=\frac{\sum_n w_n}{\sum_n w_n\,r_n^{-1}},
\end{eqnarray}
  where $w_n$ is given by Equation~\ref{eq:weight}.  We therefore
  compute bin radii in this way, in physics units, in the rest of our
  analysis.

Finally, before fitting models to the shear profiles
(Section~\ref{sec:mass}), we calculate the signal-to-noise ratio of
the tangential shear profile as follows:
\begin{eqnarray}
 (S/N)^2=\sum_{ij} \Delta
  \Sigma_{+,i}(C_{g,ij}+\tilde{C}_{s,ij}+C_{{\rm LSS},ij})^{-1}\Delta
  \Sigma_{+,j}.
\end{eqnarray}
Each cluster is detected individually at signal-to-noise ratio of $3<S/N<11$ 
(Table \ref{tab:data}).

\subsection{Mass Measurements}\label{sec:mass}

To infer galaxy cluster masses from the shear profiles, we fit a model
to the latter.  For this purpose we adopt the universal mass density
profile \citep[][hereafter NFW]{NFW96,NFW97}, that has had
considerable success in describing dark matter halo profile spanning a
wide mass range by numerical simulations based on the CDM model of
structure formation.  It has also been shown that the ensemble mass of
a sample of clusters can be recovered to good precision from this
approach, provided sufficient care is taken over the radial range over
which this model is fitted to data \cite[e.g.][]{Becker11,Bahe12}.  

The NFW profile is expressed in the form:
\begin{equation}
\rho_{\rm NFW}(r)=\frac{\rho_s}{(r/r_s)(1+r/r_s)^2},
\label{eq:rho_nfw}
\end{equation}
where $\rho_s$ is the central density parameter and $r_s$ is the scale
radius.  The three-dimensional spherical mass, $M_\Delta$, enclosed
by the radius, $r_\Delta$, inside of which the mean density is
$\Delta$ times the critical mass density, $\rho_{\rm cr}(z)$, at the
redshift, $z$, is given by
\begin{equation}
M_{\rm NFW}(<r_\Delta)=\frac{4\pi \rho_s r_\Delta^3}{c_\Delta^3}\left[
\ln(1+c_\Delta)-\frac{c_\Delta}{1+c_\Delta}\right].
\label{eq:MNFW}
\end{equation}
The NFW profile is fully specified by two parameters: $M_\Delta$ and
the halo concentration $c_\Delta=r_\Delta/r_s$.  We fit this model to
the shear profile of each cluster, taking full account of errors of
shape measurements, photometric redshifts and the uncorrelated LSS
(Section \ref{sec:fitting}).  For a given $M_\Delta$ and $c_\Delta$ we
predict the observed shear signal following the formalism described by
\citet{Wright2000}.

Measurements of $M_\Delta$ are mainly sensitive to the lensing signal
around the overdensity radii $r_\Delta$ \citep{Okabe10b}.  However the
concentration parameter, $c_\Delta$, is more strongly affected by the
lensing signal in the cluster central regions, i.e.\ cluster centric
radii of hundreds of kpc.  Our careful selection of background
galaxies ensures that contamination of our background galaxy samples
is negligible across the full radial range of our shear profiles.
However the very stringent colour cut employed in the central regions,
$\Delta\,C>0.52$, and the relatively small solid angle subtended by
these innermost bins, render them the noisiest of the entire radial
range.  To guard against obtaining results on concentration that
suffer biases due to the noisy inner profiles, we choose a binning
scheme for each cluster via the following procedure.  We fit the NFW
model to a suite of measured shear profiles that span inner radii in
the range $r_{\rm in}=50-300~h^{-1}{\rm kpc}$, outer radii in the
range $r_{\rm out}=2000-3000~h^{-1}{\rm kpc}$, and number of bins in
the range $N_{\rm bin}=4-8$.  We then compute the mean of the suite of
$M_\Delta$ values obtained from these fits, and adopt the binning
scheme that yields the value of $M_\Delta$ closest to that mean.  Note
that we allow the virial concentration parameter to be in the range
$0<c_{\rm vir}<30$ in the fits.  Also, we restrict the radial range
of the shear profile fits for A1758N to $r_{\rm out}<2100h^{-1}{\rm
  kpc}$ to avoid contamination of lensing signal by its neighbour
A1758S.  Note that we test the procedure described above using mock
observations of simulated and toy model clusters, and confirm that it
returns masses and concentrations with negligible bias
(Section~\ref{sec:sim}).

\begin{figure*}
\includegraphics[width=0.45\hsize]{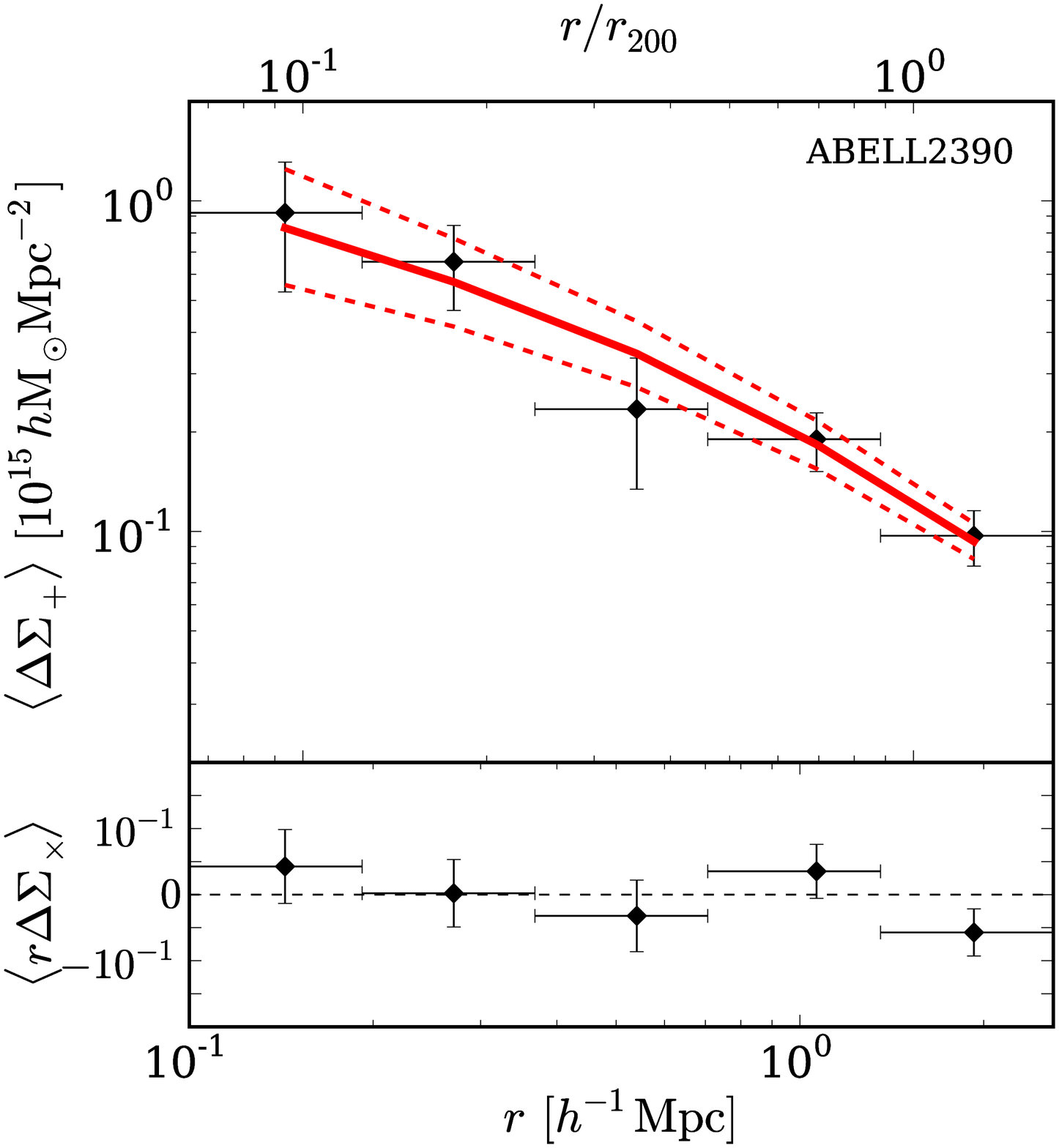}
\includegraphics[width=0.45\hsize]{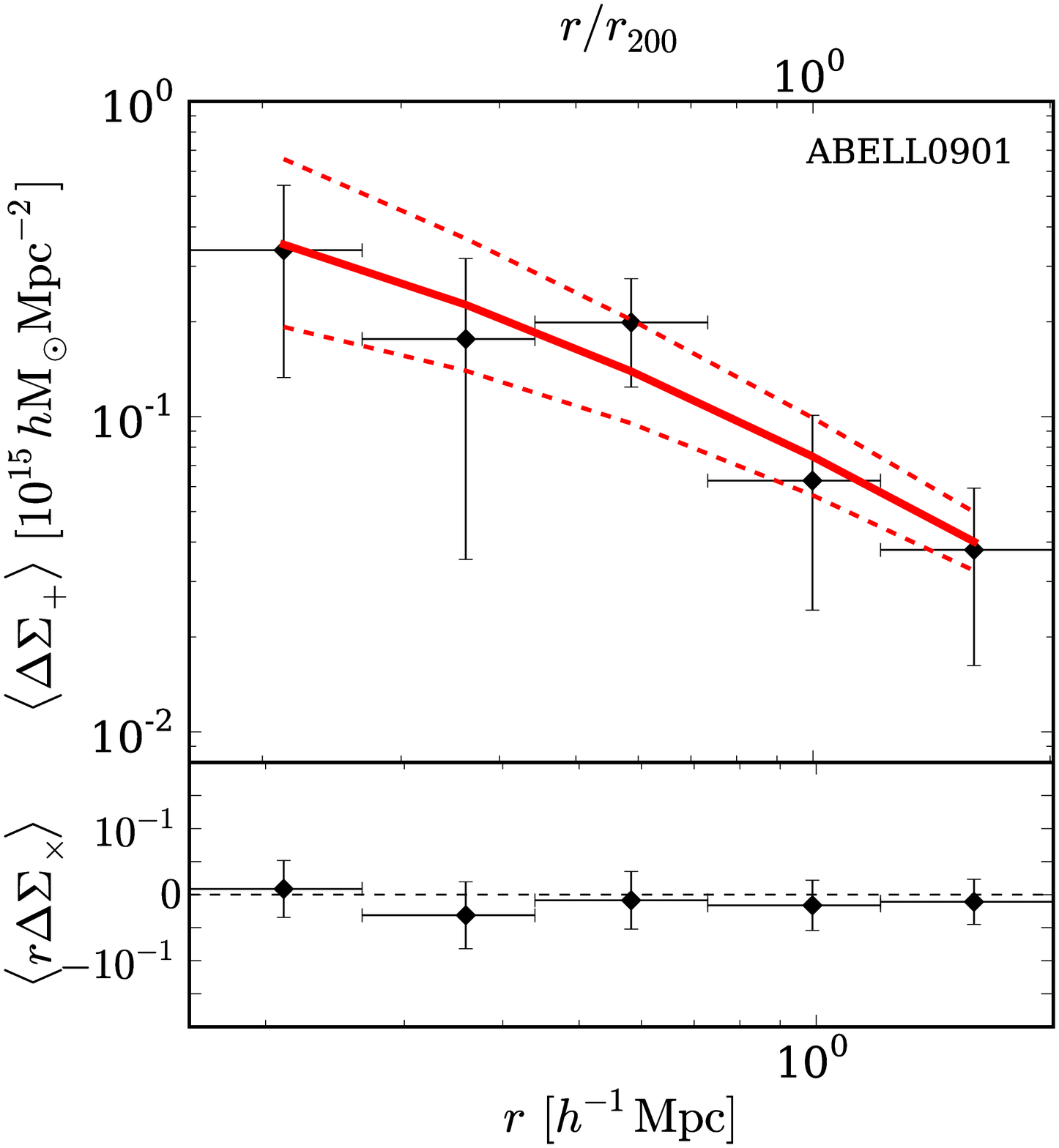}
\caption{Radial profiles of the tangential shear component (top
  panel), $\Delta \Sigma_+$, and the product of the $45$ degree
    rotated component, $\Delta \Sigma_\times$, and clustercentric
    radius, $r$} (bottom panel), , for ABELL\,2390 (left) and
  ABELL\,0901 (right), respectively.
\label{fig:g+_each}
\end{figure*}

Figure~\ref{fig:g+_each} shows the tangential distortion profiles as a
function of the projected cluster-centric radius for two example
clusters, ABELL\,2390 and ABELL\,0901.  The former is among the most
massive in the sample and the latter among the least massive.  The
tangential shear clearly decreases from the cluster centre to the
outskirts, with the less massive cluster, ABELL\,0901, presenting an
overall shear signal of approximately half that of the more massive
cluster, ABELL\,2390.  Note that the $45$ degree rotated component
times the clustercentric radius $r$, $r\Delta\Sigma_\times$, is
  consistent with zero -- i.e.\ this simple test of residual
  systematics is consistent with zero.

Table~\ref{tab:mass} lists $M_\Delta$ from our weak-lensing analysis,
defined as $M_\Delta=\Delta\rhoc(z)4\pi r_\Delta^3/3$ where $\rhoc(z)$
is the critical density of the universe at the respective cluster
redshifts, and $\Delta=\Delta_{\rm vir}$, $200$, $500$, $1000$, and
$2500$.  We also list $M_\Delta$ defined as $M_\Delta=\Delta\rho_{\rm
  m}(z)4\pi r_\Delta^3/3$, where $\rho_{\rm m}(z)=\rhoc(z)\Omega_{\rm
  M}(z)$ is the mean matter density of the universe, and $\Delta=180$
and $200$.  We denote these latter two masses as $M_{\rm 180m}$ and
$M_{\rm 200m}$ respectively.

\input{tab-mass.tex}

\subsection{Mass-Concentration Relation}\label{sec:MC}

Numerical simulations \citep[e.g.][]{Bullock01, Duffy08,
  Bhattacharya13, Diemer14, Meneghetti14, Ludlow14} predict that the
halo concentration $c_\Delta$ and the mass $M_\Delta$ for the NFW mass
model is weakly anti-correlated.  Such a correlation is naturally
explained by the hierarchical structure formation, that is, less
massive halos first form and more massive halos form through mass
accretion and mergers of smaller objects.  The characteristic central
density of more massive halos is lower as reflected by the critical
mass density of the universe at the redshift of collapse.
Measurements of cluster mass and concentration therefore provide us
with a unique opportunity to test structure formation.

\begin{figure}
  \includegraphics[width=\hsize]{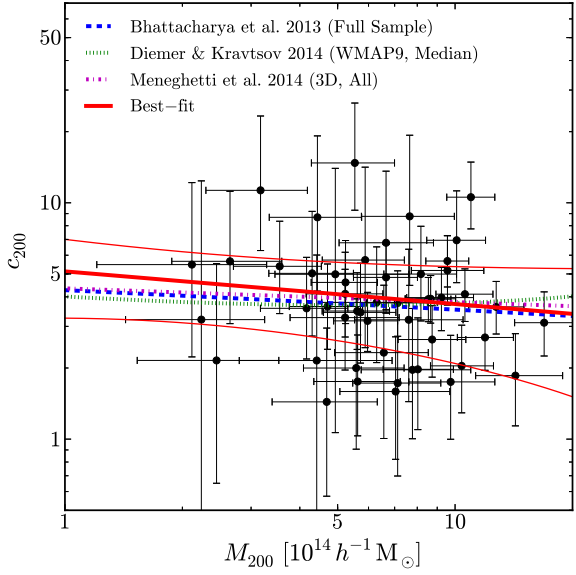}
  \caption{The observed distribution of the concentration parameters
    $c_{\rm 200}$ as a function of the cluster masses $M_{\rm 200}$ for
    50 clusters. The errors denote 68\% confidence intervals.  The thick
    and thin lines (red) are the best-fit function and the errors,
    respectively.  The dashed blue, dotted green and dotted-dashed
    magenta lines are the mean mass-concentration relation from recent
    numerical simulations of \citet{Bhattacharya13}, \citet{Diemer14}
    and \citet{Meneghetti14} at $z_l=0.23$, respectively.}
  \label{fig:MC}
\end{figure}

Our cluster sample is selected purely on X-ray luminosities without
imposing any requirement on the physical properties of the clusters.
In particular, we do not select on the dynamical state of clusters as
inferred from their X-ray morphology.  We are therefore able to
investigate the correlation between mass and concentration for a large
sample of clusters that is unbiased beyond that which is inherent
  to an X-ray selection.  A typical cluster in our sample has a
concentration of $c_{200}\simeq4$ (Figure~\ref{fig:MC}), with central
values of $c_{200}$ in the range $c_{200}\sim2-20$.  We quantify the
mass-concentration correlation with the following function:
\begin{eqnarray}
c_\Delta(M_\Delta)=c_0\left(\frac{M_\Delta}{10^{14}\hMsol}\right)^b,\label{eq:MC}
\end{eqnarray}
where $c_0$ and $b$ are the normalization of the concentration
parameter at $M_\Delta=10^{14}\hMsol$ and the slope, respectively.
This form is motivated by the studies of the numerical simulations
\citep[e.g.][]{Bullock01}.  Note that we ignore redshift evolution in
this model because the redshift range of our sample is narrow.  When
we fit this model, we take account of the correlation between the
errors on concentration and mass by calculating the error covariance
matrix, and include the intrinsic scatter of the lensing-based
concentration parameter, $\sigma_{\rm int}$.  The log-likelihood is
given by:
\begin{eqnarray}
-2\ln {\mathcal L}&=&\sum_i\ln\left(\sigma_{\ln c,i}^2+b^2\sigma_{\ln
			   M,i}^2-2b\sigma_{\ln c,\ln M,i}+\sigma_{\rm
			   int}^2\right) \nonumber \\ 
& &+\sum_i \frac{(\ln(c_i)-(a+b\ln(M_i)))^2}{\sigma_{\ln
 c,i}^2+b^2\sigma_{\ln M,i}^2-2b\sigma_{\ln c,\ln M,i}+\sigma_{\rm
 int}^2} \nonumber, 
\end{eqnarray}
where $a=\log(c_0)$ and $\sigma_{\ln c}$ and $\sigma_{\ln M}$ are the
fractional errors of the concentration and the mass, $\sigma_{\ln
    c,\ln M}$ is the error correlation, and $\sigma_{\rm int}$ is the
  intrinsic scatter in $\ln c$.  We perform the fitting at
overdensities of $\Delta=180{\rm m},200{\rm m},{\rm vir}$, and $200$.
The normalisation of our best-fit mass-concentration relation is in
excellent agreement with the results of recent numerical simulations
at $z_l=0.23$ (Table~\ref{tab:MC}; Figure~\ref{fig:MC};
\citealt{Bhattacharya13, Diemer14, Meneghetti14}).  As an aside,
  we note that these three recent independent theoretical studies
  agree both with each other and with our observational results,
  whilst older simulations showed considerable variation between their
  respective mass-concentration relations and a lower overall
  normalization \citep[e.g.][]{Duffy08,Stanek10}.  The best-fit
slopes agree with the weak-mass dependence of the concentration,
$b\simeq-0.1$, seen in simulations (Table~\ref{tab:MC}), although the
uncertainties are too large to rule out positive values of $b$.
Adding less massive clusters and increasing the number density of
background galaxies will allow improved constraints in future studies.

\begin{table}
  \caption{Best-fit parameters for the mass concentration
    relation. } \label{tab:MC}
  \begin{center}
    \begin{tabular}{cccc}
      \hline
      \hline
      $\Delta$ & $c_0$ &$b$ & $\sigma_{\rm int}$\\
      \hline
      $180{\rm m}$ & $9.24^{+3.99}_{-2.50}$ & $-0.20_{-0.15}^{+0.13}$ &
      $<0.17$ \\
      $200{\rm m}$ & $8.74^{+3.75}_{-2.49}$ & $-0.19_{-0.15}^{+0.14}$ &
      $<0.17$ \\
      Virial & $7.26^{+3.18}_{-2.07}$ & $-0.17_{-0.16}^{+0.15}$ & $<0.18$\\
      $200$ & $5.12^{+2.08}_{-1.44}$ & $-0.14_{-0.16}^{+0.16}$ & $<0.20$\\
      \hline
    \end{tabular}
  \end{center}
\end{table}

\subsection{Stacked Lensing Analysis}\label{sec:stackedWL}

\begin{figure}
  \includegraphics[width=\hsize]{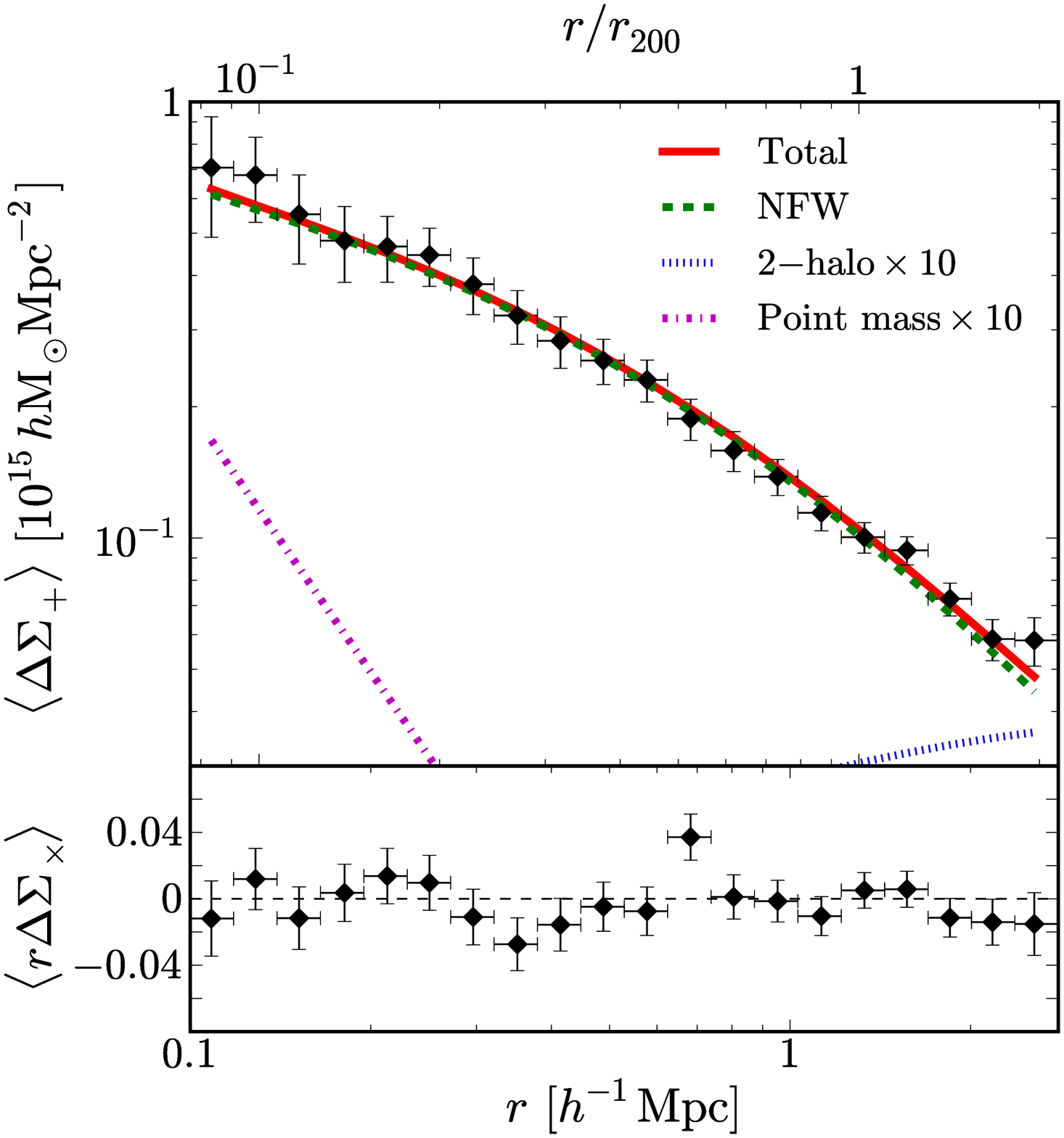}
  \caption{Stacked tangential shear profile for 50 clusters.  The
    errors are composed of $(C_{g,ii}+\tilde{C}_{s,ii}+C_{{\rm
        LSS},ii})^{1/2}$.  Thick solid red and dashed green lines are
    the total mass and the NFW model, respectively.  The dotted blue
    and dashed-dotted magenta lines are the two-halo term and point
    source multiplied by 10, respectively.}
  \label{fig:stacked_g+}
\end{figure}

Stacked lensing analysis is a powerful technique for measuring the
average density profile of a sample of clusters.  Stacking the shear
signal from a sample of clusters averages over the distribution of
internal structures and halo triaxiality, and thus overcomes the
structural biases suffered by some individual cluster mass
measurements \citep[e.g.][]{Mandelbaum06, Johnston07, Okabe10b,
  Okabe13, Umetsu11, Umetsu14, Umetsu15b,Oguri12, Leauthaud12, Miyatake13, Niikura15}.

We compute the average lensing signal in physical length unit centered
on the respective BCGs.  Note that our redshift range is narrow, and
therefore the results described below are unchanged if we instead use
comoving length units.  Moreover, we have previously tested that
adopting physical length units, and not scaling length to an
overdensity radius, yields an unbiased measurement of the stacked
shear profile of our sample \citep{Okabe13}.  The innermost radius of
the stacked shear profile is that at which the innermost bin of the
stacked profile contains a minimum of one background galaxy from each
cluster.  The outermost radius of the stacked profile is the median of
the maximum physical scale on which the field of view of the Subaru
observations fully encloses a circular aperture centered on each BCG.
Note that this simultaneously matches the angular extent of the data,
and satisfies the requirement placed on the innermost radius.  The
stacked shear profile decreases smoothly as a function of
clustercentric radius (Figure~\ref{fig:stacked_g+}), and yields a
signal-to-noise ratio of is $S/N\simeq35.6$, after taking into account
the LSS covariance matrix, $C_{\rm LSS}$.

To interpret the average mass profile from the stacked lensing
signals, we consider three mass components,
\begin{eqnarray}
  \Delta \Sigma_{\rm model}=\Delta \Sigma_{\rm pt} +\Delta \Sigma_{\rm NFW}+\Delta \Sigma_{\rm 2h},
\end{eqnarray}
where $\Delta \Sigma_{\rm pt}$ is a point mass associated with the
BCGs, $\Delta \Sigma_{\rm NFW}$ is the large-scale cluster mass
distribution that we parametetrise following NFW, and $\Delta
\Sigma_{\rm 2h}$ is the two-halo term
\citep[e.g.][]{Johnston07,Oguri11a,Oguri11b} to account for structure
adjacent to the clusters.  Note that the latter two terms were ignored
in the modeling of individual clusters because the noise level in
individual cluster shear profiles renders them insensitive to these
contributions.

We describe the contribution from the point mass, of mass $M_{\rm
  pt}$, as:
\begin{eqnarray}
  \Delta \Sigma_{\rm pt}=\frac{M_{\rm pt}}{\pi r^2},
\end{eqnarray}
and adopt a prior on $M_{\rm pt}$ based on the stellar mass for the
BCG.  The stellar mass of each BCG is estimated from the $K$-band
luminosity with a \citet{Salpeter55} initial mass function.  The prior
on the point mass then matches the mean and standard deviation of the
BCG stellar masses.  The two-halo term is computed following the
formulation of \citet{Oguri11b}.  We use the WMAP9 cosmology
\citep{WMAP09} to compute the linear power spectrum.  Given the
average mass and redshift for an ensemble of clusters, $\Delta
\Sigma_{2h}$ is proportional to ${\bar \rho_m(z_l)} b_h(M)$, where
$b_h(M)$ is the halo bias.  To estimate $b_h(M)$, we use a single
scaling relation \citep{Tinker10} which is calibrated by a large set
of numerical simulations.

\begin{figure}
\includegraphics[width=\hsize]{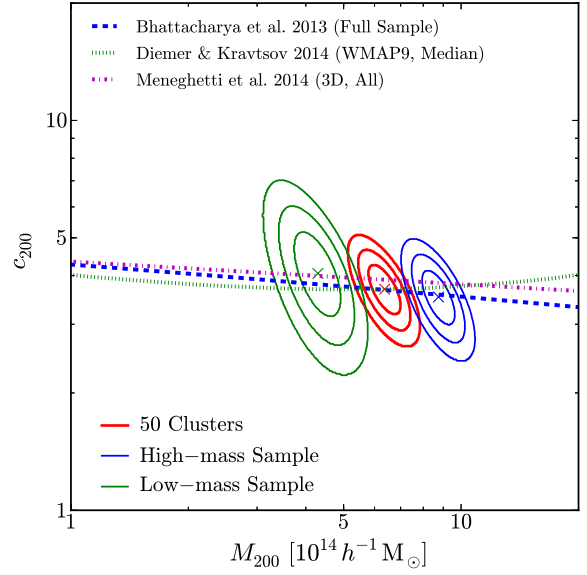}
\caption{Confidence intervals on mass and concentration from
    stacked lensing analysis. From left to right, the contours show
    the low-mass sample (thin green), the full sample (thick red), and
    the high-mass sample (thin blue).  Crosses denote the best-fit
    parameters and the contours show the 68.3\%, 95.4\%, and 99.7\%
    confidence levels.  The dashed, dotted and dashed-dotted lines are
    the same as Figure \ref{fig:MC}. }
\label{fig:MC_stacked}
\end{figure}

\begin{table}
  \caption{Best-fit parameters for stacked lensing analysis for 50 clusters.} \label{tab:Mass_stacked}
  \begin{center}
    \begin{tabular}{ccc}
      \hline
      \hline
      $\Delta$ & $M_{\Delta}$ & $c_{\Delta}$  \\
      & $10^{14}\hMsol$& \\
      \hline
      $180{\rm m}$ & $8.66_{-0.43}^{+0.45}$  & $5.57_{-0.35}^{+0.37}$ \\
      $200{\rm m}$ & $8.39_{-0.41}^{+0.43}$ & $5.32_{-0.33}^{+0.35}$ \\
      vir & $7.65_{-0.36}^{+0.38}$ & $4.68_{-0.30}^{+0.31}$\\
      200 & $6.37_{-0.27}^{+0.28}$ & $3.69_{-0.24}^{+0.26}$ \\
      \hline
    \end{tabular}
  \end{center}
\end{table}

The best-fit model describes the data very well
(Figure~\ref{fig:stacked_g+}, Table~\ref{tab:Mass_stacked}).  The
two-halo term is an order of magnitude less than the NFW model, with
an estimated halo bias of $b_h(M_{200})\simeq5.5$.  The point mass is
constrained by the upper limit, $M_{\rm pt}<6.19\times10^{11}\hMsol$.
The mass and the concentration at $\Delta=200$
(Figure~\ref{fig:MC_stacked}) is in excellent agreement with both
numerical simulations \citep{Bhattacharya13,Diemer14,Meneghetti14} and
individual cluster mass measurements (Section \ref{sec:MC}).
\cite{Okabe13} conducted a similar stacked lensing analysis using a
background galaxy catalogue based on a single colour cut, and based on
the \citet{Ilbert09} COSMOS photometric redshift catalogue.  The
stacked shear signal presented here is consistent with \cite{Okabe13}.
The measurement uncertainties on the shear signal decrease as radius
increases in this study due to radial dependence of our colour cut;
this increases the weight of the outer bins in our fit, relative to
that of \citet{Okabe13}.  Therefore, the mass and concentration from
our new stacked analysis are marginally higher and lower than
\cite{Okabe13} respectively.  

 We compare the stacked result at $\Delta=200$, shown in
  Table~\ref{tab:Mass_stacked} with the lognormal mean of the individual cluster
  mass and concentration measurements listed in Table~\ref{tab:MC},
  finding excellent agreement, with the latter being $\langle
  M_{200}\rangle=6.38\pm0.24\times10^{14}\hMsol$ and $\langle
  c_{200}\rangle=3.73\pm0.38$. 

We also divide the clusters into two sub-samples of 25 clusters based
on the virial mass measured from the individual cluster shear profile
models (Section \ref{sec:MC}), adopting $M_{\rm vir}=8\times
10^{14}\hMsol$ as the dividing line between the sub-samples.  We
calculate the stacked shear profile for both sub-samples, and fit
models, following the procedures applied to the full sample.  The
results are in excellent agreement with both numerical simulations
\citep{Bhattacharya13,Diemer14,Meneghetti14} and the best-fit
mass-concentration relation for individual cluster mass
measurements (Table~\ref{tab:MC}~\&~\ref{tab:Mass_stacked}; Figure
\ref{fig:MC_stacked}).

\begin{figure*}
  \includegraphics[width=0.45\hsize]{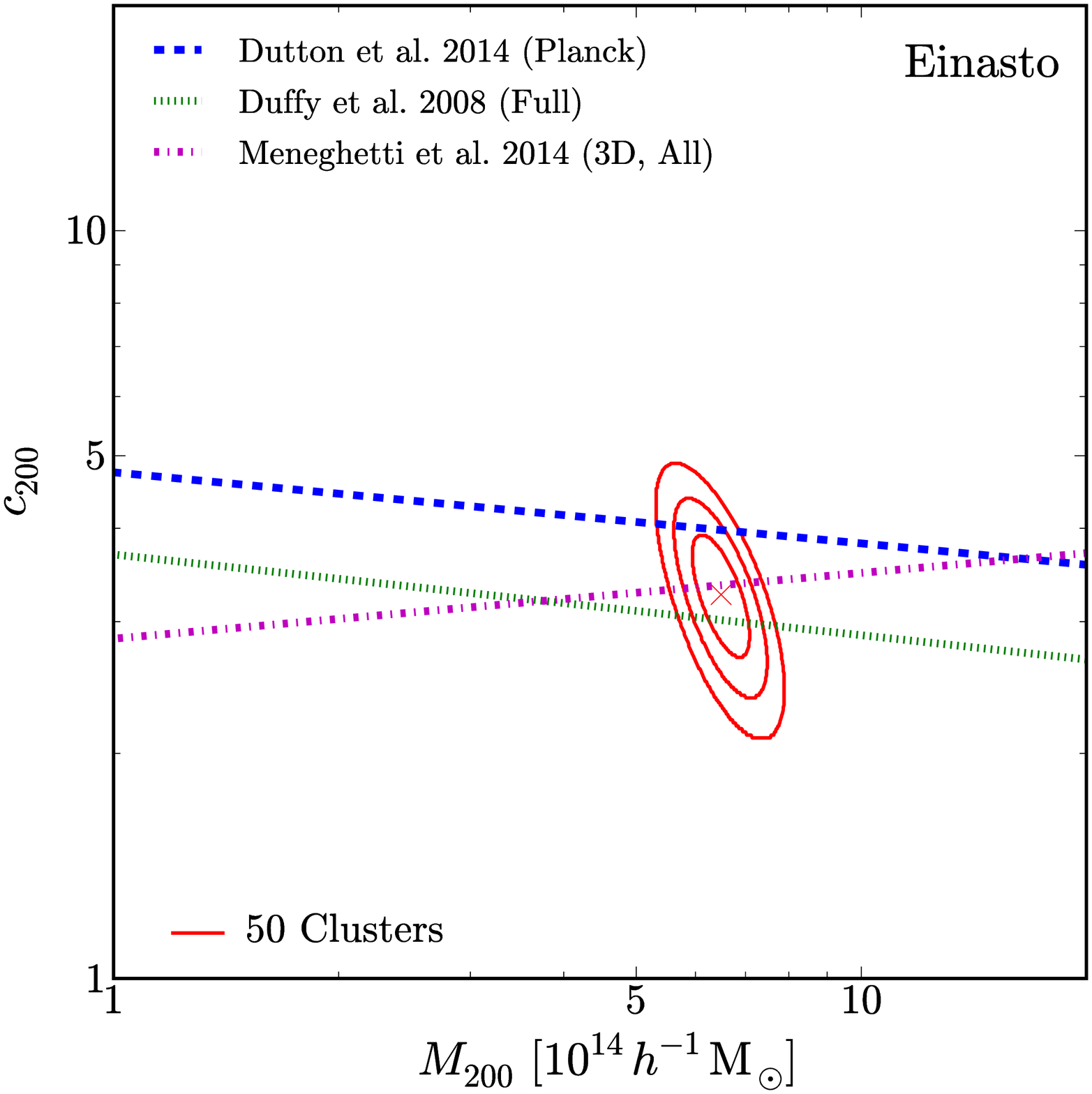}
  \includegraphics[width=0.45\hsize]{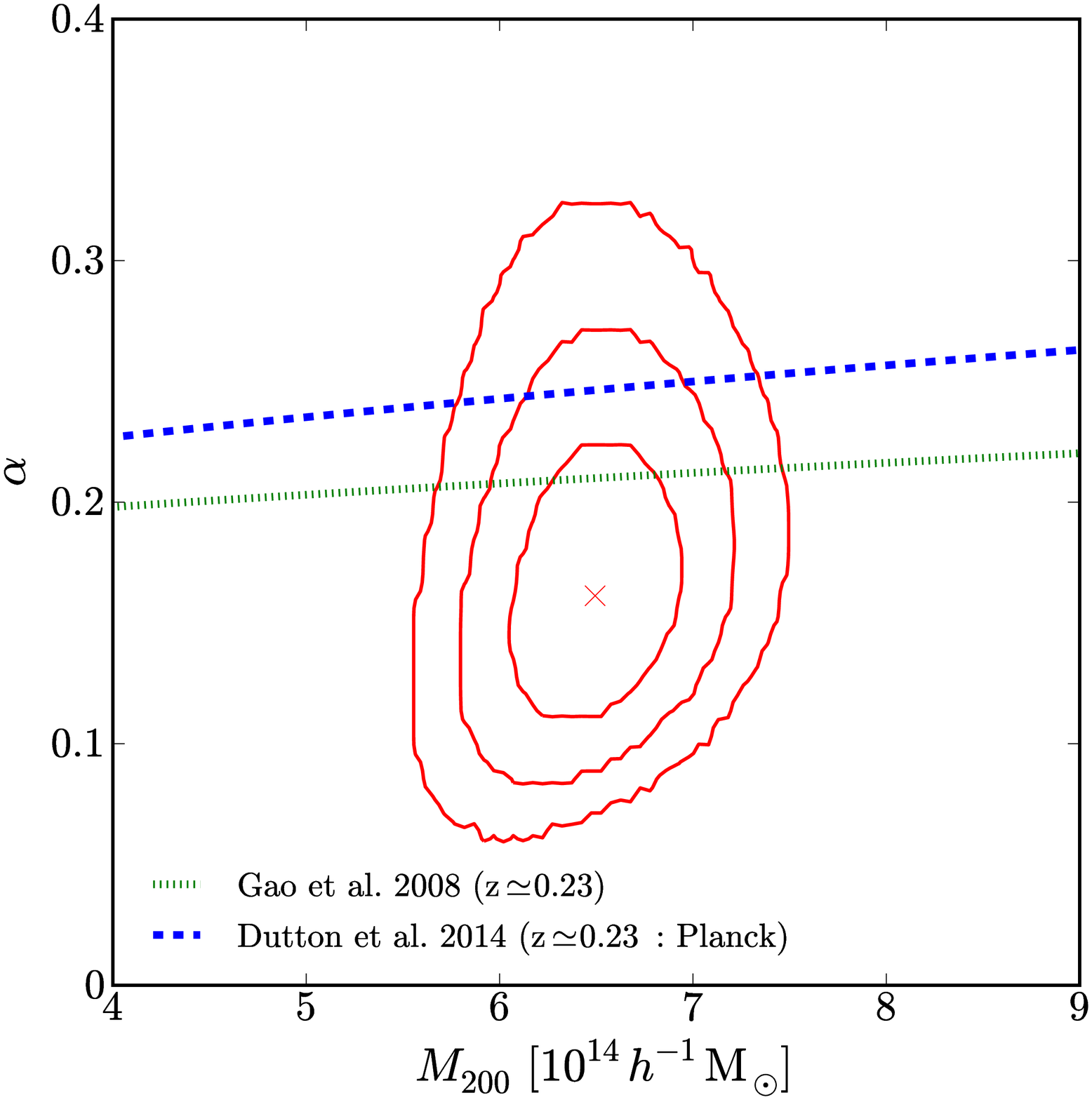}
  \caption{Mass-concentration relation (left) and $\alpha$-mass
    relation (right) for the Einasto profile obtained by the stacked
    lensing analysis.  The cross denotes the best-fit parameters and
    the contours show the 68.3\%, 95.4\%, and 99.7\% confidence
    levels. Left : blue dashed, green dotted and magenta dashed-dotted
    lines denote the mass-concentration relation derived by
    \citet{Dutton14},\citet{Duffy08}, and \citet{Meneghetti14},
    respectively.  Right: blue dashed and green dotted lines are the
    $\alpha$-mass relation from \citet{Dutton14} and \citet{Gao08},
    respectively.  }
  \label{fig:Einasto}
\end{figure*}

Some numerical simulation indicate that an \cite{Einasto65} profile
describes the spherically averaged mass density profile for simulated
halos better than the NFW profile \citep{Navarro04,Gao12,Klypin14}.
The Einasto profile has the form:
\begin{eqnarray}
\frac{d\log\rho}{d\log r}=-2\left(\frac{r}{r_{-2}}\right)^{\alpha},
\end{eqnarray}
where $r_{-2}$ is a scale radius at which the logarithmic slope is
$-2$ and $\alpha$ is a shape parameter to describe the degree of
curvature of the profile.  The Einasto profile is specified by three
parameters of $M_\Delta$, $c_\Delta=r_\Delta/r_{-2}$ and $\alpha$.  We
measure these three parameters by fitting the stacked lensing profile
for all 50 clusters.  As demonstrated by the NFW fitting, the
contribution from the point mass is negligible compared to that of the
main halo in the radial range $0.1-2.8\hMpc$.  We therefore just fit
the Einasto profile and two-halo term.  The best-fit parameters are
 $\alpha=0.161_{-0.041}^{+0.042}$,
  $M_{200}=6.49_{-0.29}^{+0.31}\times10^{14}\hMsol$ and
  $c_{200}=3.26_{-0.54}^{+0.39}$.  These constraints on $M_{200}$ and
$c_{200}$ are in excellent agreement with the NFW-based measurements
(Table~\ref{tab:Mass_stacked}), and also agree within $\sim1\sigma$
with predictions from numerical simulations (Figure
\ref{fig:Einasto}); \citealt{Duffy08, Gao12, Bhattacharya13, Diemer14,
  Meneghetti14}).  As noted above, the agreement between our results
and the predictions from 2013-2014 is excellent. More precise
observational constraints on the density profile shape of clusters,
including mass dependence of the Einasto profile parameters await
larger cluster samples, for example from the Dark Energy Survey
  (DES), Hyper Suprime-Cam survey (HSC) and the Large Synoptic Survey
  Telescope (LSST).

\section{Discussion}\label{sec:discuss}

In Section~\ref{sec:syst} we quantify the remaining systematics
in our analysis, in Section~\ref{sec:errorbudget} we summarize our
overall error budget, and in Section~\ref{sec:compare} we compare our
mass measurements with results from the literature.

\subsection{Systematics}\label{sec:syst}

In Section~\ref{sec:sim} we test the methods described in
Section~\ref{sec:fitting} and that we use in Section~\ref{sec:mass} to
fit NFW models to the observed shear profiles.  In
Section~\ref{sec:sysShear} we correct the shear signal for the small
colour selection and galaxy shape measurement biases calculated in
Sections~\ref{sec:shape}~\&~\ref{sec:bkg} and re-fit the NFW models to
the corrected shear profiles.  In Section~\ref{sec:Mass_Pz} we
calibrate the impact of using the full photometric redshift
probability distribution of the background galaxies on our mass
measurements.  In Section~\ref{sec:boost} we consider the impact of
forcing the number density profile of background galaxies to be flat.
(We emphasize again that in our analysis and results we do not assume
the number density profile to be flat.)

\subsubsection{Simulation Tests}\label{sec:sim}

The radial range over which recent cluster weak-lensing studies
\citep[e.g.][]{Israel12,Melchior14,Applegate14,Hoekstra15} have
modeled the shear profile has been motivated in part by results from
numerical simulations.  Here, we expand upon \citet{Okabe13}, to test
our individual mass measurements (Section \ref{sec:mass}) using
synthetic weak shear catalogues based on simulated clusters and toy
models.  The former have the advantage of incorporating the full
effects of the large-scale structure that surrounds massive clusters,
whilst the latter have the advantage of toy model clusters having
perfectly known properties, and the properties of the background
galaxy catalogues are matched to the observational data.  Importantly,
we calibrate the specific model fitting method that we apply to our
observational data directly on simulations whose properties match our
own sample and data.

We use mock observations of clusters from the ``Cosmo-OWLS''
  cosmological hydrodynamical simulation that reproduces a large
  number of local galaxy cluster scaling relations, within a
  $400\hMpc$ box \citep{LeBrun14,McCarthy14}.  
We use the simulations that include  
cooling, star formation, supernova feedback and AGN feedback with a 
heating temperature $\Delta T_{\rm heat}=10^8$K, known as the {\sc AGN} 8.0 model.
Weak-lensing
catalogues comprising 100 galaxies per ${\rm arcmin}^2$ were
constructed following \citet{Bahe12}.  Specifically, the mock
  observations include the effect of shape noise, cluster substructure
  and triaxiality, and correlated large-scale structure,
  and ignore uncorrelated large-scale structure and observational
  effects such as uncertainties in galaxy shape measurements and
  redshifts.  Note that ignoring uncorrelated large-scale structure is
  not expected to affect the measurement of possible biases in mass
  measurements via tests such as those described here
  \citep{Hoekstra11}.  The mass of the simulated clusters spans
  $5\times10^{14}h_{73}^{-1}M_\odot
  <M_{200}<17\times10^{14}h_{73}^{-1}M_\odot$ at a redshift of
$z_l=0.23$ with the WMAP7 cosmology of $\Omega_{m,0}=0.272$ and
  $\Omega_\Lambda=0.728$.
We randomly extracted galaxies from the
parent synthetic weak shear catalogues to match statistically the
cluster-centric number density profiles of colour-selected background
galaxies in the observational analysis (Section \ref{sec:bkg}), and
fitted NFW models to the shear profiles following exactly the
procedure laid down in Sections~\ref{sec:fitting}~\&~\ref{sec:mass}.
This was repeated for 30 realizations, each containing 23 simulated
clusters.

\begin{figure}
  \includegraphics[width=\hsize]{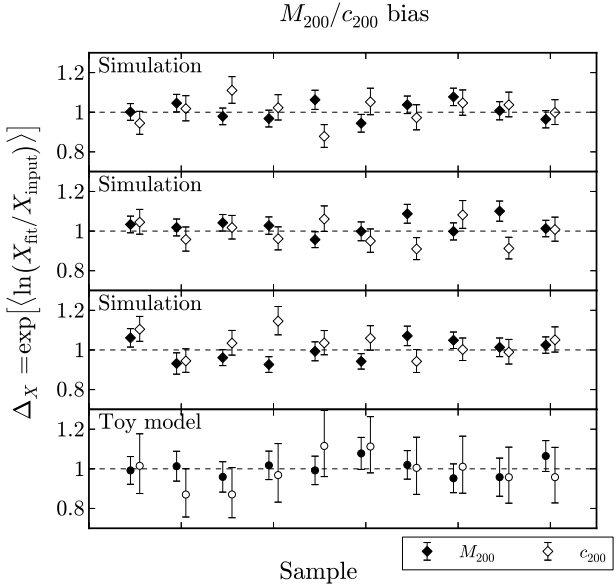}
  \caption{Calibration tests for weak-lensing mass measurements
    ($M_{200}$,$c_{200}$). The top three panels and the bottom panel
    represent the results based on the numerical simulations
    \citep[diamonds][; 23 clusters]{McCarthy11,Bahe12} and the toy models
 (circles; 50 clusters),
    respectively. The filled and open symbols denote the geometric means
    for the mass and the concentration, respectively.}
  \label{fig:fit_sys}
\end{figure}

To quantify the mass measurement, we define $\Delta_X$ in terms of the
geometric mean: $\Delta_X=\exp\left(\langle \ln(X_{\rm fit}/X_{\rm
  input})\rangle\right)-1$, following \cite{Umetsu14}, and similar to
the methods of \cite{Becker11}.  Here, $X_{\rm fit}$ is the best-fit
mass or concentration.  We recover the input $M_{200}$ and $c_{200}$
from the numerical simulations with negligible bias.  The mean
  bias on the mass and concentration measurements across the full
  suite of realizations of the simulations is $\ls1$ per cent.  The
  scatter between measurements of the bias on mass using individual
  realizations is $4.8$ per cent, which is comparable with the
  measurement uncertainty of $4.4$ per cent on the bias from an
  individual realization.  Likewise the realization-to-realization
  scatter in bias on concentration is $6.2$ per cent, with a typical
  measurement uncertainty on an individual realization of $6.3$ per
  cent (upper three panels of Figure~\ref{fig:fit_sys}).

We repeat this test using cluster density profile models based on
analytic NFW halos, and construct synthetic background galaxy
catalogues that match the observed catalogues as closely as possible.
For each of 50 analytic cluster profiles we adopt the observed
positions of background galaxies and randomly draw a galaxy from the
full background galaxy sample across all 50 clusters, thus
simultaneously randomising the galaxy orientations, and matching
statistically the source redshift distribution.  The NFW parameters
are randomly chosen from the measured values for our cluster sample.
We compute the synthetic shear profile for each of these 50 analytic
clusters 10 times and fit an NFW model following the procedures laid
down in Sections~\ref{sec:fitting}~\&~\ref{sec:mass}.  Again, we
recover the input masses and concentrations with $\ls1$ per cent
bias.  The scatter between realizations is $4.3$ and $8.9$ per cent on
mass and concentration respectively, and the typical measurement
uncertainty on individual realizations is $7.7$ and $14.5$ per cent on
mass and concentration respectively (lowest panel of
Figure~\ref{fig:fit_sys}).

In summary, we conclude that our shear profile fitting algorithm, a
key feature of which is the adaptive choice of binning scheme,
recovers the mean mass of our sample with negligible bias.

\subsubsection{Shear Calibration and Background Selection}\label{sec:sysShear}

In Sections~\ref{sec:shape}~\&~\ref{sec:bkg} we developed methods to
measure the shape of faint galaxies and select faint red galaxies as
background galaxies with small systematic biases of $3$ and $1$ per
cent respectively; both acting in the sense that we slightly
under-estimate cluster mass.  Here we estimate how these biases
propagate through to the actually cluster mass measurements.

The shape measurement bias is expressed as the multiplicative shear
calibration factor, $m=-0.03$, following the STEP programme. Given
$m$, we therefore correct the measured tangential shear signal by
$\Delta\Sigma_+\rightarrow\Delta\Sigma_+ (1+m)^{-1}$ and repeat the
tangential shear fitting described in Section~\ref{sec:fitting}.  We
express the comparison between the original masses
(Table~\ref{tab:mass}) and the corrected masses we define $\Delta_M$
in terms of the geometric mean:
$\Delta_M=\exp\left(\langle\ln(M_\Delta^{\rm orig}/M_\Delta^{{\rm
    corr}})\rangle\right)-1)$.  We find that the corrected masses are
$\sim3-5\%$ higher than the original masses, the range of values
reflecting the non-linearity of tangential shear profiles (Table
\ref{tab:sys}).

Turning to the colour selection of background galaxies, the right
panel of Figure \ref{fig:dilution} shows that the contamination levels
are $1$ per cent at $r\simlt1.3\hMpc$ and well below this level
at $r\simgt1.3\hMpc$.  We therefore boost the shear signals at
$r\simlt1.3\hMpc$ and re-derive the cluster masses, again following
Section~\ref{sec:fitting}.  Expressing the comparison in the same
manner as above, we found that the masses corrected for
contamination are within $\ls1$ per cent of the original masses
(Table \ref{tab:sys}).

For completeness, we combine these two shear correction terms in
quadrature to give an effective multiplicative bias of
  $m=-0.032$, and evaluate the mass measurement bias of the combined
shear calibration and contamination effects.  As expected, the bias is
mainly attributed to the shear calibration, with the combined
correction yielding results indistinguishable from the pure shear
calibration correction.  Individual cluster masses based on the
corrected tangential shear profile are given in
Appendix~\ref{sec:app1}.

\subsubsection{Mass Estimates with photometric redshift P(z)}\label{sec:Mass_Pz}

\begin{figure*}
  \includegraphics[width=\hsize]{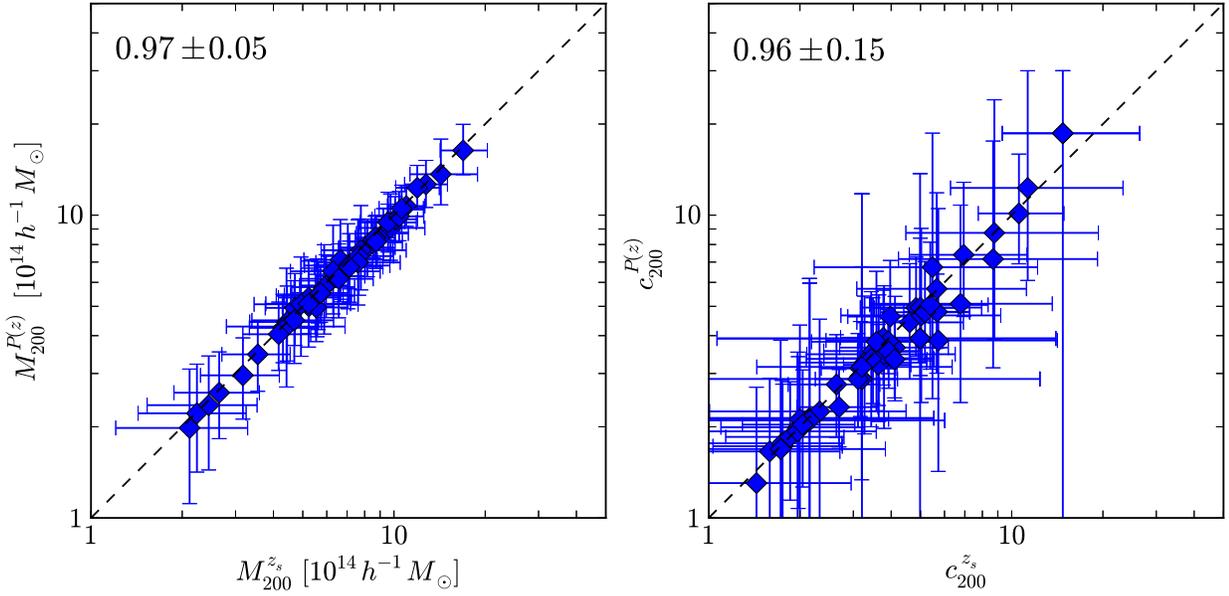}
  \caption{A comparison of the mass (left) and the concentration (right) at $\Delta=200$,
    estimated by the single redshift, ($z_s$; Section \ref{sec:mass})
    and the probability density function ($P(z)$; Section \ref{sec:Mass_Pz}).}
  \label{fig:Mzs_vs_MPz}
\end{figure*}

In Section \ref{sec:photometry} we adopted as the redshift of each
faint galaxy in our sample, the median of the stacked posterior
probability distribution of the nearest 100 neighbours in the
$(V-i)-i$ space of the COSMOS catalogue.  We therefore essentially
adopted a point estimate of the redshift of each of our galaxies.
However it is well known that the photometric redshift probability
distribution of galaxies can be asymmetric, and present multiple
peaks.  The full photometric redshift probability density function,
$P(z)$, fully describes such implicit systematic uncertainties.
Indeed some recent studies have used the full $P(z)$ for some clusters
in their weak-lensing sample \cite[e.g.][]{Applegate14}.  

Here, we test whether our method that ignores the full $P(z)$
available from the COSMOS survey suffers any significant bias.  In a
similar vein to Section~\ref{sec:photometry}, we estimate the full
$P(z)$ probability function of individual galaxies as an ensemble
average of $P^{\rm COSMOS}(z)$ for 100 neighbouring COSMOS galaxies in
the colour-magnitude plane,
\begin{eqnarray}
 P(z)=\frac{1}{N_{\rm nei}}\sum_j^{N_{\rm nei}} P_j^{\rm COSMOS}(z).
\end{eqnarray}
Given the probability function, the tangential shear component can
then be calculated as follows:
\begin{eqnarray}
 \langle \Delta \Sigma_+ \rangle (r_i)=\frac{\sum_n\int_{z_l}^\infty
  g_{+,n}\Sigma_{\rm cr,n}(z_s) w_n(z_s) P(z_s) dz_s}{\sum_n\int_{z_l}^\infty
  w_n(z_s) P(z_s) dz_s}. \label{eq:g+_Pz}
\end{eqnarray}
The errors for the shape noise ($C_{g}$) and the photometric redshift
($\tilde{C}_{s}$) are estimated as:
\begin{eqnarray}
 C_{g,ij}=\frac{1}{2}\frac{\sum_n\int_{z_l}^\infty
  \sigma_{g,n}^2\Sigma_{\rm cr,n}^2(z_s) w_n^2(z_s) P(z_s) dz_s}{(\sum_n\int_{z_l}^\infty
  w_n(z_s) P(z_s) dz_s)^2}\delta_{ij},\nonumber
\end{eqnarray}
and
\begin{eqnarray}
 \tilde{C}_{s,ij}=\frac{\sum_n\int_{z_l}^\infty
  (g_{+,n}\Sigma_{\rm cr,n}(z_s)w_n(z_s)  -\langle \Delta \Sigma_+ \rangle)^2 P(z_s) dz_s}{\sum_n\int_{z_l}^\infty
  P(z_s) dz_s}\delta_{ij},\nonumber
\end{eqnarray}
respectively. 

We select the same background galaxies used in our main analysis and
compute the tangential shear profiles using Equation~(\ref{eq:g+_Pz}).
The radial bins are chosen using the same method as in our main
analysis.  The difference in the tangential shear components estimated
by the single source redshift and $P(z)$ is calculated using:
\begin{eqnarray}
  \Delta_+^2=\frac{1}{N_{\rm bin}}\sum_i \frac{(\langle \Delta
    \Sigma_+^{z_s} \rangle-\langle \Delta \Sigma_+^{P(z)}
    \rangle)^2}{(\sigma_+^{z_s})^2+(\sigma_+^{P(z)})^2},
\end{eqnarray}
where $\sigma_+^2=C_g+\tilde{C}_s$. The deviation is, on average,
  $\Delta_+\simeq0.05$, with smaller deviations of $\Delta_+\sim0.02$
  at small radii ($r<0.3\hMpc$) and larger deviations of
  $\Delta_+\sim0.07$ at large radii ($r>1\hMpc$).  This is because the
  errors at small radii are larger, due to a the relatively small
  number of background galaxies in bins of smaller solid angle.  When
  we ignore the errors, the average deviations are still negligible,
  $\Delta_+\sim0.01$ at $r<0.3\hMpc$ and $\Delta_+\sim0.002$ at
  $r>1\hMpc$.   We also compare the best-fit mass and concentration
parameters (Figure~\ref{fig:Mzs_vs_MPz}).  The two measurements are in
excellent agreement, with geometric means of $0.97\pm0.05$ and
 $0.96\pm0.15$ for $M_{200}$ and $c_{200}$, respectively.  We
also made a background galaxy catalogue using the $P(z)$ function and
our radius-dependent colour-cut (Section~\ref{sec:bkg}), and found
that again the mass measurements do not change significantly
($0.96\pm0.05$ and $0.96\pm0.15$ for $M_{200}$ and $c_{200}$).
We conclude that with the current sample and data we are unable to
detect any systematic difference between the mass and
  concentration measurements based on the mean of COSMOS point
estimates of the redshift of individual galaxies and the full COSMOS
$P(z)$ function.

\subsubsection{Boost factor}\label{sec:boost}

\begin{figure*}
  \includegraphics[width=\hsize]{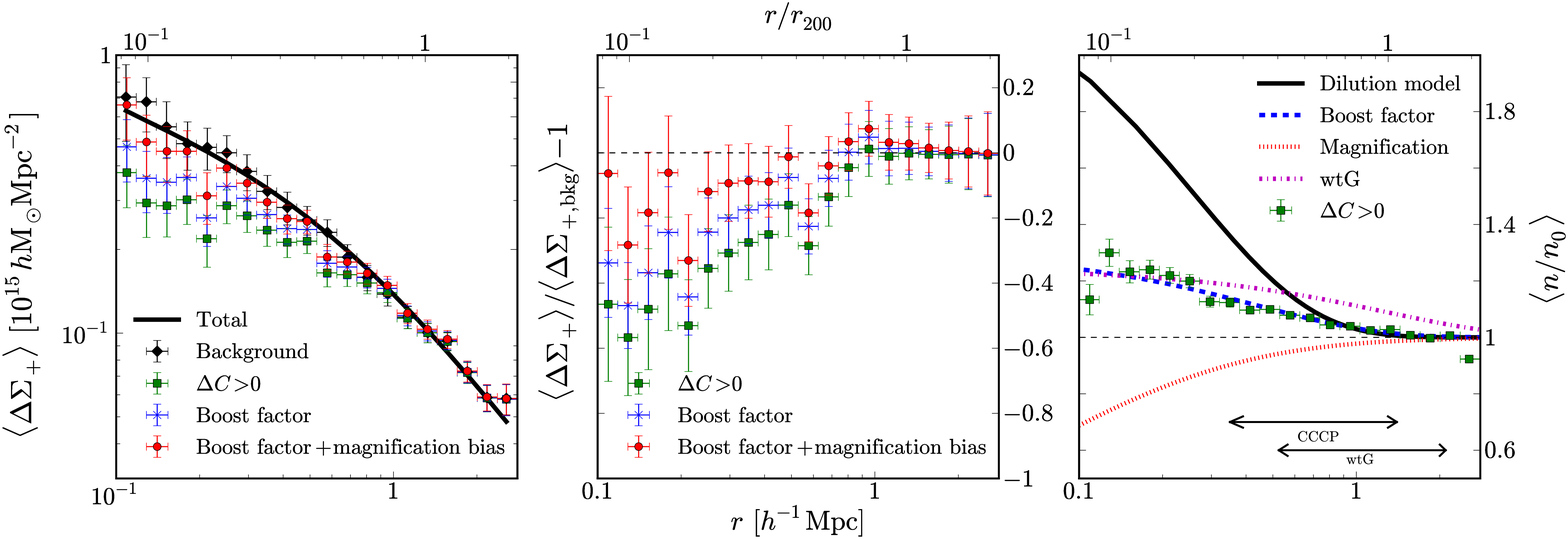}
  \caption{Systematic mass measurement errors caused by the boost
    factor. Left: the stacked tangential shear profiles. Back diamonds
    and green squares denote the profile computed by our background
    catalogue and galaxies with $\Delta C>0$, respectively.  Blue
    crosses and red circles are the profile corrected by the
    boost-factor and the boost-factor taking into account the
    magnification bias, respectively. Middle: the differential
    fraction of lensing signals from the signals for our background
    catalogue. The lensing signal corrected by the boost factor, with
    the assumption of the uniform background distribution, is
    significantly underestimated. Right:the normalized number density
    profile. Green squares shows a clear excess of the density profile
    of galaxies with $\Delta C>0$. The best-fit model is represented
    by the dashed blue lines. The dotted red line denotes the depleted
    number density profile due to the magnification bias.  The solid
    black line is the dilution model (Section \ref{sec:bkg}). The
    dotted-dashed magenta line is the boost factor for
    MACS\,J0417.5$-$1154 from the WtG project \citep{Applegate14},
    showing a shallow distribution.  }
  \label{fig:boost}
\end{figure*}

Imperfect selection of background galaxies leads to under-estimated
weak-shear signals because background galaxy samples are contaminated
by faint cluster galaxies, the shapes of which present no lensing
signal due to the cluster.  The number density of cluster galaxies
decreases with increasing projected cluster-centric radius.  Thus at
fixed selection method, the ratio of cluster galaxies to background
galaxies, $f_{\rm mem}$, decreases with increasing projected
cluster-centric radius, and thereby dilutes the shear signal more at
smaller radii than at larger radii.  Our approach is to vary the
colour cut used to select background galaxies as a function of
cluster-centric radius, simply exploiting the declining number density
of cluster galaxies as a function of radius, without invoking any
physical assumptions.  An alternative
\citep[e.g.][]{Kneib03,Applegate14,Hoekstra15} is to correct the
measured shear signal by a factor $1+f_{\rm mem}$, and thus to assume
that the observed number density profile of background galaxies is
flat.  This correction is referred to as a {\it boost factor}.
However, the assumption of a flat observed number density profile of
background galaxies ignores magnification bias
\citep[e.g.][]{Broadhurst95,Umetsu14} -- i.e.\ the depletion or
enhancement of the number density of background galaxies due to
lensing magnification.

We compare the boost-factor method with our methods that do not
  invoke a boost factor and instead rely on selection of red galaxies
  to achieve $\le1$ per cent dilution across all radii.  For this
purpose, we construct background galaxy catalogues that suffer
dilution by selecting red galaxies with a positive colour-offset from
the red-sequence, $\Delta C>0$, and compute the stacked shear profile
for this diluted sample of galaxies.  As expected, the amplitude of
this shear profile is suppressed relative to the shear profile from
our main analysis, with the suppression increasing to $\sim50\%$ at
$\sim100\hkpc$ from the cluster centre (left and central panels of
Figure~\ref{fig:boost}).  The suppression of the signal due to
dilution appears to be negligible at $\simgt1\hMpc$.  The
contaminating population of faint cluster galaxies is seen clearly as
an excess of galaxies at $\ls1\hMpc$ in the stacked number density
profile of galaxies selected as having $\Delta C>0$ (right panel of
Figure~\ref{fig:boost}).  In other words, the excess of number density
profile is negligible beyond $r_{200}$. The evidence indicates an
internal consistency that the number density excess and the
stacked-lensing mass estimate are consistent with each other.  Note
that the number density profile in Figure~\ref{fig:boost} is
calculated after masking the solid angle subtended by bright galaxies
($i'<20$) out to elliptical radii a factor of $3$ than the elliptical
shape parameters of SExtractor -- i.e.\ corresponding to the isophotal
limit of detected objects.  We also fully consider the finite
field-of-view in the number density calculation.  To quantify the
contaminating population we fit the function $f_{\rm
  mem}=A\exp\left(-r/r_0\right)$ to the measured number density
profile (dashed blue curve in right panel of Figure~\ref{fig:boost}).
We use this model to boost the measured shear signal by a factor
$1+f_{\rm mem}$ (blue crosses in left and central panels of
Figure~\ref{fig:boost}).  It is clear that the boosted shear signal
underestimates the lensing signal that we detect by $\sim40\%$ on
small scales, and $\gs10\%$ at all radii interior to $\sim1\hMpc$.
Lens magnification is an obvious culprit for this apparent deficit of
signal in the boost-factor-corrected shear profile.

The galaxy-count is depleted by the magnification bias, as expressed by 
\begin{eqnarray}
 N(r;<m) =N_0(r;<m) \mu(r)^{2.5s-1},
\end{eqnarray}
where $\mu^{-1}=(1-\kappa)^2-|\gamma|^2$ is the lensing magnification
which expands the area of sky and enhances the flux of galaxies, and
$s=d\log_{10} N_0(m)/dm$ is a logarithmic count slope.  Given the
best-fit mass model derived by the stacked shear analysis (Table
\ref{tab:Mass_stacked}), we calculate the number density profile
expected from the magnification bias, assuming $s=0.15$
\citep{Umetsu14}.  This calculation shows clearly that the expected
number density profile of background galaxies is {\it not} flat,
showing a decline interior to $\sim1\hMpc$ (dotted red curve in right
panel of Figure~\ref{fig:boost}).  This indicates that the assumption
of a flat background galaxy number density profile is incorrect, even
on scales comparable with $r_{500}$.  

Next, we boost the $\Delta C>0$ shear profile (blue squares in left
and central panels of Figure~\ref{fig:boost}) by both the boost
factor, and the expected number density profile of background galaxies
from the magnification bias calculation discussed in the proceeding
two paragraphs.  This boost factor and magnification bias corrected
shear profile comes closer to recovering our measured shear profile,
although it remains $\sim10\%$ lower interior to $\sim1\hMpc$ (red
points in left and central panels of Figure~\ref{fig:boost}).
Clearly, the number density profile based on an imperfect background
selection is tightly coupled with the dilution effect and the
magnification bias.  It is therefore very difficult to break the
degeneracy between the dilution effect and the magnification bias
using the imperfect background catalogue.  We also mention that the
boost-factor gives rise to systematics in the source redshift because
member galaxies in background catalogue have inadequate redshifts.
Overall, our analysis in this Section indicates that application
  of a boost factor without consideration of lens magnification may
  cause systematic biases even on quite large scales up to
  $\sim\,r_{500}$.

\subsection{Error budget}\label{sec:errorbudget}

\begin{table*}
\caption{Systematic errors on the mean mass, $\langle M_\Delta
 \rangle$, for 50 clusters, in units of a percentage. } \label{tab:sys}
\begin{center}
\begin{tabular}{ccccccc}
\hline
\hline
Name & expected
     & $M_{\rm vir}$
     & $M_{200}$
     & $M_{500}$
     & $M_{1000}$
     & $M_{2500}$ \\
\hline
Shear calibration  & $-3$
       & $-4.1$
       & $-4.2$
       & $-2.7$
       & $-4.8$
       & $-5.3$\\
Colour selection     & $-1$
     & $0.2$
     & $0.0$
     & $-0.5$
     & $-0.9$
     & $-1.7$\\
Shear+colour calibration  & $-3.2$
       & $-4.1$
       & $-4.2$
       & $-2.7$
       & $-4.8$
       & $-5.3$\\
Radial bins    & $-$
     & $\pm0.9$
     & $\pm1.0$
     & $\pm1.4$
     & $\pm2.3$
     & $\pm4.4$ \\
\hline
\end{tabular}
\end{center}
\end{table*}

We summarize the contributions to our overall error budget.  The
largest systematic bias in our measurements is caused by imperfections
in our faint galaxy shape measurements, with a STEP-like
multiplicative bias of $m\simeq-0.03$ that is independent of galaxy
size.  We tuned our faint galaxy selection method to yield a
contamination level of $\le1$ per cent, thus again giving a very small
systematic bias.  This low level of contamination renders the
background galaxy samples essential pure, and thus not requiring
correction.  Nevertheless, for completeness, we correct the shear
profiles for these two bias terms and obtain ``corrected masses'' (see
Appendix) that are $\sim3-5$ per cent higher than the ``uncorrected
masses'' in Table~\ref{tab:mass}.  In particular, after correction,
our measurements of $M_{500}$ increase by just $2.7$ per cent on
average.

We have also investigated several other errors, none of which
contribute to systematic biases.  First, the typical uncertainty on
the distance ratio $D_{ls}/D_s$ for each individual background galaxy,
computed from the COSMOS photometric redshift catalogue following the
methods described in Section~\ref{sec:photometry} is
$\sigma_\beta\simeq13-28$ per cent.  These uncertainties are included
in the covariance matrix when fitting the NFW model to the respective
shear profiles (Section~\ref{sec:fitting}).  Second, when exploring
different shear profile binning schemes, and applying them to the
observational shear measurements, we found that the $M_{500}$ values
for individual clusters scatter by $\pm1.4$ per cent.  Third, when
applying our shear profile binning scheme to simulated data we
recovered the true masses with $\ls1$ per cent bias and $\sim5$ per
cent scatter between different realizations of the simulations
(Section~\ref{sec:sim}).  Fourth, we tested whether using a point
estimator of the photometric redshift of a galaxy in the COSMOS
catalogue introduced any systematic bias with respect to using the
full $P(z)$ distribution.  We found that the latter yields masses
$4\pm6$ per cent lower than the former.  Given the poor precision to
which have been able to measure this possible bias, we ignore it here.
However this deserves more detailed investigation in future larger
surveys.  Whilst we have sought to minimize the assumptions in
  our analysis pipeline, we would like to draw attention to the
  dependence of our results on the COSMOS field.  We have assumed that
  the galaxy population probed by observations of this field are
  representative of the universe as a whole.  This assumption is
  relevant to the {\sc shera}-based shape measurement tests described
  in Section~\ref{sec:shapetest}, and aspects of our analysis that
  rely on the COSMOS photometric redshift distribution
  (Sections~\ref{sec:photometry}~\&~\ref{sec:bkg}).  Several
  many-filter and deep COSMOS-like calibration fields would be very
  beneficial for future deep and wide lensing surveys including LSST.

\subsection{Comparison of mass measurements with the literature} \label{sec:compare}

\begin{figure*}
  \includegraphics[width=0.45\hsize]{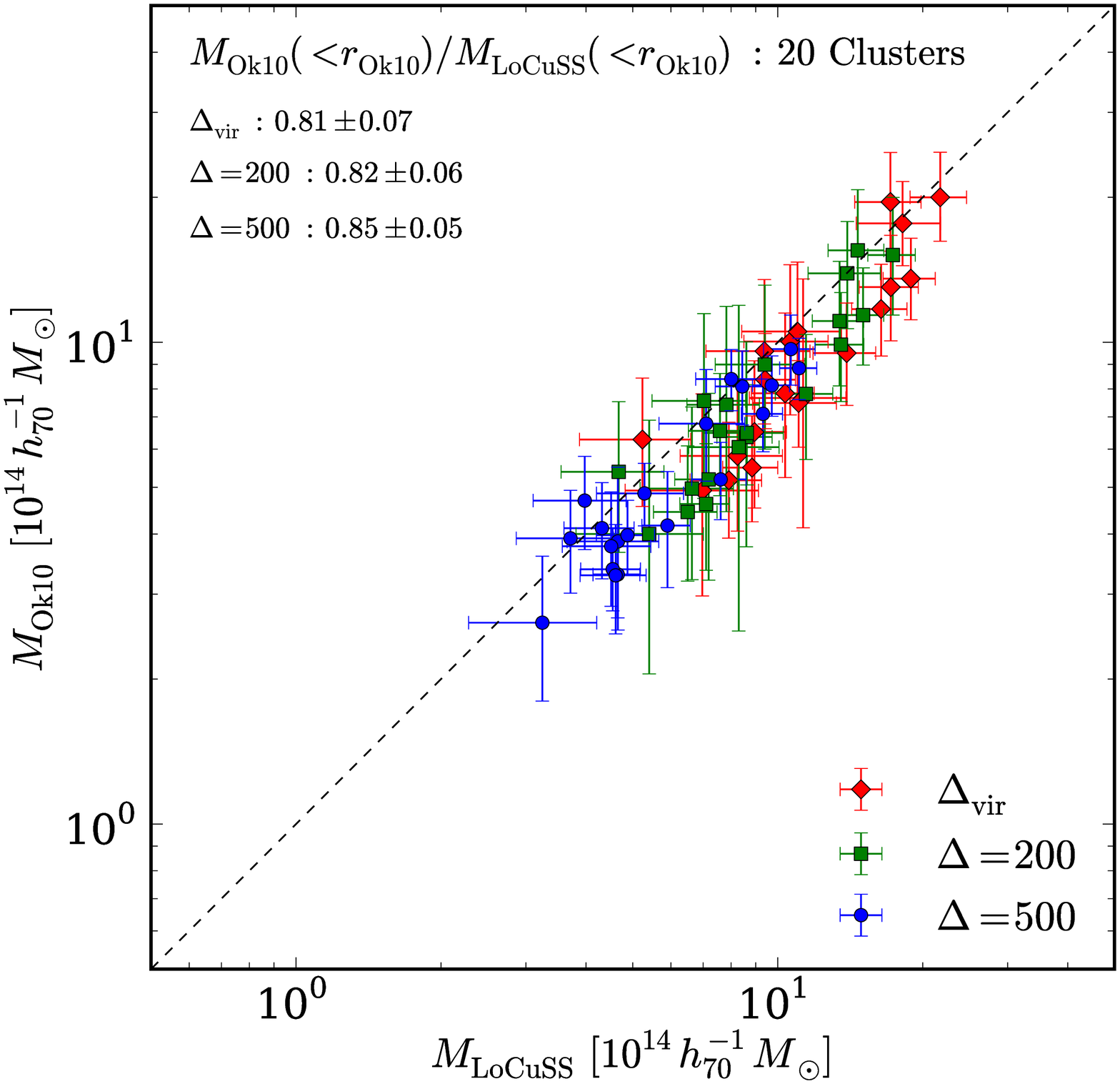}
  \includegraphics[width=0.45\hsize]{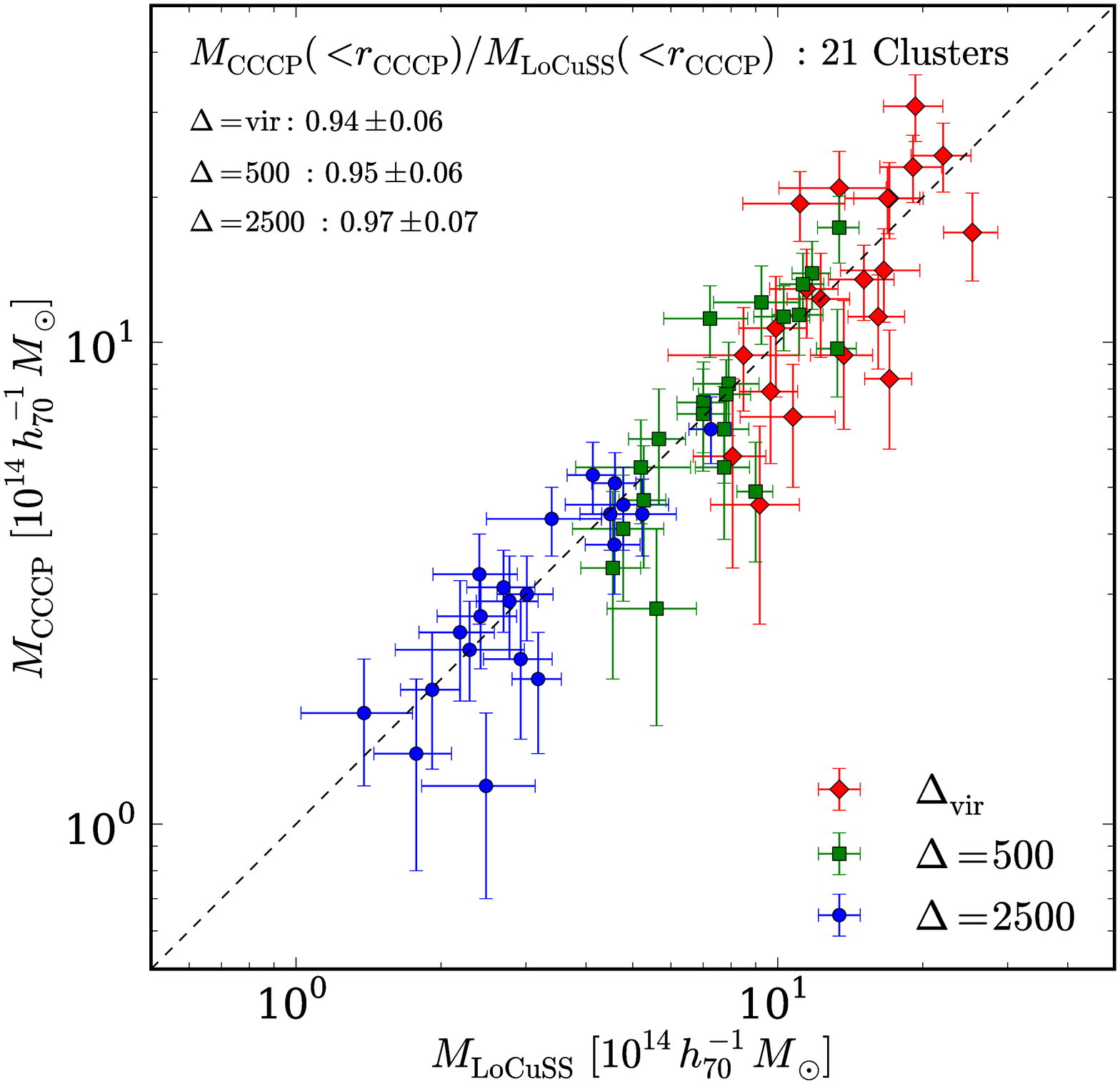}
  \includegraphics[width=0.45\hsize]{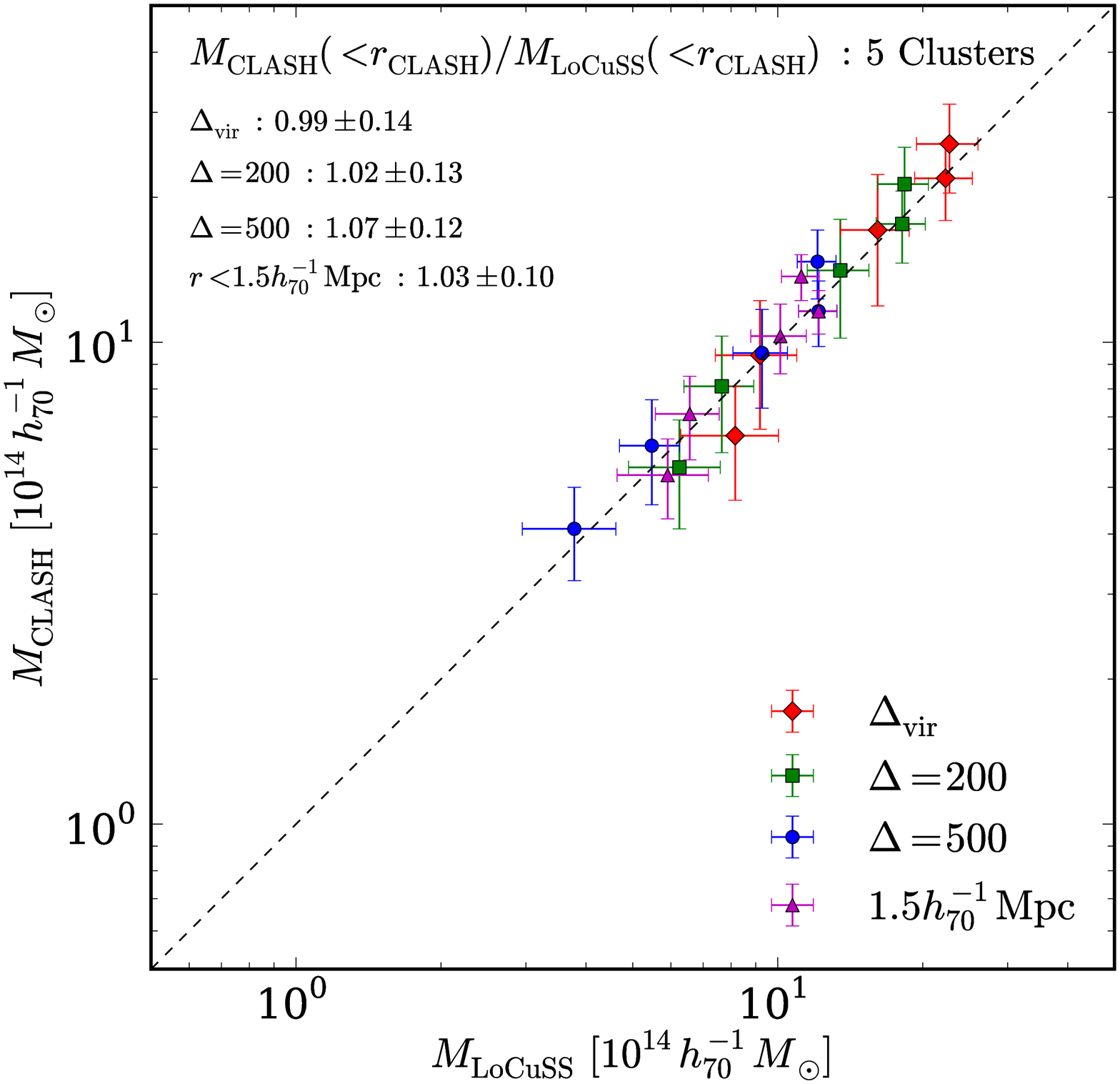}
  \includegraphics[width=0.45\hsize]{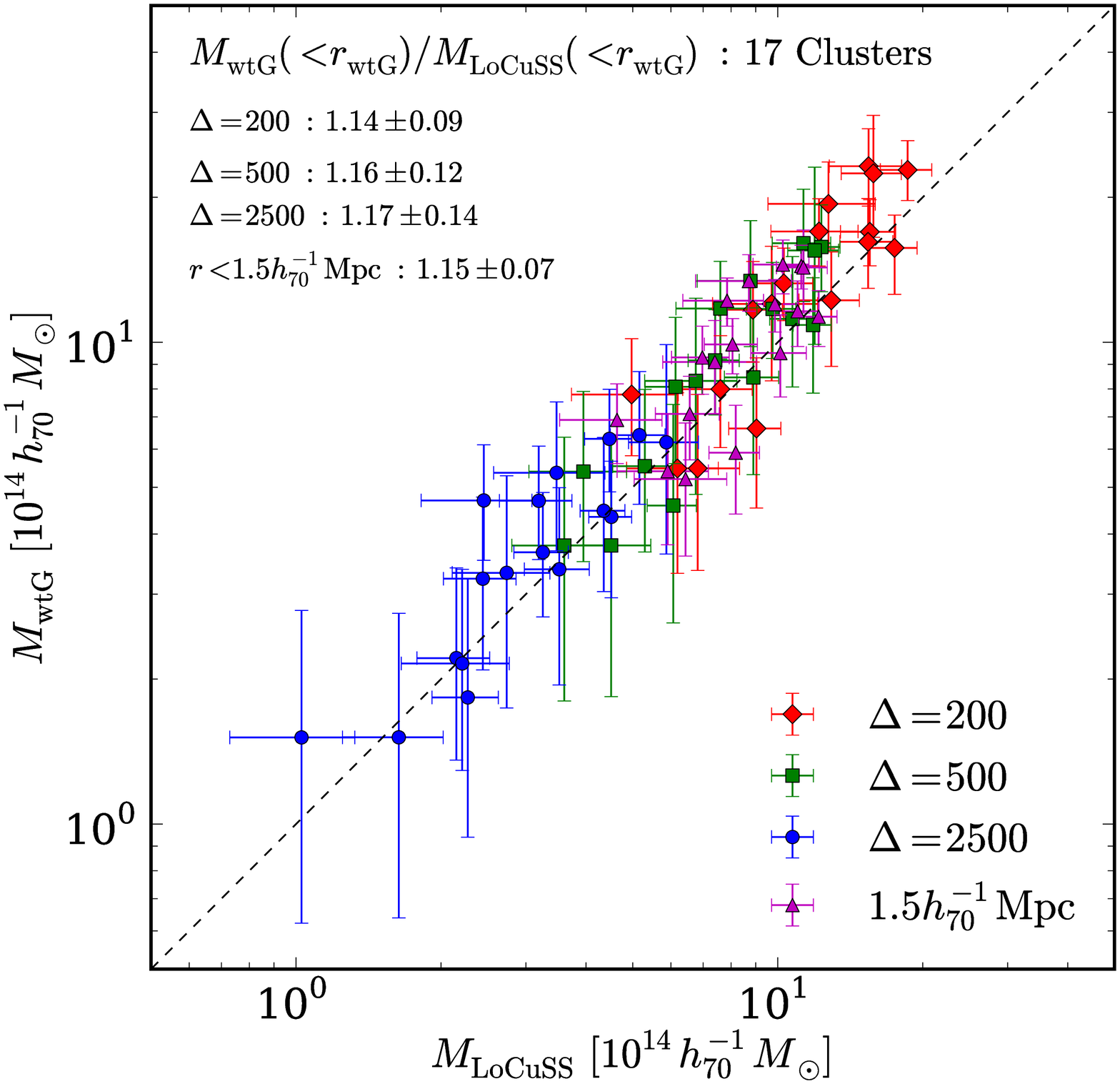}
  \caption{Mass comparisons of different projects: the top-left
    \citep[Ok10;][]{Okabe10b}, top-right \citep[CCCP][]{Hoekstra15},
    bottom-left \citep[CLASH][]{Umetsu14} and bottom-right
    \citep[WtG][]{Applegate14}. Masses are measured at the same radii in
    order to avoid the aperture-induced errors. The mass ratios for the
    overlap sample are computed by the geometric means.}
  \label{fig:compare}
\end{figure*}

\begin{table*}
  \caption{Comparison with previous studies, $M_{\rm
      pre}(<r_\Delta^{\rm pre})/M_{\rm LoCuSS}(<r_\Delta^{\rm
      pre})$. All masses are the NFW spherical masses measured within
    the overdensity radii $r_\Delta$ defined by previous studies
    (Ok10,CCCP, CLASH and WtG) or $1.5h_{70}^{-1}\Mpc$, in order to
    avoid the aperture-induced-errors in the mass estimates.  We also
    summarize the mass-measurement method and approach to correction
    of biases for each project.  ``Method'' denotes the tangential
    shear fitting ($g_+$) and a joint fitting using tangential shear
    profiles and the magnification bias ($g_+$ \& $\mu$).
    ``Calibration factor'' is the shear-calibration factor, with
    ``Yes'' indicating that such a factor was applied to the shear
    signal before fitting mass models, and ``No'' indicating
    otherwise.  ``Boost factor'' is the correction factor by the
    number density caused by imperfect background selection -- Yes/No
    indicates whether or not this factor was calculated and applied to
    the data. $c_\Delta$ states whether the concentration parameter
    was a free parameter in the fit, or fixed, or scaling with the
    mass. ``Radial bins'' gives the choice of radial binning
      scheme for the fitting of the shear profile.  ``N'' is the
    number of clusters in common between this study and each of the
    other studies. \label{tab:com}}
  \begin{center}
{\small 
    \begin{tabular}{c@{\hspace{-0.002em}}cc@{\hspace{-0.0002em}}cc@{\hspace{-0.002em}}ccccccc}
      \hline
      \hline
 Name & {\small Method}
      & {\small Calibration}
      & {\small Boost}
      & {\small Radial}
      & $c_\Delta$
      & {\small N}
      & Virial
      & $200$
      & $500$
      & $2500$
      & $1.5h_{70}^{-1}$Mpc \\
      &
      & {\small factor}
      & {\small factor}
      & {\small bins}
      & 
      &
      &
      &
      &
      &
      &\\
      \hline
{\small This paper} & $g_+$
      & No
      & No
      & Adaptive
      & Free
      & $-$
      & $-$
      & $-$
      & $-$
      & $-$
      & $-$ \\
      Ok10 & $g_+$ 
      & No
      & No
      & Fixed
      & Free
      & 20
      & {\small $0.81\pm0.07$}
      & {\small$0.82\pm0.06$}
      & {\small$0.85\pm0.05$}
      & {\small$0.91\pm0.08$}
      & {\small$-$} \\
{\small CCCP} & $g_+$
      & Yes
      & Yes
      & Fixed
      & Scaling
      & 21    
      & {\small$0.94\pm0.06$}
      & {\small$-$ }
      & {\small$0.95\pm0.06$}
      & {\small$0.97\pm0.07$}
      & {\small$-$} \\
{\small CLASH} & $g_+$ \& $\mu$ 
      & Yes 
      & No
      & Fixed
      & Free
      & 5
      & {\small$0.99\pm0.14$}
      & {\small$1.02\pm0.13$}
      & {\small$1.07\pm0.12$}
      & {\small$-$}
      & {\small$1.03\pm0.10$} \\
{\small WtG}  & $g_+$
      & Yes
      & Yes
      & Fixed
      & Fixed
      & 17
      & {\small$-$}
      & {\small$1.14\pm0.09$}
      & {\small$1.16\pm0.12$ }
      & {\small$1.17\pm0.14$}
      & {\small$1.15\pm0.07$} \\
      \hline
    \end{tabular}
}
  \end{center}
\end{table*}

Ongoing projects, including LoCuSS, CCCP \citep{Hoekstra15}, CLASH
\citep{Umetsu14} and WtG \citep{Applegate14}, conducted cluster
weak-lensing analyses.  Comparison of the cluster mass measurements
between these different surveys is of paramount importance for cluster
cosmology experiments.  There are important differences between the
approach taken by each survey to the cluster mass measurements.  For
example, CCCP and WtG implement the boost factor method (Section
\ref{sec:boost}) to correct their shear signal.  Another key
difference is the prior adopted on the concentration parameter of the
NFW halo model that is fitted to the data.  For example, WtG fix the
concentration parameter at $c_{200}=4$, whilst we allow it to be a
free parameter in our fits.  We summarize the key differences between
the respective analysis methods in Table \ref{tab:com}, and highlight
key points in the following sections.  Throughout these sections, when
we compare cluster masses between two surveys we do so within the same
radii so as to avoid errors caused by aperture mis-match.  Also, all
comparisons are done without applying the colour-selection and
shear-calibration corrections to our shear measurements, discussed in
Section~\ref{sec:sysShear}.

\subsubsection{LoCuSS -- Okabe et al.\ (2010)}

\cite{Okabe10b} conducted weak-lensing analysis for 30 clusters using
Suprime-Cam data.  As some of clusters lacked $V-$ band data, we
measured masses only for 22 clusters, defining as background galaxies
those galaxies with colours are redder or bluer than those of the
red-sequence.  The red and blue colour cuts were chosen by eye based
on the run of lensing signal with colour offset from the red sequence.
We didn't adopt any correction factor inherent in the shear
calibration and the profiles of the background number density.  The
error on the tangential shear measurements only took into account
shape noise.  We treated the mass and the concentration as free
parameters when fitting the NFW model to the tangential shear
profiles.  The masses derived from our new analysis presented in this
article are $20\%$ to $9\%$ higher than obtained in 2010, with larger
differences found for masses measured in larger radii (Table
\ref{tab:com}; Figure \ref{fig:compare}).  Our previous mass
measurements were under-estimated due to issues in shear calibration
and imperfect background galaxy selection.

\subsubsection{CCCP -- Hoekstra et al.\ (2015)}\label{sec:CCCP}

The Canadian Cluster Comparison Project \citep[CCCP][]{Hoekstra15}
carried out weak-lensing mass measurement for 52 clusters using the
Canada-France-Hawaii Telescope (CFHT).  Their sample is defined by
clusters with an {\it ASCA} temperature of $k_BT_X>5\,{\rm keV}$ in
the range of $0.15<z<0.55$.  They used $B-$ and $R-$ bands for the
first 20 clusters with the CFH12k camera and $g'-$ and $r'-$ bands for
the other 32 clusters with the Megacam (hereafter we refer to the
redder band as $r$-band for both instruments).  They used solely their
$r$-band data to select galaxies at $22<r<25$ as background galaxies.
The number density profile of these galaxies was found to increase
towards the cluster centres.  Hoekstra et al.\ modelled this excess of
galaxies on a cluster-by-cluster basis, and boosted the lensing
signals by a factor of $1+f_{\rm contam}(r)$ assuming a flat
background number density profile of background galaxies, where
$f_{\rm contam}(r)$ is the fraction of contaminating galaxies obtained
from their models.  This boost factor increases their masses by $1-2$
per cent.  They adopt the mass-concentration relation \citep{Dutton14} 
for their mass estimates.

There are 21 clusters in common between the CCCP and LoCuSS samples.
On average, the CCCP masses are $\sim3\%-6\%$ lower than our masses
(Table \ref{tab:com}; Figure \ref{fig:compare}), however this
difference is not statistically significant.  We note that the radial
range of their model fits is $0.5-2h_{70}^{-1}{\rm Mpc}$ -- i.e. it
includes scales on which we expect the magnification bias to affect
the slope of the number density profile of background galaxies (Figure
\ref{fig:boost}; see also Ziparo et al.\ 2015).  Nevertheless, we see
good agreement between LoCuSS and CCCP mass measurements.

\subsubsection{CLASH -- Umetsu et al.\ (2014)}\label{sec:CLASH}

The Cluster Lensing And Supernova survey with Hubble
\citep[CLASH;][]{Umetsu14} conducted a joint shear-and-magnification
weak-lensing analysis of a sample of 20 galaxy clusters at
$0.19\simlt\,z\simlt0.69$, using imaging through multiple filters with
Subaru/Suprime-Cam.  They measure galaxy ellipticities using the
KSB$+$ method, and calibrate the isotropic PSF correction for galaxies
detected with high signal-to-noise ratio -- i.e.\ using methods
similar to our own.  They also employ a correction factor $1/0.95$ to
account for residual shear calibration.  Background galaxies are
selected in a colour-colour plane, typically based on the
$B_JR_Cz'$-band filters, following \cite{Medezinski10}.  They do not
employ a boost factor to compensate for contamination of their
background galaxy catalogues.  The halo concentration for the NFW
model is treated as a free parameter.  We compare the CLASH and LoCuSS
masses for 5 clusters in common, obtaining excellent agreement
(Figure~\ref{fig:compare} and Table~\ref{tab:com}).  Recently, Umetsu
et al.\ (2015) have published joint strong plus weak-lensing mass
measurements of the CLASH sample.  Their strong+weak-lensing masses
are in a similar excellent agreement with our measurements.

\subsubsection{WtG -- Applegate et al.\ (2014)} \label{sec:WtG}

The Weighing the Giants programme \citep[WtG;][]{Applegate14} have
conducted weak-lensing mass measurements for 51 X-ray luminous galaxy
clusters in the redshift of $0.15\simlt z\simlt0.7$, using data from
Subaru/Suprime-Cam.  They calibrated their shape measurement pipeline
using the STEP simulations and used this calibration to correct their
faint galaxy shape measurements down to 25th magnitude.  Note that
their shape measurements suffered significant noise bias at fainter
magnitudes.  We compare our masses with WtG masses based on
colour-selected galaxies, and therefore concentrate further discussion
of their methods on those measurements.

WtG define two catalogues of background galaxies to compute tangential
shear profiles.  The first catalogue is used to calculate the
contamination correction, namely the boost factor, which is shown by
blue points in Figure~4 of \citet{Applegate14}, and the second
catalogue is used to compute the tangential shear profile represented
by orange points in the left panel of their Figure~4.  The first
catalogue is defined by employing magnitude and size cuts and
exclusions of stars and galaxies that lie on the red-sequence of
cluster galaxies.  They found an apparent excess in the number density
profiles at small radii because of imperfect background selection.
The excess is described by $f_{\rm mem}(r)=N_{\rm mem}/(N_{\rm
  mem}+N_{\rm bkg})=f_{500}\exp(1-r/r_{500,X})$ which is the ratio of
the member galaxies divided by the total number (member and
background) of galaxies selected.  Here, they assume a constant number
density profile of background galaxies, and thus ignore the potential
effects of magnification bias.  All clusters are fitted simultaneously
to estimate $f_{500}$ and $r_{500,X}$.  The normalization $f_{500}$
for the first catalogue is $(8.6\pm0.9)\%$.  Figure~4 in
\cite{Applegate14} shows that $r_{500,X}\sim1.4-2h_{70}^{-1}{\rm Mpc}$
which is very large.  For example, \cite{Martino14} obtain a median of
$r_{500,X}\simeq1h_{70}^{-1}{\rm Mpc}$ and values in the range
$0.8-1.6h_{70}^{-1}{\rm Mpc}$ for the same sample that we study here,
that has considerable overlap with the WtG sample.  Whilst
cluster-by-cluster comparison of $r_{500,X}$ measurements is required
to be certain, the large values of this parameter used by WtG suggests
that they assume a rather shallow radial density distribution when
they model contamination of the background galaxy samples (right
  panel of Figure~\ref{fig:boost}).  Their shallow number density
profile indicates that cluster members are distributed beyond the
cluster virial radii, which conflicts with our results that the number
density of cluster members is negligible in these regions
(Section~\ref{sec:boost}).  Their number density result therefore
appears to be inconsistent with their weak-lensing mass measurements.
The shallow number density profile found by the WtG may therefore be a
source of systematic bias in their mass measurements.

Next, they draw the second catalogue from the first catalogue with
additional conditions of robust shape measurements and lensing cut
($S/N>3$ and $r_h>1.15r_h*$).  The tangential shear profile is
computed from the second catalogue.  The excess of the number density
profile, $f'_{500}=4.8\pm1.6\%$, is less than that obtained by the
first catalogue.  The tangential shear profile using the second
catalogue is corrected by the boost factor that was calculated using
the first catalogue.  We here explicitly describe their definition of
the corrected lensing signal, as follows, $g_+^{\rm 2nd}\rightarrow
g_+^{\rm 2nd}/(1-f_{\rm mem}^{\rm 1st}(r))$.  Here, 1st and 2nd
denotes the quantities computed by the first and second catalogues,
respectively.  This inconsistency may affect the precision of the WtG
mass measurements.

We emphasize that the boost factor cannot recover the correct lensing
signal because of the magnification bias, as demonstrated in Sections
\ref{sec:bkg} and \ref{sec:boost}, notwithstanding the fitting range,
$0.75-3h_{70}^{-1}{\rm Mpc}$, that WtG adopt.  
Finally, they assumed $c_{200}=4$ for the NFW model for
all clusters at all redshifts.  Numerical simulations
\citep[e.g.][]{Bullock01, Duffy08, Bhattacharya13, Diemer14,
  Meneghetti14, Ludlow14} show that the concentration depends on both
the halo mass and redshift.

WtG masses are $\sim15\%$ higher than our masses, independent of
overdensity, albeit at $\sim1-2\sigma$ significance
(Table~\ref{tab:com}; Figure \ref{fig:compare}).  Note that we exclude
A1758N from this comparison because the WtG adopt a radial fit range
that extends into the companion cluster A1758S.  When we follow the
WtG method -- i.e.\ restrict the radial range of the fit to
$0.75-3\,h_{70}^{-1}{\rm Mpc}$ and adopt $c_{200}=4$, the statistical
significance of the disagreement increases slightly, with geometric
means at $\Delta=200,500,2500$ and $1.5\,h_{70}^{-1}{\rm Mpc}$ of
$1.18\pm0.09$, $1.19\pm0.12$,$1.19\pm0.14$, and $1.18\pm0.07$,
respectively.  We note that the difference between our mass
measurements and those of WtG is significant at $\gs2\sigma$ on the
scale preferred by WtG, $1.5\,h_{70}^{-1}\Mpc$.  

In summary, we have identified several strong assumptions and
inconsistencies in the WtG analysis: the inconsistent calculation for
the boost factor, the shallow number density profile leading to high
boost factor, ignoring the magnification bias, and the fixed
concentration $c_{200}=4$.  We expect that the tension between our
respective mass measurements will be caused by one or more of these
issues.

\subsubsection{CCCP, WtG and LoCuSS}\label{sec:3projects}

\begin{figure}
  \includegraphics[width=\hsize]{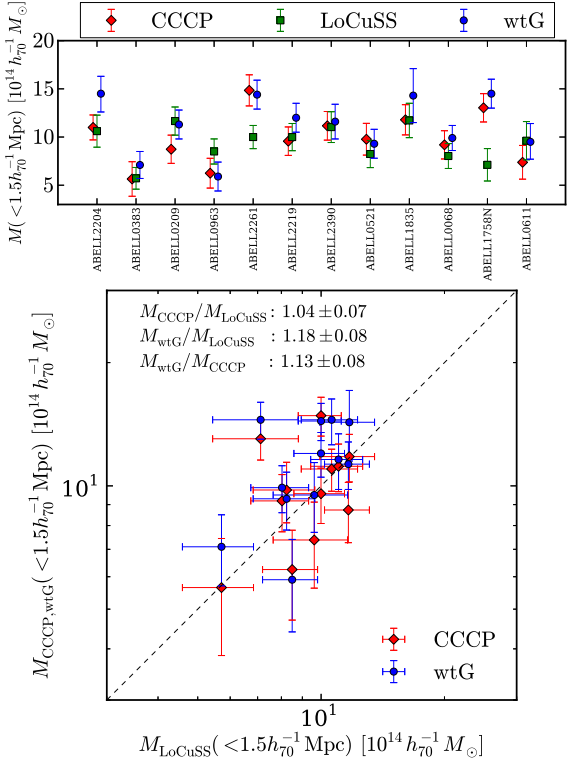}
  \caption{Mass comparisons for 12 cluster overlapped among the three
    projects of CCCP\citep{Hoekstra15}, WtG \citep{Applegate14} and
    LoCuSS.  Masses are measured at $1.5h_{70}^{-1}$Mpc, derived by
    the WtG fitting method; the radial range is
    $0.75-3\,h_{70}^{-1}$Mpc and $C_{200}=4$. The bottom panel shows a
    comparison of CCCP masses (red diamonds) and WtG masses (blue
    circles). The top panel shows individual masses of 12 clusters.
    Red diamonds, green squares and blue circles denote CCCP, LoCuSS
    and WtG masses, respectively. }
  \label{fig:compare2}
\end{figure}

The cleanest comparison between surveys is between the respective mass
measurements for 12 clusters in common between all of CCCP
\citep{Hoekstra15}, WtG \citep{Applegate14} and LoCuSS (this article).
We follow the WtG fitting method, that is, we fix the NFW model
concentration parameter at $c_{200}=4$, and fit the NFW model to the
observed shear profile in the radial range is $0.75-3h_{70}^{-1}$Mpc.
We compare three spherical NFW masses within $1.5h_{70}^{-1}$Mpc.
Based on these conditions, the mass comparison approximately
corresponds to a comparison of lensing signal at the fixed radial
range.  This like-for-like comparison confirms the results discussed
in the preceding sections: for the subsamples of clusters in
  the LoCuSS sample that have also been studied by CCCP and WtG, the
  CCCP and LoCuSS masses are in good agreements with each other, while
  the WtG masses are $\sim2\sigma$ higher than both CCCP and LoCuSS
  masses (Figure~\ref{fig:compare2}).

\subsubsection{Sensitivity of conclusions to ``sample selection''}

An important caveat on the conclusions discussed in
Sections~\ref{sec:CCCP}--\ref{sec:3projects} is that they are strictly
only applicable to the specific sub-samples of clusters that have been
observed by other surveys that are in common with our sample.  We
therefore consider whether our conclusions are supported by
inter-survey comparisons in the literature -- i.e.\ by comparison of
masses of clusters in the overlap between other pairs of surveys, thus
excluding our LoCuSS sample selection and mass measurements from the
discussion.

\citet{Hoekstra15} find that WtG masses from \citet{Applegate14} are
$\sim6-8$ per cent higher than their CCCP measurements at
$\sim2\sigma$ significance, depending on the redshift distribution
adopted by the two surveys.  This is consistent with our conclusions,
however it is intriguing to note that WtG masses agree very well with
CCCP masses when the latter are computed by deprojecting aperture mass
estimates.  \citeauthor{Hoekstra15} also compare their mass
measurements with CLASH measurements of 6 clusters in common between
these surveys, finding that CLASH masses from \citet{Umetsu14} exceed
CCCP masses by $12\pm5$ per cent.  Umetsu et al.\ (2015) also find
similar results (a $16\pm10$ per cent excess) for five clusters in
common between their joint strong plus weak-lensing analysis of the
CLASH sample and the \citeauthor{Hoekstra15} weak-lensing masses.  The
CCCP/CLASH comparison appears to be at odds with our finding that
CLASH, CCCP, and LoCuSS are all consistent, and underlines our caveat
that conclusions based on mass comparison between surveys may be
sensitive to the specific overlapping sub-samples considered.
However, the overlap between CLASH and WtG numbers 17 clusters,
i.e.\ a number of clusters comparable with the LoCuSS/CCCP and
LoCuSS/WtG samples discussed in
Sections~\ref{sec:CCCP}~\&~\ref{sec:WtG}.  Both \citet{Umetsu14} and
Umetsu et al.\ (2015) find that WtG masses exceed the CLASH masses by
$\sim7-10$ per cent, albeit at $\sim1\sigma$ significance.

In summary, a consistent picture emerges from comparisons between
LoCuSS, CLASH, CCCP, and WtG surveys when the samples in common
between the surveys number $\gs10$ objects: the WtG cluster mass
calibration exceeds that of other surveys by $\sim6-15$ per cent at
$\sim1-2\sigma$ significance per pair of surveys considered.  Further
inter-comparison of mass measurements based on all of the methods used
by the respective surveys, preferably using an enlarged and common
sample of clusters, should help to clarify the remaining differences
between the surveys.

\section{Summary}\label{sec:summary}

We observed an almost complete $L_X$-selected sample of 50 galaxy
clusters at $0.15<z<0.3$ through $V$- and $i'$-band filters with the
Suprime-CAM instrument on the Subaru 8.2-m telescope.  We used these
data to measure the weak gravitational shear signal, and thus to infer
the total mass and concentration of each cluster, and the sample as a
whole via a stacking analysis.  The size of our sample and typical
statistical precision on weak-lensing mass measurements of $30$ per
cent motivates our goal of controlling systematic biases in our
analysis at the $\sim30/\sqrt{50}\simeq4$ per cent level.

The recent literature identifies the dominant systematic uncertainty
in weak-lensing analysis of cluster samples as contamination of
background galaxy samples by faint cluster galaxies.  We extend our
background galaxy selection method based on the dependence of
gravitational shear signal on the colours of galaxies \citep{Okabe13}
to incorporate the empirical fact that, at fixed colour cut,
contamination by faint cluster members is a declining function of
clustercentric radius.  This allows us to define colour cuts that are
a function of clustercentric radius such that we achieve negligible
contamination (we required $\le1$ per cent contamination) whilst
achieving a number density of 13 galaxies per square arcminute.  We
stress that our approach to selecting background galaxies neither
assumes a form for the cluster mass distribution nor assumes a radial
distribution for the background galaxies.  We show that this latter
point is important because gravitational magnification modifies the
observed radial distribution of background galaxies on scales as large
as $\gs1\hMpc$ from the cluster centres and thus complicates methods
that assume the radial distribution of background galaxies is flat
(see also Ziparo et al.\ 2015 for a detailed discussion).

The dominant systematic bias in our analysis is the accuracy of our
faint galaxy shape measurements.  We test our modified KSB$+$ method
on large simulated datasets that incorporate smooth analytic galaxy
templates based on Sersic profiles, and realistic galaxies based on
high resolution imaging with the {\it Hubble Space Telescope}.
The key feature of our shape measurement code is that we use a very
high signal-to-noise ratio cut when selecting galaxies to model the
isotropic PSF correction.  The precision that we achieve in this model
as a result of this cut is important because it significantly reduces
the noise bias that is typically seen at faint flux levels in other
studies \citep[e.g.][]{Applegate14,Hoekstra15}.  We show that our
STEP-like multiplicative shape measurement bias is of order 3 per
cent, and is not a strong function of apparent magnitude and galaxy
size.  This is the largest systematic bias that we have
  identified in our analysis.

Controlling contamination of our background galaxy sample at the per
cent level across a wide range of clustercentric radii (down to
$\sim200\hkpc$) affords us the opportunity to fit for both mass and
concentration parameter when modeling the weak shear signal.  This is
important because it eliminates the possibility of systematic biases
caused by fixing the shape of the density profile or adopting a mass
concentration relation from simulations.  We are also interested in
measuring the mass-concentration relation of cluster-scale dark matter
halos.  To further mitigate the possibility of biases when modeling
the shear signal, we apply a range of binning schemes (number of bins,
inner and outer fit radii) to the data that are well motivated by the
physical properties of clusters and previous tests of weak-lensing
methods \citep[e.g.][]{Meneghetti10,Becker11,Bahe12}.  For each
cluster we adopt the mass and concentration measurement of the binning
scheme that yields the measurements that are the closest to the mean
of the measurements from the range of binning schemes that we explore.
This is a new method, and thus we test it on simulations including
full hydrodynamical numerical simulations from cosmo-OWLS
\citep{LeBrun14,McCarthy14}, based on an AGN model that reproduces a
large number of local X-ray-SZ-optical scaling relations. 
Our tests show that we recover the ensemble mass and concentration of the
simulated systems with sub per cent accuracy and of order 5 per cent
scatter between different realisations of the simulated lensing
observations.  Issues relating to the binning and modeling of the
shear signal are therefore sub-dominant in our analysis.

We now summarise our main science results.  We measured weak-lensing
mass for individual clusters by fitting tangential shear profile with
the spherical NFW model and investigated the mass-concentration
relation.  The best-fit mass-concentration relation is in excellent
agreement with recent numerical simulations
\citep{Bhattacharya13,Diemer14,Meneghetti14}.  We also measured the
average mass density profile for the NFW and Einasto profiles by
stacked lensing analysis, considering the point mass and the two-halo
term.  The best-fit NFW model of the stacked signal agrees well
with the mean of the individual mass measurements, supporting
that the stacked lensing analysis recover the average of individual
mass measurements.  The Einasto profile also agrees with numerical
simulations, albeit large scatter between numerical simulations.

We compared our lensing masses with masses from other projects (CCCP,
CLASH and WtG).  The philosophy of weak-lensing analyses for the four
projects include some strong differences.  Our mass measurements agree
within $1\sigma$ with the CLASH \citep{Umetsu14} and CCCP
\citep{Hoekstra15} surveys, with whom we have 5 and 21 clusters in
common respectively.  Our mass measurements are $\sim15$ per cent
lower than WtG measurements for the 17 clusters in common, at
$\sim1-2\sigma$ significance, depending on the mass measurement
aperture.  The fairest comparison between CCCP, WtG and LoCuSS is for
the 12 clusters in common between all three surveys and based on a
common mass measurement aperture and assumption on density profile
shape.  Adopting $c_{200}=4$ and measuring mass within
$1.5h_{70}^{-1}\Mpc$ (i.e.\ matching to WtG), confirms that LoCuSS and
CCCP mass calibrations are consistent with each other and WtG is in
tension at $2\sigma$ higher mass than the other two surveys.  

To guard against possible sensitivity of this conclusion to the
specific samples under investigation here, we also reviewed recent
studies in the literature that compare other pairs of surveys (and
thus exclude LoCuSS).  We find that WtG are in tension with masses
$\sim6-15$ per cent higher than other surveys at $\sim1-2\sigma$
significance in all pairwise comparisons that include $\gs10$ objects.
We discuss a range of possible causes of this tension, including
strong assumptions on the radial distribution of background galaxies
and the halo concentration parameter, and several inconsistencies in
the construction of the WtG background galaxy catalogues.  We expect
that the tension is likely caused by one or more of these factors.

In summary, we have controlled systematic biases in our weak-lensing
analysis of a large sample of clusters at $0.15\le z\le0.3$ at the 4
per cent level.  Therefore, as far as we can tell, systematics
do not dominate our results.  Our methods have numerous innovative
features that set us apart from contemporary surveys.  In that context
it is encouraging that out of four surveys including our own, three
agree within $1\sigma$ and one is in tension with the other three
  at just $\sim1-2\sigma$.  This represents important progress
towards convergence on the mass calibration of galaxy clusters for
cosmological surveys.  In a companion article \citep{Smith16} we
compare our weak-lensing mass measurements with estimates based on the
assumption that the intracluster medium is in hydrostatic equilibrium,
and discuss the recently reported tension between the {\it Planck}
results from the primary CMB and galaxy cluster counts.

  Looking to the future, we consider the development of additional
   deep, i.e.\ $\gs26$th magnitude, COSMOS-like photometric
   calibration fields to be vital for further progress on cluster mass
   calibration.  This will be particularly important to make secure
   progress on the calibration of clusters at higher redshifts,
   especially given that upcoming surveys concentrate on a relatively
   small number (typically 4-6) of photometric filters.  Further
   testing of faint galaxy shape measurement techniques in the high
   shear regime and down to faint photometric limits, preferrably in
   collaboration with experts in the cosmic shear community, will also
   be very helpful.  Overall, the emerging consensus between surveys
   at $z<0.3$ that use very different methods, encourages us that the
   future for cluster cosmology is bright, and we look forward to
   further progress from ongoing/future optical/near-infrared surveys
   including KIDS, DES, HSC, {\it Euclid}, and LSST.

\section*{Acknowledgments}

We thank the referee, Douglas Clowe, for helpful comments.  We
thank warmly our colleagues within the LoCuSS collaboration for their
support, encouragement, and advice, especially Felicia Ziparo, Keiichi
Umetsu, Pasquale Mazzotta, Dan Marrone, Alexis Finoguenov, Sarah
Mulroy, Arif Babul, Eiichi Egami, Chris Haines, Gus Evrard, James
Taylor, and Toshifumi Futamase.  We thank Masamune Oguri for making
his simulated Suprime-CAM observations available to us.  We
acknowledge Ayumu Terukina, Yuki Okura, and Masahiro Takada for
helpful discussions and advice.  We gratefully acknowledge assistance
with some of the Subaru observations described in this article from
Chris Haines, Mathilde Jauzac, and Paul May.  We thank Amandine
Le~Brun, Ian McCarthy, Yannick Bah\'e, and collaborators for providing
the shear information for clusters in their cosmo-OWLS simulation.  We
also thank Olivier Ilbert, Peter Capak and colleagues for making the
COSMOS-UltraVISTA photometric redshift catalogue and $P(z)$
distributions available to us prior to publication.  We acknowledge
stimulating and cordial discussions with Douglas Applegate, Anja von
der Linden, Adam Mantz, and Henk Hoekstra.  We also thank Henk for
kindly providing ``WtG-like'' mass measurements from his CCCP analysis
that we discuss in Section~\ref{sec:3projects}.  NO is supported by a
Grant-in-Aid from the Ministry of Education, Culture, Sports, Science,
and Technology of Japan (26800097).  This work was supported by
``World Premier International Research Center Initiative (WPI
Initiative)`` and the Funds for the Development of Human Resources in
Science and Technology under MEXT, Japan, and Core Research for
Energetic Universe in Hiroshima University (the MEXT program for
promoting the enhancement of research universities, Japan).  GPS
acknowledges support from the Royal Society and the Science and
Technology Facilities Council.

\bibliographystyle{mn2e}
\bibliography{my}


\appendix

\section{Radial Position for Tangential Shear Profile} \label{sec:app0}

  We consider how to compute the mean tangential shear averaged
  over background galaxies located in the $i$-th bin of a shear
  profile, spanning the radial range $r_1<r_i<r_2$.  It is important
  to compute these bin radii accurately so at to minimize systematic
  biases in cluster mass measurements.  In the limit of a uniform
  sheet of background galaxies with no intrinsic ellipticity and an
  infinite number density of background galaxies, the average
  tangential shear in the $i$-th bin can be calculated analytically in
  the continuum limit:
  \begin{eqnarray}
    \langle g_+\rangle_i=\frac{\int_{r_1}^{r_2}\,g_+(r)\,r\,dr}{\int_{r_1}^{r_2}r\,dr}.\label{eq:Appg+}
  \end{eqnarray}
  If one assumes a singular isothermal sphere (SIS) model,
  $g_+\simeq\gamma_+=A\,r^{-1}$, then in the weak limit, the average
  tangential shear in $i$-th bin becomes:
  \begin{eqnarray}
    \langle g_+\rangle_i=\frac{A}{r_{\rm mid}},
  \end{eqnarray}
  where $r_{\rm mid}=(r_1+r_2)/2$ -- i.e.\ the average tangential
  shear corresponds exactly with that of a SIS at a radius that is the
  mean of the inner and outer radius of the bin.  For completeness, we
  also use the SIS model to demonstrate that the correct way to
  calculate the radius at which to place a binned shear measurement is
  to weight by the shear signal, thus:
  \begin{eqnarray}
    r_i=\frac{\int_{r_1}^{r_2}g_+\,r^2\,dr}{\int_{r_1}^{r_2}g_+\,r\,dr},
    \label{eq:Appgwgt}
  \end{eqnarray}
  which yields $r_i=r_{\rm mid}$ for the SIS model, as above.  So far
  we have used the SIS model as a well-motivated illustration in the
  sense that tangential shear profiles of clusters do not deviate
  significantly from a power law slope of $r^{-1}$.  It is therefore
  interesting to note the convenience of using $r_{\rm mid}$ because
  it is trivial to compute and is independent of the details of the
  trial mass distribution when computing the likelihood of a model,
  under the assumptions described above.

  However the observed density profiles of galaxy clusters do
  deviate from the SIS model \citep[e.g.][]{Okabe13}.  It is therefore
  important to check the expected amplitude of systematic biases
  incurred if $r_{\rm mid}$ is adopted for clusters that do not have a
  SIS density profile.  We therefore consider a power-law profile,
  $g_+=A\,r^{-\alpha}$, in which case integrating Equation~\ref{eq:Appg+}
  yields an average tangential shear in a radial bin of:
  \begin{eqnarray}
    \langle g_+\rangle_i=\frac{2\,A\,(r_2^{2-\alpha}-r_1^{2-\alpha})}{(2-\alpha)\,(r_2^2-r_1^2)}.
  \end{eqnarray}
  The slope, $\alpha$, for the NFW model is equivalent to $\sim0.5$ on
  small scales and $1.5$ on large scales.  To estimate the possible
  systematic bias caused by adopting $r_{\rm mid}$ as the radius of a
  binned shear measurement, we define $r_i$ as:
  \begin{eqnarray}
    r_i=\frac{(2-\alpha)(r_2^2-r_1^2)}{2(r_2^{2-\alpha}-r_1^{2-\alpha})},
  \end{eqnarray}
  and calculate the deviation of $r_i$ from $r_{\rm mid}$
  for the cases $\alpha=0.5$ and $\alpha=1.5$.  We find that the
  deviation of $r_i$ from $r_{\rm mid}$ is less than 1 per cent
  in these cases.  We therefore conclude that, in the idealised case of
  uniform distribution of background galaxies, infinite number density
  of background galaxies, and no intrinsic ellipticity, $r_{\rm mid}$
  is a sufficiently accurate radius for binned shear measurements,
  if the goal is to control systematic biases at the per cent level.

\begin{figure}
  \begin{center}
    \includegraphics[width=\hsize]{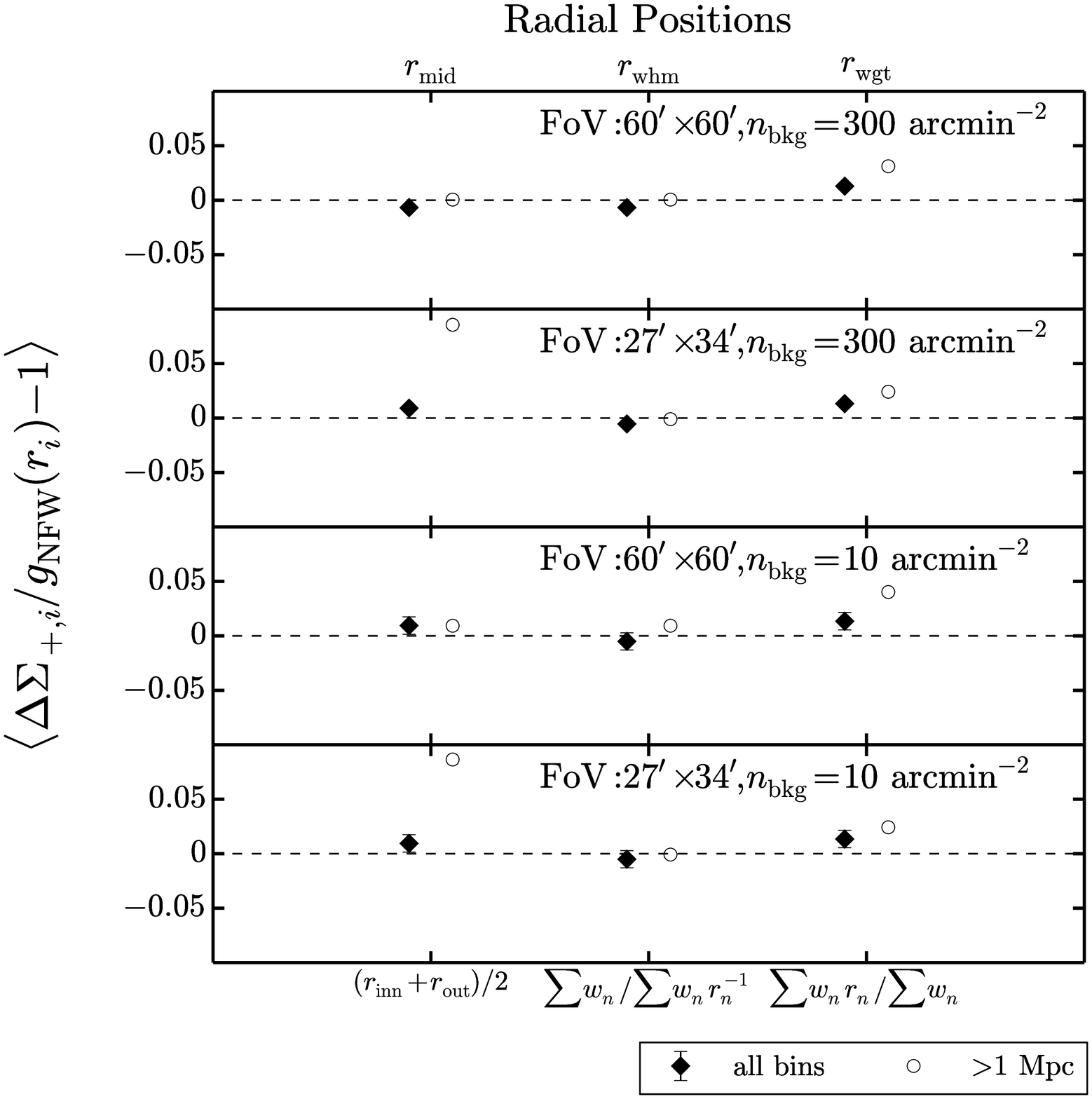}
  \end{center}
  \caption{The deviation between the averaged lensing strength and the
    input tangential shear measured at three different radial
    definitions.  The black diamonds and while circles are the average
    over all radial bins and $>1{\rm Mpc}$, respectively.  The top and
    lower middle panels show large and small numbers of
    background galaxies in the large FoV.  The upper middle
    panel is based on a large number of background galaxies with
    the Suprime-Cam's FoV, corresponding to stacked lensing analysis.
    The bottom panel is based on a small number of background
    galaxies with the Suprime-Cam's FoV, corresponding to individual
    cluster lensing analysis.  }
  \label{fig:AppReff}
\end{figure}

 In reality, background galaxies are intrinsically elliptical,
  have a finite number density, and non-uniform distribution on the
  sky.  We therefore now consider how to relate the idealized case to
  a real observation.  We begin by discretizing
  Equations~\ref{eq:Appg+}~\&~\ref{eq:Appgwgt}:
  \begin{eqnarray}
    \langle g_+\rangle_i=\frac{\sum_n g_{+,n}\,w_n}{\sum_n w_n},
    \label{eq:Appg+2}
  \end{eqnarray}
  \begin{eqnarray}
    r_i=\frac{\sum_n g_{+,n}\,w_n\,r_n}{\sum_n g_{+,n}\,w_n}.
    \label{eq:AppRsignal}
  \end{eqnarray}
  Note, when comparing Equations~\ref{eq:Appg+2}~\&~\ref{eq:g+} that
  here we assume for simplicity that all background galaxies are at
  the same redshift.  The previous discussion of the idealized case
  encourages the adoption of $g_+\propto\,r^{-1}$ in the realistic
  case, which yields a definition of $r_i$ as the weighted harmonic
  mean radius of the background galaxies in a given bin:
  \begin{eqnarray}
    r_{\rm whm}=\frac{\sum_n w_n}{\sum_n w_n r_n^{-1}}.
  \end{eqnarray}
  This definition has the important advantage that it does not rely on
  knowledge of the cluster density profile in order to compute the
  shear profile, and is therefore simple and stable to implement.  For
  example, one alternative would be to use the measured $g_+$ for each
  galaxy in Equations~\ref{eq:Appg+2}~\&~\ref{eq:AppRsignal}.  However
  the intrinsic ellipticity of galaxies results in this approach
  producing very noisy results, in some cases returning $r_i<0$.  A
  second alternative is to recompute the shear profile at every step
  when fitting a model to the shear profile, with the radii of the
  bins adjusted to reflect the shape and amplitude of the density
  profile at each step.  This approach is computationally expensive,
  and given the level of accuracy required, it is not currently
  justified.

  We test three alternative definitions of $r_i$ for tangential
  shear profiles in order to validate the arguments set out above for
  a realistic case.  We conduct a Monte-Carlo simulation using a mock
  catalogue of 1000 clusters.  We assume a spherical NFW mass model.
  The cluster masses are uniformly and randomly distributed in the log
  in the range $2\times10^{14}\hMsol< M_{200}< 2\times10^{15}\hMsol$.
  The concentration of each cluster is calculated from the respective
  cluster mass via the mass-concentration \citep{Meneghetti14}, taking
  into account intrinsic scatter. The cluster redshifts are uniformly
  and randomly distributed in $0.15<z<0.3$.  The background galaxy
  redshifts are fixed $z_s=1$.  We fixed the number of radial bins
  $N_{\rm bin}=6$ for the tangential shear profile in the range
  $100-2800h_{70}^{-1}\kpc$.  We calculate the deviation between the
  average measured tangential shear and the input tangential shear,
  measured at $r_{\rm mid}$ and $r_{\rm whm}$, as defined above.  We
  also employ a third definition of radius for comparison, the
  area-weighted radius, defined as follows:
  \begin{eqnarray}
    r_{\rm wgt}=\frac{\sum_n w_n r_n}{\sum_n w_n},
  \end{eqnarray}
  that has been used by several groups \cite[e.g.][]{Niikura15}.  The
  deviation is averaged over all radial bins and over those bins at
  $r>1h_{70}^{-1}\Mpc$.

  The Monte Carlo simulations consider four observing scenarios.
  First, a large field-of-view (FoV; $60'\times60'$) and large number
  of background galaxies ($n_{\rm bkg}=300~{\rm arcmin}^{-2}$), which
  is approximately equivalent to the idealized case discussed above
  (top panel of Figure \ref{fig:AppReff}).  Second is the FoV of the
  Suprime-Cam ($24'\times37'$) and $n_{\rm bkg}=300~{\rm
    arcmin}^{-2}$, corresponding to the stacked lensing analysis
  (Section~\ref{sec:stackedWL}; upper middle panel of
  Figure~\ref{fig:AppReff}).  Third is the large FoV and the small
  number of background galaxies ($n_{\rm bkg}=10~{\rm arcmin}^{-2}$);
  lower middle panel of Figure~\ref{fig:AppReff}.  Fourth is the FoV
  of the Suprime-cam and $n_{\rm bkg}=10~{\rm arcmin}^{-2}$,
  representing our analysis of individual clusters
  (Section~\ref{sec:mass}; bottom panel of Figure~\ref{fig:AppReff}.

  As expected, based on the discussion above, in the ideal case
  (top panel) case the lensing signal measured at both $r_{\rm mid}$
  and $r_{\rm whm}$ recover the input values accurately, while the
  signal measured at $r_{\rm wgt}$ is biased high by $\sim1$ per cent.
  In the case that represents our stacked weak-lensing analysis (upper
  middle panel) the shear measured at $r_{\rm mid}$ suffers $\sim9$
  per cent bias on large scales due to the finite FoV impinging on the
  outer annuli.  On the other hand, the performance of $r_{\rm whm}$
  and $r_{\rm wgt}$ are invariant to the size of the FoV, drawing
  attention to how these radii automatically take account of
  departures from azimuthal symmetry in the bin definitions.  Similar
  results are seen in the bottom panel.  Overall, the weighted
  harmonic mean radius ($r_{\rm whm}$) well describes the average
  tangential shear signal in all setup conditions, while the
  area-weighted radius $r_{\rm wgt}$ gives biases of a few per cent in
  all cases, and $r_{\rm mid}$ is vulnerable to the accuracy of
  corrections for annuli that are not fully covered by the
  observational data.  We therefore conclude that the weighted
  harmonic mean radius is the lowest bias radius at which to place the
  binned tangential shear measurements for the tangential shear
  profiles in individual and stacked weak-lensing analyses.

\section{Mass Estimates with correction factors} \label{sec:app1}

  Cluster mass calibration is of primary importance for
  cluster-based cosmology.  Although we do not apply any correction
  factor in our analysis, the systematic uncertainty ($m=-0.03$;
  Section~\ref{sec:sysShear}) inherent in the shear calibration and
  the 1 per cent contamination in our background galaxy selection
  might be not negligible for some scientific motivations like
  precision cosmology.  We therefore tabulate cluster masses
  determined by fitting the tangential shear profile corrected with
  the systematic uncertainty in Table~\ref{tab:mass2}.  Note that
  \citet{Smith16} based their analysis on the masses listed in the
  submitted version of this paper that are slightly different from
  those listed in Table~\ref{tab:mass2}.  \citeauthor{Smith16}'s
  result of $\beta_{\rm P}=0.95\pm0.04$ and $\beta_{\rm
    X}=0.95\pm0.05$ is unchanged when the updated masses in
  Table~\ref{tab:mass2} are used.

\input{tab-mass-corr.tex}

\bsp

\label{lastpage}

\end{document}

%% file: tab-sample.tex
\begin{table*}
\caption{Cluster sample.$^{1)}$:cluster name. $^{2)}$: cluster
 redshift.$^{3)}$:the band name for shape measurement.$^{4)}$:the band
 combinations for the color. $^{5)}$:the seeing size for the WL
 bands.$^{6)}$:the magnitude range for the red-band. $^{7)}$:the number
 density for background galaxies after the color selection.$^{8)}$:the
 signal-to-noise ratio in the tangential shear profile.} \label{tab:data}
\begin{center}
\begin{tabular}{lccccccc}
\hline
\hline
Name$^{1)}$ & Redshift$^{2)}$
     & WL$^{3)}$
     & color$^{4)}$
     & Seeing$^{5)}$
     & mag$^{6)}$
     & $n_{\rm bkg}$$^{7)}$
     & $S/N$$^{8)}$ \\
     &
     & 
     & 
     & [arcsec]
     & [ABmag]
     & [${\rm arcmin}^{-2}$]
     &  \\
\hline
ABELL2697 &   $0.2320$
          & $i'$
          & $i'V$
          &  $0.77$
          &  $21-26.0$
          &  $15.7$
          &  $6.5$\\
ABELL0068 &   $0.2546$
          & $i'$
          & $i'V$
          &  $0.73$
          &  $21-25.9$
          &  $19.0$
          &  $8.1$\\
ABELL2813 &   $0.2924$
          & $i'$
          & $i'V$
          &  $0.77$
          &  $21-26.1$
          &  $10.2$
          &  $6.7$\\
ABELL0115 &   $0.1971$
          & $i'$
          & $i'V$
          &  $0.71$
          &  $21-24.3$
          &  $7.2$
          &  $3.9$\\
ABELL0141 &   $0.2300$
          & $i'$
          & $i'V$
          &  $0.71$
          &  $21-26.2$
          &  $17.7$
          &  $5.7$\\
ZwCl0104.4+0048 &   $0.2540$
          & $i'$
          & $i'V$
          &  $0.65$
          &  $21-25.8$
          &  $15.5$
          &  $2.8$\\
ABELL0209 &   $0.2060$
          & $i'$
          & $i'V$
          &  $0.63$
          &  $21-24.9$
          &  $13.7$
          &  $9.4$\\
ABELL0267 &   $0.2300$
          & $i'$
          & $i'V$
          &  $0.61$
          &  $21-25.6$
          &  $20.1$
          &  $7.2$\\
ABELL0291 &   $0.1960$
          & $i'$
          & $i'V$
          &  $0.71$
          &  $21-25.9$
          &  $15.8$
          &  $5.9$\\
ABELL0383 &   $0.1883$
          & $i'$
          & $i'V$
          &  $0.67$
          &  $21-25.9$
          &  $19.8$
          &  $7.6$\\
ABELL0521 &   $0.2475$
          & $i'$
          & $i'V$
          &  $0.61$
          &  $21-24.8$
          &  $15.1$
          &  $6.2$\\
ABELL0586 &   $0.1710$
          & $i'$
          & $i'V$
          &  $0.83$
          &  $21-25.6$
          &  $6.7$
          &  $6.3$\\
ABELL0611 &   $0.2880$
          & $i'$
          & $i'V$
          &  $0.79$
          &  $21-25.8$
          &  $8.0$
          &  $5.7$\\
ABELL0697 &   $0.2820$
          & $i'$
          & $i'V$
          &  $0.73$
          &  $21-26.1$
          &  $12.8$
          &  $6.5$\\
ZwCl0857.9+2107 &   $0.2347$
          & $i'$
          & $i'V$
          &  $0.85$
          &  $21-26.0$
          &  $9.2$
          &  $2.6$\\
ABELL0750 &   $0.1630$
          & $i'$
          & $i'V$
          &  $0.71$
          &  $21-25.8$
          &  $13.4$
          &  $6.2$\\
ABELL0773 &   $0.2170$
          & $i'$
          & $i'V$
          &  $0.57$
          &  $21-26.5$
          &  $21.4$
          &  $10.5$\\
ABELL0781 &   $0.2984$
          & $i'$
          & $i'V$
          &  $0.87$
          &  $21-26.1$
          &  $8.4$
          &  $4.2$\\
ZwCl0949.6+5207 &   $0.2140$
          & $i'$
          & $i'V$
          &  $0.83$
          &  $21-26.2$
          &  $9.9$
          &  $5.1$\\
ABELL0901 &   $0.1634$
          & $i'$
          & $i'V$
          &  $0.73$
          &  $21-26.2$
          &  $12.9$
          &  $4.0$\\
ABELL0907 &   $0.1669$
          & $i'$
          & $i'V$
          &  $0.73$
          &  $21-26.4$
          &  $13.1$
          &  $8.8$\\
ABELL0963 &   $0.2050$
          & $I_{\rm c}$
          & $I_{\rm c}V$
          &  $0.75$
          &  $21-25.7$
          &  $15.0$
          &  $7.2$\\
ZwCl1021.0+0426 &   $0.2906$
          & $i'$
          & $i'V$
          &  $0.61$
          &  $21-26.3$
          &  $20.4$
          &  $6.7$\\
ABELL1423 &   $0.2130$
          & $i'$
          & $i'V$
          &  $0.71$
          &  $21-26.0$
          &  $16.5$
          &  $6.0$\\
ABELL1451 &   $0.1992$
          & $i'$
          & $i'V$
          &  $0.65$
          &  $21-26.4$
          &  $21.3$
          &  $10.2$\\
RXCJ1212.3-1816 &   $0.2690$
          & $i'$
          & $i'V$
          &  $0.91$
          &  $21-26.1$
          &  $10.2$
          &  $2.1$\\
ZwCl1231.4+1007 &   $0.2290$
          & $i'$
          & $i'V$
          &  $0.75$
          &  $21-26.0$
          &  $8.7$
          &  $4.4$\\
ABELL1682 &   $0.2260$
          & $i'$
          & $i'V$
          &  $0.69$
          &  $21-26.2$
          &  $21.9$
          &  $10.8$\\
ABELL1689 &   $0.1832$
          & $i'$
          & $i'V$
          &  $0.87$
          &  $21-25.7$
          &  $8.9$
          &  $10.6$\\
ABELL1758N &   $0.2800$
          & $R_{\rm c}$
          & $i'B$
          &  $0.69$
          &  $21-25.9$
          &  $13.7$
          &  $4.8$\\
ABELL1763 &   $0.2279$
          & $i'$
          & $i'V$
          &  $0.77$
          &  $21-25.5$
          &  $11.7$
          &  $9.6$\\
ABELL1835 &   $0.2528$
          & $i'$
          & $i'V$
          &  $0.89$
          &  $21-25.2$
          &  $8.4$
          &  $7.9$\\
ABELL1914 &   $0.1712$
          & $R_{\rm c}$
          & $i'g'$
          &  $0.61$
          &  $21-26.1$
          &  $11.7$
          &  $6.7$\\
ZwCl1454.8+2233 &   $0.2578$
          & $i'$
          & $i'V$
          &  $0.81$
          &  $21-25.2$
          &  $6.9$
          &  $3.1$\\
ABELL2009 &   $0.1530$
          & $i'$
          & $i'V$
          &  $0.77$
          &  $21-26.1$
          &  $12.8$
          &  $6.1$\\
ZwCl1459.4+4240 &   $0.2897$
          & $R_{\rm c}$
          & $i'V$
          &  $0.57$
          &  $21-26.0$
          &  $12.3$
          &  $3.7$\\
RXCJ1504.1-0248 &   $0.2153$
          & $i'$
          & $i'V$
          &  $0.79$
          &  $21-24.6$
          &  $6.7$
          &  $5.2$\\
ABELL2111 &   $0.2290$
          & $i'$
          & $i'V$
          &  $0.89$
          &  $21-25.3$
          &  $8.3$
          &  $3.4$\\
ABELL2204 &   $0.1524$
          & $i'$
          & $i'V$
          &  $0.81$
          &  $21-24.9$
          &  $8.6$
          &  $7.9$\\
ABELL2219 &   $0.2281$
          & $i'$
          & $i'V$
          &  $0.71$
          &  $21-26.0$
          &  $17.2$
          &  $7.7$\\
RXJ1720.1+2638 &   $0.1640$
          & $i'$
          & $i'V$
          &  $0.71$
          &  $21-24.3$
          &  $7.5$
          &  $4.7$\\
ABELL2261 &   $0.2240$
          & $i'$
          & $i'V$
          &  $0.61$
          &  $21-26.0$
          &  $21.3$
          &  $10.8$\\
RXCJ2102.1-2431 &   $0.1880$
          & $i'$
          & $i'V$
          &  $0.71$
          &  $21-25.9$
          &  $14.2$
          &  $4.0$\\
RXJ2129.6+0005 &   $0.2350$
          & $i'$
          & $i'V$
          &  $0.85$
          &  $21-25.8$
          &  $11.6$
          &  $4.6$\\
ABELL2390 &   $0.2329$
          & $R_{\rm c}$
          & $i'V$
          &  $0.65$
          &  $21-26.2$
          &  $12.3$
          &  $8.9$\\
ABELL2485 &   $0.2472$
          & $i'$
          & $i'V$
          &  $0.67$
          &  $21-25.6$
          &  $17.0$
          &  $5.9$\\
ABELL2537 &   $0.2966$
          & $i'$
          & $i'V$
          &  $0.99$
          &  $21-25.9$
          &  $4.9$
          &  $5.2$\\
ABELL2552 &   $0.2998$
          & $R_{\rm c}$
          & $i'V$
          &  $0.77$
          &  $21-26.0$
          &  $9.9$
          &  $4.3$\\
ABELL2631 &   $0.2779$
          & $R_{\rm c}$
          & $i'V$
          &  $0.65$
          &  $21-25.9$
          &  $12.4$
          &  $4.5$\\
ABELL2645 &   $0.2510$
          & $i'$
          & $i'V$
          &  $0.67$
          &  $21-26.0$
          &  $11.9$
          &  $4.8$\\
\hline
\end{tabular}
\end{center}
\end{table*}

%% file: tab-mass.tex
\begin{table*}
\caption{Mass measures for individual clusters on $\langle M_\Delta \rangle_{\rm cog}$} \label{tab:mass}
\begin{center}
\begin{tabular}{lccccccc}
\hline
\hline
Name & $M_{\rm vir}$
     & $M_{200}$
     & $M_{500}$
     & $M_{1000}$
     & $M_{2500}$
     & $M_{180{\rm m}}$
     & $M_{200{\rm m}}$ \\
     & $10^{14}\hMsol$
     &  $10^{14}\hMsol$
     & $10^{14}\hMsol$
     &  $10^{14}\hMsol$
     & $10^{14}\hMsol$
     &  $10^{14}\hMsol$
     & $10^{14}\hMsol$ \\
\hline
ABELL2697 &   $10.37_{-2.23}^{+3.26}$
          & $8.03_{-1.48}^{+1.88}$
          & $4.62_{-0.84}^{+0.84}$
          & $2.72_{-0.80}^{+0.73}$
          & $1.12_{-0.53}^{+0.61}$
          & $12.29_{-2.92}^{+4.65}$
          & $11.76_{-2.73}^{+4.25}$ \\
ABELL0068 &   $7.78_{-1.47}^{+1.76}$
          & $6.65_{-1.16}^{+1.35}$
          & $4.78_{-0.71}^{+0.78}$
          & $3.51_{-0.48}^{+0.50}$
          & $2.12_{-0.36}^{+0.34}$
          & $8.64_{-1.71}^{+2.09}$
          & $8.39_{-1.64}^{+2.00}$ \\
ABELL2813 &   $9.47_{-2.02}^{+2.53}$
          & $8.17_{-1.61}^{+1.91}$
          & $5.90_{-0.97}^{+1.06}$
          & $4.37_{-0.70}^{+0.72}$
          & $2.66_{-0.65}^{+0.58}$
          & $10.43_{-2.35}^{+3.02}$
          & $10.14_{-2.25}^{+2.87}$ \\
ABELL0115 &   $9.49_{-2.94}^{+4.56}$
          & $7.04_{-1.97}^{+2.66}$
          & $3.77_{-1.04}^{+1.14}$
          & $2.07_{-0.77}^{+0.77}$
          & $0.75_{-0.40}^{+0.51}$
          & $11.60_{-3.84}^{+6.60}$
          & $11.06_{-3.60}^{+6.04}$ \\
ABELL0141 &   $5.67_{-1.23}^{+1.45}$
          & $4.71_{-0.95}^{+1.08}$
          & $3.19_{-0.60}^{+0.65}$
          & $2.22_{-0.48}^{+0.48}$
          & $1.21_{-0.38}^{+0.38}$
          & $6.42_{-1.46}^{+1.78}$
          & $6.22_{-1.39}^{+1.69}$ \\
ZwCl0104.4+0048 &   $2.71_{-1.09}^{+1.66}$
          & $2.24_{-0.81}^{+1.01}$
          & $1.47_{-0.74}^{+0.56}$
          & $0.99_{-0.70}^{+0.48}$
          & $0.51_{-0.53}^{+0.46}$
          & $3.08_{-1.32}^{+2.46}$
          & $2.98_{-1.26}^{+2.21}$ \\
ABELL0209 &   $15.42_{-2.54}^{+3.14}$
          & $12.75_{-1.91}^{+2.27}$
          & $8.64_{-1.05}^{+1.15}$
          & $5.99_{-0.65}^{+0.68}$
          & $3.26_{-0.49}^{+0.46}$
          & $17.57_{-3.07}^{+3.90}$
          & $17.01_{-2.93}^{+3.70}$ \\
ABELL0267 &   $7.28_{-1.33}^{+1.58}$
          & $5.97_{-1.08}^{+1.16}$
          & $3.92_{-0.59}^{+0.64}$
          & $2.63_{-0.42}^{+0.43}$
          & $1.35_{-0.31}^{+0.30}$
          & $8.33_{-1.60}^{+1.95}$
          & $8.04_{-1.53}^{+1.85}$ \\
ABELL0291 &   $7.47_{-1.92}^{+2.68}$
          & $5.63_{-1.28}^{+1.61}$
          & $3.12_{-0.67}^{+0.71}$
          & $1.77_{-0.52}^{+0.50}$
          & $0.68_{-0.30}^{+0.35}$
          & $9.05_{-2.51}^{+3.80}$
          & $8.64_{-2.35}^{+3.50}$ \\
ABELL0383 &   $6.27_{-1.42}^{+1.83}$
          & $5.23_{-1.07}^{+1.30}$
          & $3.64_{-0.60}^{+0.65}$
          & $2.59_{-0.37}^{+0.39}$
          & $1.48_{-0.33}^{+0.29}$
          & $7.11_{-1.71}^{+2.30}$
          & $6.90_{-1.64}^{+2.18}$ \\
ABELL0521 &   $6.76_{-1.36}^{+1.58}$
          & $5.61_{-1.05}^{+1.18}$
          & $3.77_{-0.65}^{+0.69}$
          & $2.59_{-0.51}^{+0.51}$
          & $1.39_{-0.40}^{+0.40}$
          & $7.65_{-1.62}^{+1.93}$
          & $7.40_{-1.54}^{+1.83}$ \\
ABELL0586 &   $7.69_{-2.05}^{+2.99}$
          & $6.65_{-1.61}^{+2.15}$
          & $5.04_{-0.98}^{+1.12}$
          & $3.92_{-0.64}^{+0.69}$
          & $2.60_{-0.68}^{+0.53}$
          & $8.53_{-2.41}^{+3.72}$
          & $8.32_{-2.32}^{+3.54}$ \\
ABELL0611 &   $10.89_{-2.12}^{+2.46}$
          & $9.23_{-1.71}^{+1.92}$
          & $6.37_{-1.09}^{+1.17}$
          & $4.51_{-0.80}^{+0.82}$
          & $2.54_{-0.60}^{+0.58}$
          & $12.14_{-2.45}^{+2.91}$
          & $11.77_{-2.35}^{+2.77}$ \\
ABELL0697 &   $12.62_{-3.14}^{+4.87}$
          & $9.74_{-2.13}^{+2.90}$
          & $5.39_{-1.00}^{+1.08}$
          & $3.05_{-0.81}^{+0.76}$
          & $1.17_{-0.51}^{+0.56}$
          & $14.91_{-4.01}^{+6.72}$
          & $14.22_{-3.73}^{+6.14}$ \\
ZwCl0857.9+2107 &   $3.12_{-1.20}^{+1.58}$
          & $2.45_{-0.91}^{+1.09}$
          & $1.45_{-0.75}^{+0.69}$
          & $0.88_{-0.59}^{+0.57}$
          & $0.37_{-0.42}^{+0.45}$
          & $3.67_{-1.47}^{+2.14}$
          & $3.52_{-1.39}^{+1.97}$ \\
ABELL0750 &   $7.66_{-2.34}^{+3.98}$
          & $6.30_{-1.74}^{+2.71}$
          & $4.31_{-0.95}^{+1.20}$
          & $3.02_{-0.54}^{+0.60}$
          & $1.67_{-0.46}^{+0.39}$
          & $8.78_{-2.85}^{+5.14}$
          & $8.51_{-2.72}^{+4.86}$ \\
ABELL0773 &   $11.11_{-1.45}^{+1.67}$
          & $9.56_{-1.14}^{+1.28}$
          & $7.05_{-0.70}^{+0.75}$
          & $5.33_{-0.48}^{+0.50}$
          & $3.37_{-0.39}^{+0.37}$
          & $12.31_{-1.70}^{+2.00}$
          & $12.00_{-1.63}^{+1.91}$ \\
ABELL0781 &   $8.17_{-2.25}^{+3.01}$
          & $6.57_{-1.65}^{+1.97}$
          & $3.97_{-1.19}^{+1.11}$
          & $2.46_{-1.10}^{+0.93}$
          & $1.09_{-0.69}^{+0.77}$
          & $9.40_{-2.77}^{+4.06}$
          & $9.02_{-2.60}^{+3.72}$ \\
ZwCl0949.6+5207 &   $5.02_{-1.34}^{+1.72}$
          & $4.44_{-1.10}^{+1.32}$
          & $3.48_{-0.73}^{+0.79}$
          & $2.79_{-0.53}^{+0.56}$
          & $1.96_{-0.53}^{+0.44}$
          & $5.47_{-1.54}^{+2.06}$
          & $5.35_{-1.49}^{+1.97}$ \\
ABELL0901 &   $3.11_{-0.97}^{+1.26}$
          & $2.65_{-0.77}^{+0.95}$
          & $1.95_{-0.50}^{+0.57}$
          & $1.48_{-0.36}^{+0.39}$
          & $0.93_{-0.31}^{+0.29}$
          & $3.48_{-1.14}^{+1.54}$
          & $3.39_{-1.10}^{+1.47}$ \\
ABELL0907 &   $18.97_{-4.66}^{+8.04}$
          & $14.28_{-2.99}^{+4.59}$
          & $8.07_{-1.17}^{+1.37}$
          & $4.66_{-0.78}^{+0.75}$
          & $1.85_{-0.61}^{+0.62}$
          & $23.03_{-6.19}^{+11.51}$
          & $22.06_{-5.83}^{+10.61}$ \\
ABELL0963 &   $8.59_{-1.56}^{+1.87}$
          & $7.13_{-1.20}^{+1.38}$
          & $4.87_{-0.72}^{+0.78}$
          & $3.41_{-0.53}^{+0.54}$
          & $1.88_{-0.42}^{+0.41}$
          & $9.77_{-1.88}^{+2.31}$
          & $9.46_{-1.79}^{+2.20}$ \\
ZwCl1021.0+0426 &   $6.11_{-1.22}^{+1.43}$
          & $5.24_{-0.96}^{+1.09}$
          & $3.73_{-0.57}^{+0.61}$
          & $2.72_{-0.41}^{+0.42}$
          & $1.61_{-0.37}^{+0.34}$
          & $6.76_{-1.42}^{+1.72}$
          & $6.56_{-1.36}^{+1.63}$ \\
ABELL1423 &   $5.05_{-1.26}^{+1.67}$
          & $4.30_{-0.97}^{+1.19}$
          & $3.11_{-0.57}^{+0.62}$
          & $2.30_{-0.43}^{+0.43}$
          & $1.41_{-0.45}^{+0.37}$
          & $5.64_{-1.49}^{+2.09}$
          & $5.48_{-1.43}^{+1.98}$ \\
ABELL1451 &   $10.27_{-1.42}^{+1.64}$
          & $8.54_{-1.08}^{+1.22}$
          & $5.88_{-0.63}^{+0.67}$
          & $4.15_{-0.42}^{+0.43}$
          & $2.32_{-0.31}^{+0.30}$
          & $11.67_{-1.71}^{+2.01}$
          & $11.31_{-1.63}^{+1.92}$ \\
RXCJ1212.3-1816 &   $2.44_{-1.08}^{+1.48}$
          & $2.12_{-0.91}^{+1.17}$
          & $1.55_{-0.63}^{+0.75}$
          & $1.17_{-0.49}^{+0.54}$
          & $0.73_{-0.40}^{+0.37}$
          & $2.69_{-1.22}^{+1.74}$
          & $2.62_{-1.18}^{+1.67}$ \\
ZwCl1231.4+1007 &   $7.20_{-2.09}^{+2.91}$
          & $5.58_{-1.49}^{+1.82}$
          & $3.23_{-1.03}^{+1.01}$
          & $1.91_{-0.86}^{+0.83}$
          & $0.79_{-0.49}^{+0.66}$
          & $8.53_{-2.65}^{+4.09}$
          & $8.17_{-2.49}^{+3.75}$ \\
ABELL1682 &   $10.35_{-1.57}^{+1.85}$
          & $8.66_{-1.21}^{+1.38}$
          & $5.97_{-0.69}^{+0.75}$
          & $4.21_{-0.44}^{+0.46}$
          & $2.36_{-0.33}^{+0.31}$
          & $11.69_{-1.87}^{+2.25}$
          & $11.33_{-1.79}^{+2.14}$ \\
ABELL1689 &   $12.34_{-1.78}^{+2.07}$
          & $10.98_{-1.46}^{+1.66}$
          & $8.80_{-0.98}^{+1.07}$
          & $7.22_{-0.69}^{+0.72}$
          & $5.28_{-0.47}^{+0.47}$
          & $13.41_{-2.04}^{+2.40}$
          & $13.15_{-1.97}^{+2.32}$ \\
ABELL1758N &   $6.76_{-1.64}^{+2.08}$
          & $5.88_{-1.32}^{+1.53}$
          & $4.35_{-0.97}^{+0.98}$
          & $3.29_{-1.02}^{+0.87}$
          & $2.09_{-1.02}^{+0.90}$
          & $7.42_{-1.91}^{+2.57}$
          & $7.22_{-1.83}^{+2.42}$ \\
ABELL1763 &   $20.69_{-3.71}^{+4.97}$
          & $16.92_{-2.70}^{+3.42}$
          & $11.06_{-1.36}^{+1.51}$
          & $7.40_{-0.84}^{+0.86}$
          & $3.77_{-0.73}^{+0.68}$
          & $23.71_{-4.55}^{+6.31}$
          & $22.89_{-4.32}^{+5.94}$ \\
ABELL1835 &   $11.50_{-2.05}^{+2.43}$
          & $10.09_{-1.63}^{+1.88}$
          & $7.68_{-1.01}^{+1.09}$
          & $5.99_{-0.72}^{+0.74}$
          & $4.00_{-0.68}^{+0.64}$
          & $12.57_{-2.38}^{+2.88}$
          & $12.27_{-2.28}^{+2.75}$ \\
ABELL1914 &   $11.04_{-2.22}^{+2.87}$
          & $8.73_{-1.59}^{+1.92}$
          & $5.48_{-0.87}^{+0.94}$
          & $3.51_{-0.63}^{+0.64}$
          & $1.66_{-0.47}^{+0.47}$
          & $12.98_{-2.79}^{+3.78}$
          & $12.51_{-2.65}^{+3.55}$ \\
ZwCl1454.8+2233 &   $5.61_{-2.29}^{+4.55}$
          & $4.42_{-1.62}^{+2.47}$
          & $2.62_{-1.01}^{+1.03}$
          & $1.58_{-0.91}^{+0.78}$
          & $0.68_{-0.66}^{+0.67}$
          & $6.54_{-2.85}^{+6.80}$
          & $6.28_{-2.69}^{+6.10}$ \\
ABELL2009 &   $10.29_{-3.15}^{+5.75}$
          & $7.78_{-2.03}^{+3.19}$
          & $4.47_{-0.87}^{+1.02}$
          & $2.63_{-0.69}^{+0.64}$
          & $1.07_{-0.49}^{+0.56}$
          & $12.49_{-4.18}^{+8.40}$
          & $11.97_{-3.93}^{+7.74}$ \\
ZwCl1459.4+4240 &   $3.51_{-1.01}^{+1.20}$
          & $3.17_{-0.87}^{+1.01}$
          & $2.56_{-0.64}^{+0.71}$
          & $2.11_{-0.50}^{+0.53}$
          & $1.56_{-0.38}^{+0.38}$
          & $3.75_{-1.11}^{+1.34}$
          & $3.68_{-1.08}^{+1.30}$ \\
RXCJ1504.1-0248 &   $6.10_{-1.45}^{+1.73}$
          & $5.53_{-1.25}^{+1.46}$
          & $4.58_{-0.93}^{+1.04}$
          & $3.87_{-0.71}^{+0.77}$
          & $2.99_{-0.49}^{+0.51}$
          & $6.54_{-1.56}^{+1.94}$
          & $6.42_{-1.55}^{+1.89}$ \\
ABELL2111 &   $5.78_{-1.92}^{+4.87}$
          & $4.93_{-1.48}^{+2.68}$
          & $3.56_{-0.85}^{+0.97}$
          & $2.63_{-1.09}^{+0.67}$
          & $1.60_{-1.14}^{+0.64}$
          & $6.43_{-2.28}^{+6.93}$
          & $6.26_{-2.18}^{+6.35}$ \\
ABELL2204 &   $11.32_{-2.36}^{+3.07}$
          & $9.56_{-1.83}^{+2.29}$
          & $6.95_{-1.11}^{+1.27}$
          & $5.17_{-0.70}^{+0.76}$
          & $3.18_{-0.46}^{+0.45}$
          & $12.77_{-2.81}^{+3.75}$
          & $12.43_{-2.71}^{+3.59}$ \\
ABELL2219 &   $13.38_{-2.60}^{+3.55}$
          & $10.40_{-1.75}^{+2.15}$
          & $6.06_{-0.91}^{+0.94}$
          & $3.60_{-0.81}^{+0.74}$
          & $1.50_{-0.57}^{+0.60}$
          & $15.83_{-3.38}^{+4.92}$
          & $15.17_{-3.16}^{+4.53}$ \\
RXJ1720.1+2638 &   $6.45_{-1.94}^{+2.85}$
          & $5.23_{-1.45}^{+1.96}$
          & $3.46_{-0.82}^{+0.97}$
          & $2.34_{-0.53}^{+0.58}$
          & $1.22_{-0.40}^{+0.38}$
          & $7.48_{-2.37}^{+3.66}$
          & $7.23_{-2.26}^{+3.46}$ \\
ABELL2261 &   $14.85_{-2.55}^{+3.29}$
          & $11.93_{-1.80}^{+2.18}$
          & $7.52_{-0.84}^{+0.91}$
          & $4.85_{-0.56}^{+0.55}$
          & $2.32_{-0.50}^{+0.47}$
          & $17.20_{-3.19}^{+4.28}$
          & $16.56_{-3.01}^{+4.01}$ \\
RXCJ2102.1-2431 &   $4.16_{-1.06}^{+1.29}$
          & $3.55_{-0.85}^{+0.99}$
          & $2.60_{-0.56}^{+0.61}$
          & $1.95_{-0.40}^{+0.43}$
          & $1.21_{-0.32}^{+0.30}$
          & $4.65_{-1.23}^{+1.54}$
          & $4.53_{-1.19}^{+1.48}$ \\
RXJ2129.6+0005 &   $6.34_{-2.00}^{+2.99}$
          & $4.69_{-1.29}^{+1.63}$
          & $2.42_{-0.85}^{+0.79}$
          & $1.28_{-0.65}^{+0.64}$
          & $0.44_{-0.30}^{+0.45}$
          & $7.74_{-2.67}^{+4.60}$
          & $7.35_{-2.48}^{+4.11}$ \\
ABELL2390 &   $12.61_{-2.14}^{+2.53}$
          & $10.60_{-1.67}^{+1.91}$
          & $7.37_{-0.99}^{+1.06}$
          & $5.25_{-0.64}^{+0.67}$
          & $2.99_{-0.47}^{+0.44}$
          & $14.18_{-2.52}^{+3.05}$
          & $13.75_{-2.42}^{+2.91}$ \\
ABELL2485 &   $6.90_{-1.51}^{+1.91}$
          & $5.72_{-1.13}^{+1.33}$
          & $3.83_{-0.64}^{+0.69}$
          & $2.63_{-0.54}^{+0.50}$
          & $1.40_{-0.48}^{+0.45}$
          & $7.82_{-1.83}^{+2.42}$
          & $7.56_{-1.74}^{+2.27}$ \\
ABELL2537 &   $8.55_{-2.27}^{+2.90}$
          & $7.65_{-1.90}^{+2.31}$
          & $6.00_{-1.27}^{+1.42}$
          & $4.82_{-0.94}^{+1.00}$
          & $3.40_{-0.88}^{+0.78}$
          & $9.21_{-2.55}^{+3.37}$
          & $9.01_{-2.47}^{+3.22}$ \\
ABELL2552 &   $9.14_{-2.74}^{+4.33}$
          & $7.61_{-2.08}^{+2.88}$
          & $5.01_{-1.18}^{+1.32}$
          & $3.38_{-1.06}^{+0.94}$
          & $1.74_{-0.90}^{+0.83}$
          & $10.30_{-3.27}^{+5.61}$
          & $9.94_{-3.11}^{+5.20}$ \\
ABELL2631 &   $9.26_{-2.51}^{+3.73}$
          & $7.13_{-1.66}^{+2.07}$
          & $3.93_{-1.25}^{+1.10}$
          & $2.22_{-1.08}^{+1.05}$
          & $0.84_{-0.56}^{+0.89}$
          & $10.96_{-3.30}^{+5.57}$
          & $10.45_{-3.06}^{+4.98}$ \\
ABELL2645 &   $5.00_{-1.27}^{+1.53}$
          & $4.16_{-0.99}^{+1.15}$
          & $2.81_{-0.64}^{+0.69}$
          & $1.95_{-0.52}^{+0.52}$
          & $1.05_{-0.40}^{+0.41}$
          & $5.65_{-1.49}^{+1.86}$
          & $5.46_{-1.43}^{+1.76}$ \\
\hline
\end{tabular}
\end{center}
\end{table*}

%% file: tab-mass-corr.tex
\begin{table*}
  \caption{Mass measures for individual clusters on $\langle M_\Delta
    \rangle_{\rm cog}$ determined by lensing signals with corrections
    of the shear calibration and the contamination of member galaxies;
    $g_+/(1+m)$ where $m=-3.2\%$. See the details in Section
    \ref{sec:sysShear} } \label{tab:mass2}
\begin{center}
\begin{tabular}{cccccccc}
\hline
\hline
Name & $M_{\rm vir}$
     & $M_{200}$
     & $M_{500}$
     & $M_{1000}$
     & $M_{2500}$
     & $M_{180{\rm m}}$
     & $M_{200{\rm m}}$ \\
     & $10^{14}\hMsol$
     &  $10^{14}\hMsol$
     & $10^{14}\hMsol$
     &  $10^{14}\hMsol$
     & $10^{14}\hMsol$
     &  $10^{14}\hMsol$
     & $10^{14}\hMsol$ \\
\hline
ABELL2697 &   $10.82_{-2.29}^{+3.32}$
          & $8.39_{-1.52}^{+1.92}$
          & $4.86_{-0.85}^{+0.85}$
          & $2.87_{-0.80}^{+0.73}$
          & $1.18_{-0.54}^{+0.62}$
          & $12.81_{-2.99}^{+4.71}$
          & $12.26_{-2.79}^{+4.31}$ \\
ABELL0068 &   $8.09_{-1.49}^{+1.78}$
          & $6.93_{-1.18}^{+1.37}$
          & $4.98_{-0.72}^{+0.79}$
          & $3.67_{-0.48}^{+0.51}$
          & $2.22_{-0.36}^{+0.35}$
          & $8.98_{-1.73}^{+2.12}$
          & $8.73_{-1.66}^{+2.02}$ \\
ABELL2813 &   $9.84_{-2.05}^{+2.56}$
          & $8.50_{-1.63}^{+1.94}$
          & $6.15_{-0.98}^{+1.07}$
          & $4.56_{-0.70}^{+0.72}$
          & $2.78_{-0.65}^{+0.58}$
          & $10.84_{-2.38}^{+3.05}$
          & $10.54_{-2.28}^{+2.90}$ \\
ABELL0115 &   $9.92_{-3.01}^{+4.64}$
          & $7.38_{-2.02}^{+2.72}$
          & $3.98_{-1.06}^{+1.16}$
          & $2.19_{-0.78}^{+0.78}$
          & $0.81_{-0.41}^{+0.51}$
          & $12.12_{-3.93}^{+6.68}$
          & $11.55_{-3.69}^{+6.12}$ \\
ABELL0141 &   $5.90_{-1.24}^{+1.47}$
          & $4.91_{-0.97}^{+1.10}$
          & $3.34_{-0.61}^{+0.65}$
          & $2.32_{-0.48}^{+0.48}$
          & $1.27_{-0.38}^{+0.38}$
          & $6.68_{-1.48}^{+1.80}$
          & $6.47_{-1.42}^{+1.71}$ \\
ZwCl0104.4+0048 &   $2.82_{-1.11}^{+1.68}$
          & $2.33_{-0.82}^{+1.03}$
          & $1.54_{-0.71}^{+0.57}$
          & $1.05_{-0.70}^{+0.48}$
          & $0.54_{-0.53}^{+0.46}$
          & $3.20_{-1.35}^{+2.46}$
          & $3.10_{-1.28}^{+2.22}$ \\
ABELL0209 &   $16.06_{-2.59}^{+3.18}$
          & $13.28_{-1.95}^{+2.30}$
          & $9.02_{-1.07}^{+1.17}$
          & $6.27_{-0.66}^{+0.69}$
          & $3.42_{-0.49}^{+0.46}$
          & $18.28_{-3.12}^{+3.95}$
          & $17.71_{-2.98}^{+3.74}$ \\
ABELL0267 &   $7.58_{-1.36}^{+1.60}$
          & $6.22_{-1.03}^{+1.18}$
          & $4.10_{-0.60}^{+0.65}$
          & $2.76_{-0.42}^{+0.44}$
          & $1.42_{-0.31}^{+0.31}$
          & $8.67_{-1.63}^{+1.97}$
          & $8.37_{-1.55}^{+1.87}$ \\
ABELL0291 &   $7.81_{-1.96}^{+2.73}$
          & $5.89_{-1.31}^{+1.65}$
          & $3.28_{-0.67}^{+0.72}$
          & $1.87_{-0.52}^{+0.51}$
          & $0.72_{-0.31}^{+0.36}$
          & $9.44_{-2.57}^{+3.85}$
          & $9.02_{-2.40}^{+3.55}$ \\
ABELL0383 &   $6.52_{-1.45}^{+1.85}$
          & $5.44_{-1.09}^{+1.32}$
          & $3.80_{-0.61}^{+0.67}$
          & $2.71_{-0.38}^{+0.40}$
          & $1.55_{-0.33}^{+0.29}$
          & $7.39_{-1.74}^{+2.32}$
          & $7.17_{-1.67}^{+2.20}$ \\
ABELL0521 &   $7.03_{-1.38}^{+1.60}$
          & $5.85_{-1.07}^{+1.20}$
          & $3.94_{-0.66}^{+0.70}$
          & $2.72_{-0.51}^{+0.51}$
          & $1.46_{-0.40}^{+0.40}$
          & $7.96_{-1.64}^{+1.95}$
          & $7.70_{-1.57}^{+1.85}$ \\
ABELL0586 &   $7.97_{-2.08}^{+3.00}$
          & $6.91_{-1.64}^{+2.17}$
          & $5.24_{-1.00}^{+1.14}$
          & $4.08_{-0.65}^{+0.70}$
          & $2.72_{-0.67}^{+0.53}$
          & $8.84_{-2.44}^{+3.71}$
          & $8.63_{-2.36}^{+3.53}$ \\
ABELL0611 &   $11.34_{-2.16}^{+2.50}$
          & $9.61_{-1.74}^{+1.95}$
          & $6.65_{-1.10}^{+1.18}$
          & $4.71_{-0.81}^{+0.83}$
          & $2.65_{-0.60}^{+0.58}$
          & $12.64_{-2.49}^{+2.95}$
          & $12.25_{-2.39}^{+2.81}$ \\
ABELL0697 &   $13.18_{-3.22}^{+4.95}$
          & $10.20_{-2.19}^{+2.97}$
          & $5.67_{-1.02}^{+1.10}$
          & $3.23_{-0.81}^{+0.76}$
          & $1.25_{-0.52}^{+0.57}$
          & $15.56_{-4.10}^{+6.79}$
          & $14.84_{-3.83}^{+6.21}$ \\
ZwCl0857.9+2107 &   $3.26_{-1.22}^{+1.60}$
          & $2.56_{-0.93}^{+1.10}$
          & $1.52_{-0.76}^{+0.70}$
          & $0.92_{-0.60}^{+0.57}$
          & $0.40_{-0.43}^{+0.45}$
          & $3.82_{-1.50}^{+2.18}$
          & $3.67_{-1.42}^{+2.00}$ \\
ABELL0750 &   $7.95_{-2.37}^{+3.99}$
          & $6.55_{-1.78}^{+2.73}$
          & $4.50_{-0.97}^{+1.22}$
          & $3.16_{-0.55}^{+0.62}$
          & $1.76_{-0.46}^{+0.39}$
          & $9.12_{-2.89}^{+5.13}$
          & $8.84_{-2.77}^{+4.85}$ \\
ABELL0773 &   $11.55_{-1.47}^{+1.69}$
          & $9.94_{-1.16}^{+1.30}$
          & $7.34_{-0.71}^{+0.76}$
          & $5.56_{-0.49}^{+0.50}$
          & $3.53_{-0.39}^{+0.37}$
          & $12.80_{-1.72}^{+2.02}$
          & $12.47_{-1.66}^{+1.93}$ \\
ABELL0781 &   $8.52_{-2.30}^{+3.06}$
          & $6.86_{-1.69}^{+2.01}$
          & $4.17_{-1.18}^{+1.12}$
          & $2.59_{-1.10}^{+0.93}$
          & $1.16_{-0.71}^{+0.77}$
          & $9.79_{-2.82}^{+4.12}$
          & $9.39_{-2.65}^{+3.78}$ \\
ZwCl0949.6+5207 &   $5.21_{-1.36}^{+1.74}$
          & $4.61_{-1.11}^{+1.34}$
          & $3.61_{-0.74}^{+0.81}$
          & $2.90_{-0.53}^{+0.56}$
          & $2.05_{-0.53}^{+0.44}$
          & $5.67_{-1.56}^{+2.08}$
          & $5.55_{-1.51}^{+1.99}$ \\
ABELL0901 &   $3.23_{-0.98}^{+1.27}$
          & $2.75_{-0.78}^{+0.96}$
          & $2.03_{-0.51}^{+0.58}$
          & $1.54_{-0.37}^{+0.40}$
          & $0.98_{-0.31}^{+0.29}$
          & $3.61_{-1.15}^{+1.55}$
          & $3.52_{-1.11}^{+1.48}$ \\
ABELL0907 &   $19.79_{-4.77}^{+8.14}$
          & $14.93_{-3.07}^{+4.69}$
          & $8.47_{-1.21}^{+1.41}$
          & $4.92_{-0.78}^{+0.75}$
          & $1.96_{-0.62}^{+0.63}$
          & $24.02_{-6.36}^{+11.55}$
          & $22.99_{-5.96}^{+10.69}$ \\
ABELL0963 &   $8.94_{-1.59}^{+1.90}$
          & $7.42_{-1.22}^{+1.40}$
          & $5.08_{-0.73}^{+0.79}$
          & $3.57_{-0.53}^{+0.54}$
          & $1.97_{-0.42}^{+0.41}$
          & $10.15_{-1.90}^{+2.34}$
          & $9.84_{-1.82}^{+2.22}$ \\
ZwCl1021.0+0426 &   $6.35_{-1.24}^{+1.45}$
          & $5.45_{-0.98}^{+1.10}$
          & $3.88_{-0.58}^{+0.62}$
          & $2.84_{-0.42}^{+0.43}$
          & $1.69_{-0.37}^{+0.34}$
          & $7.02_{-1.44}^{+1.73}$
          & $6.81_{-1.38}^{+1.65}$ \\
ABELL1423 &   $5.24_{-1.28}^{+1.68}$
          & $4.47_{-0.99}^{+1.21}$
          & $3.24_{-0.58}^{+0.63}$
          & $2.41_{-0.43}^{+0.43}$
          & $1.47_{-0.45}^{+0.37}$
          & $5.85_{-1.51}^{+2.10}$
          & $5.69_{-1.45}^{+1.99}$ \\
ABELL1451 &   $10.69_{-1.44}^{+1.66}$
          & $8.89_{-1.10}^{+1.23}$
          & $6.14_{-0.64}^{+0.68}$
          & $4.34_{-0.42}^{+0.44}$
          & $2.44_{-0.31}^{+0.30}$
          & $12.13_{-1.73}^{+2.03}$
          & $11.76_{-1.66}^{+1.94}$ \\
RXCJ1212.3-1816 &   $2.54_{-1.10}^{+1.50}$
          & $2.20_{-0.92}^{+1.19}$
          & $1.62_{-0.64}^{+0.76}$
          & $1.22_{-0.49}^{+0.54}$
          & $0.76_{-0.40}^{+0.38}$
          & $2.80_{-1.24}^{+1.76}$
          & $2.72_{-1.20}^{+1.68}$ \\
ZwCl1231.4+1007 &   $7.51_{-2.13}^{+2.97}$
          & $5.83_{-1.52}^{+1.86}$
          & $3.39_{-1.03}^{+1.02}$
          & $2.01_{-0.87}^{+0.84}$
          & $0.83_{-0.51}^{+0.66}$
          & $8.89_{-2.70}^{+4.15}$
          & $8.52_{-2.54}^{+3.80}$ \\
ABELL1682 &   $10.78_{-1.60}^{+1.87}$
          & $9.02_{-1.23}^{+1.40}$
          & $6.23_{-0.71}^{+0.76}$
          & $4.41_{-0.45}^{+0.47}$
          & $2.48_{-0.33}^{+0.31}$
          & $12.16_{-1.90}^{+2.28}$
          & $11.79_{-1.82}^{+2.17}$ \\
ABELL1689 &   $12.80_{-1.80}^{+2.09}$
          & $11.39_{-1.48}^{+1.68}$
          & $9.13_{-1.00}^{+1.09}$
          & $7.50_{-0.70}^{+0.73}$
          & $5.49_{-0.47}^{+0.47}$
          & $13.90_{-2.06}^{+2.43}$
          & $13.63_{-1.99}^{+2.34}$ \\
ABELL1758N &   $7.02_{-1.67}^{+2.10}$
          & $6.11_{-1.33}^{+1.55}$
          & $4.50_{-0.98}^{+0.99}$
          & $3.41_{-1.02}^{+0.88}$
          & $2.15_{-1.02}^{+0.91}$
          & $7.71_{-1.94}^{+2.60}$
          & $7.51_{-1.86}^{+2.45}$ \\
ABELL1763 &   $21.55_{-3.78}^{+5.03}$
          & $17.63_{-2.76}^{+3.48}$
          & $11.56_{-1.39}^{+1.55}$
          & $7.76_{-0.84}^{+0.87}$
          & $3.97_{-0.73}^{+0.68}$
          & $24.67_{-4.63}^{+6.38}$
          & $23.83_{-4.40}^{+6.00}$ \\
ABELL1835 &   $11.94_{-2.08}^{+2.46}$
          & $10.48_{-1.66}^{+1.90}$
          & $7.98_{-1.02}^{+1.11}$
          & $6.24_{-0.73}^{+0.75}$
          & $4.18_{-0.68}^{+0.64}$
          & $13.04_{-2.41}^{+2.91}$
          & $12.73_{-2.32}^{+2.78}$ \\
ABELL1914 &   $11.51_{-2.27}^{+2.91}$
          & $9.12_{-1.63}^{+1.96}$
          & $5.73_{-0.89}^{+0.96}$
          & $3.69_{-0.64}^{+0.64}$
          & $1.76_{-0.48}^{+0.47}$
          & $13.51_{-2.84}^{+3.82}$
          & $13.02_{-2.70}^{+3.59}$ \\
ZwCl1454.8+2233 &   $5.85_{-2.34}^{+4.59}$
          & $4.62_{-1.66}^{+2.53}$
          & $2.75_{-1.01}^{+1.04}$
          & $1.67_{-0.92}^{+0.79}$
          & $0.72_{-0.67}^{+0.68}$
          & $6.83_{-2.92}^{+6.77}$
          & $6.55_{-2.75}^{+6.10}$ \\
ABELL2009 &   $10.73_{-3.22}^{+5.81}$
          & $8.13_{-2.09}^{+3.26}$
          & $4.70_{-0.89}^{+1.04}$
          & $2.78_{-0.69}^{+0.65}$
          & $1.14_{-0.50}^{+0.57}$
          & $13.01_{-4.28}^{+8.42}$
          & $12.46_{-4.02}^{+7.78}$ \\
ZwCl1459.4+4240 &   $3.64_{-1.02}^{+1.21}$
          & $3.29_{-0.89}^{+1.02}$
          & $2.65_{-0.65}^{+0.72}$
          & $2.19_{-0.50}^{+0.54}$
          & $1.62_{-0.38}^{+0.39}$
          & $3.89_{-1.13}^{+1.35}$
          & $3.81_{-1.10}^{+1.31}$ \\
RXCJ1504.1-0248 &   $6.33_{-1.47}^{+1.75}$
          & $5.74_{-1.27}^{+1.48}$
          & $4.75_{-0.94}^{+1.05}$
          & $4.02_{-0.72}^{+0.78}$
          & $3.10_{-0.49}^{+0.51}$
          & $6.78_{-1.59}^{+1.96}$
          & $6.66_{-1.58}^{+1.90}$ \\
ABELL2111 &   $5.99_{-1.95}^{+4.57}$
          & $5.12_{-1.50}^{+2.64}$
          & $3.71_{-0.86}^{+0.99}$
          & $2.75_{-1.00}^{+0.67}$
          & $1.68_{-1.12}^{+0.64}$
          & $6.67_{-2.31}^{+6.36}$
          & $6.49_{-2.22}^{+5.86}$ \\
ABELL2204 &   $11.78_{-2.40}^{+3.09}$
          & $9.95_{-1.87}^{+2.32}$
          & $7.24_{-1.13}^{+1.30}$
          & $5.40_{-0.71}^{+0.77}$
          & $3.33_{-0.46}^{+0.45}$
          & $13.27_{-2.86}^{+3.78}$
          & $12.92_{-2.75}^{+3.62}$ \\
ABELL2219 &   $13.96_{-2.67}^{+3.62}$
          & $10.87_{-1.79}^{+2.20}$
          & $6.36_{-0.91}^{+0.95}$
          & $3.80_{-0.81}^{+0.74}$
          & $1.59_{-0.58}^{+0.61}$
          & $16.50_{-3.45}^{+4.99}$
          & $15.81_{-3.23}^{+4.60}$ \\
RXJ1720.1+2638 &   $6.72_{-1.98}^{+2.88}$
          & $5.45_{-1.48}^{+1.99}$
          & $3.62_{-0.84}^{+0.99}$
          & $2.46_{-0.54}^{+0.59}$
          & $1.28_{-0.40}^{+0.38}$
          & $7.78_{-2.41}^{+3.69}$
          & $7.52_{-2.30}^{+3.49}$ \\
ABELL2261 &   $15.47_{-2.60}^{+3.34}$
          & $12.45_{-1.84}^{+2.22}$
          & $7.87_{-0.86}^{+0.93}$
          & $5.09_{-0.57}^{+0.56}$
          & $2.45_{-0.50}^{+0.47}$
          & $17.89_{-3.25}^{+4.34}$
          & $17.24_{-3.07}^{+4.06}$ \\
RXCJ2102.1-2431 &   $4.32_{-1.07}^{+1.30}$
          & $3.69_{-0.86}^{+1.00}$
          & $2.71_{-0.56}^{+0.62}$
          & $2.03_{-0.41}^{+0.43}$
          & $1.27_{-0.32}^{+0.30}$
          & $4.83_{-1.25}^{+1.56}$
          & $4.70_{-1.20}^{+1.49}$ \\
RXJ2129.6+0005 &   $6.64_{-2.05}^{+3.06}$
          & $4.92_{-1.33}^{+1.68}$
          & $2.55_{-0.85}^{+0.80}$
          & $1.36_{-0.67}^{+0.65}$
          & $0.47_{-0.31}^{+0.46}$
          & $8.10_{-2.74}^{+4.68}$
          & $7.69_{-2.54}^{+4.20}$ \\
ABELL2390 &   $13.12_{-2.17}^{+2.56}$
          & $11.04_{-1.70}^{+1.93}$
          & $7.69_{-1.00}^{+1.08}$
          & $5.49_{-0.65}^{+0.68}$
          & $3.13_{-0.47}^{+0.44}$
          & $14.74_{-2.56}^{+3.09}$
          & $14.30_{-2.45}^{+2.94}$ \\
ABELL2485 &   $7.18_{-1.54}^{+1.93}$
          & $5.96_{-1.15}^{+1.35}$
          & $4.00_{-0.65}^{+0.70}$
          & $2.75_{-0.53}^{+0.51}$
          & $1.47_{-0.48}^{+0.45}$
          & $8.13_{-1.86}^{+2.45}$
          & $7.86_{-1.77}^{+2.30}$ \\
ABELL2537 &   $8.88_{-2.30}^{+2.93}$
          & $7.94_{-1.92}^{+2.34}$
          & $6.23_{-1.29}^{+1.44}$
          & $5.01_{-0.95}^{+1.01}$
          & $3.54_{-0.87}^{+0.78}$
          & $9.56_{-2.59}^{+3.40}$
          & $9.36_{-2.50}^{+3.26}$ \\
ABELL2552 &   $9.52_{-2.80}^{+4.37}$
          & $7.93_{-2.13}^{+2.92}$
          & $5.24_{-1.20}^{+1.34}$
          & $3.54_{-1.05}^{+0.95}$
          & $1.84_{-0.91}^{+0.83}$
          & $10.71_{-3.34}^{+5.63}$
          & $10.34_{-3.17}^{+5.23}$ \\
ABELL2631 &   $9.66_{-2.58}^{+3.82}$
          & $7.46_{-1.70}^{+2.12}$
          & $4.14_{-1.24}^{+1.11}$
          & $2.35_{-1.10}^{+1.05}$
          & $0.90_{-0.59}^{+0.90}$
          & $11.42_{-3.38}^{+5.66}$
          & $10.90_{-3.14}^{+5.06}$ \\
ABELL2645 &   $5.20_{-1.29}^{+1.55}$
          & $4.34_{-1.01}^{+1.16}$
          & $2.94_{-0.65}^{+0.70}$
          & $2.04_{-0.52}^{+0.52}$
          & $1.11_{-0.41}^{+0.41}$
          & $5.87_{-1.52}^{+1.88}$
          & $5.68_{-1.45}^{+1.78}$ \\
\hline
\end{tabular}
\end{center}
\end{table*}